\documentclass[floatfix,aps,pre,preprint,groupedaddress]{revtex4}

\usepackage{graphicx}
\usepackage{enumerate}
\usepackage{paralist}

\usepackage{booktabs}
\usepackage{subfigure}

\usepackage{amsfonts}
\usepackage{amssymb}
\usepackage{amsmath}
\DeclareGraphicsExtensions{.pdf,.png}
\usepackage{tabularx}
\graphicspath{{/}}

\renewcommand{\vec}[1]{\ensuremath\boldsymbol{#1}}
\newcommand{\Pe}{\ensuremath \textnormal{Pe}}	
\newcommand{\Cn}{\ensuremath \textnormal{Cn}}	

\usepackage{multirow}

\begin{document}

\begin{abstract}

	Phase separation and coarsening is a phenomenon commonly seen in binary
	physical and chemical systems that occur in nature. 
	Often times, thermal fluctuations, modeled as stochastic noise, are present in the system and the 
	phase segregation process occurs on a surface.
	In this work, the segregation process is modeled via the Cahn-Hilliard-Cook model,
	which is a fourth-order parabolic stochastic system.
	Coarsening is analyzed on two sample 
	surfaces: a unit sphere and a dumbbell using a 
	variety and a statistical analysis of the growth rate is performed.
	The influence of noise level and mobility is also investigated.
	It is also shown that a log-normal distribution fits the results well.
	
\end{abstract}

\title{Stochastic Phase Segregation on Surfaces}
\author{Prerna Gera}
\affiliation{Department of Mechanical and Aerospace Engineering, University at Buffalo,Buffalo, New York 14260-4400}
\author{David Salac}%
\email[Corresponding author: ]{davidsal@buffalo.edu}
\affiliation{Department of Mechanical and Aerospace Engineering, University at Buffalo,Buffalo, New York 14260-4400}
\date{\today}%
\maketitle

\section{Introduction}

Domains on curved surfaces are found in numerous industrial and biomedical
applications such as chemical reactors~\cite{takagi2011surfactant}, enhanced oil recovery~\cite{morrow2001recovery}, and pulmonary
functions~\cite{goerke1998pulmonary}. These domains have the 
potential to change the dynamics of these system
significantly. For example, surfactants on bubbles or droplets can reduce the velocity of the 
rising bubble~\cite{ceccio2010friction} or they can prevent the 
coalescence of multiple bubbles~\cite{takagi2011surfactant}. 
The effect of surface molecules can also
be seen in the area of biology such as the cell membrane~\cite{laradji2004dynamics}. 
The cell membrane is composed of multiple components
including saturated lipids, unsaturated lipids and cholesterol. The saturated
lipid molecules combine with the cholesterol to form lipid domains, with
these lipid domains being more ordered and stable than the surrounding
membrane~\cite{shaw2006lipid}. Due to the nature of the domains, experimental
visualization is difficult with artifacts and errors influencing the
accuracy of the experimental results. Using numerical tools and mathematical
modeling to investigate the dynamics of surface domains can provide important
information not obtainable experimentally. In addition to biological membranes,
other interesting phase dynamics on a curved surface includes crystal
growth~\cite{peczak1993monte}, phase separation within thin
films~\cite{sens2000inclusions}, and phase separation patterns in diblock
polymers~\cite{tang2005phase}. With this motivation, the goal of this work is to
study the phase segregation dynamics on a smooth curved surface. 

The Cahn-Hilliard (CH) equation is a popular model to capture and investigate phase segregation
dynamics in a multi-component system. 
It describes the temporal evolution of an order parameter
that defines the phase or domain, with the driving force given by
energy minimization under the assumption of quantity conservation. Pioneered by
Cahn \& Hilliard in 1958~\cite{cahn1958free}, the equation has been used to model many physical systems
including binary alloys~\cite{cahn1958free}, polymers and ceramics~\cite{cogswell2010phase}, droplet
breakup~\cite{jacqmin1996energy}, liquid-liquid jets pinching
off~\cite{longmire1999comparison}, multicomponent lipid
vesicles~\cite{funkhouser2014dynamics}, and for the tracking of tumor
growth~\cite{wise2008three}.

In 1970 Cook proposed to make the system more realistic by including internal
thermal fluctuations, which are represented by a conserved noise source term.
This extension is more commonly known as the Cahn-Hilliard-Cook (CHC)
model~\cite{COOK1970}.  The first numerical work to study the CHC equation
was done by Langer~\cite{langer1975new}. This work was compared against
theoretical results and it was concluded that the thermal fluctuations play an
important role in the early stage of phase dynamics~\cite{langer1975new}. To
understand the early stages better, Grant et~al. developed a perturbation theory
for a long range force limit and utilized numerical methods to investigate the model~\cite{grant1985theory}. 
Rodgers et~al. studied convergence of the solution, the growth characteristics of the domain
formation and the effect of noise~\cite{rogers1988numerical}. The impact of noise in the CHC has also been
analyzed by several other works, see Refs. ~\cite{ibanes2000dynamics,garcia1998phase,zheng2015parallel} for examples.
Using this
model, the influence of non-equilibrium lipid transport on a membrane~\cite{fan2010influence}, dendritic
branching~\cite{karma1999phase}, nucleation in
micro-structures~\cite{shen2007effect}, and the dynamics of solvent based organic
cells~\cite{wodo2012modeling}, have also been investigated.

The Cahn-Hilliard-Cook model can be solved computationally
using a variety of numerical methods including finite
differences~\cite{saylor2007diffuse,furihata2001stable,choo1998conservative},
finite
elements~\cite{elliott1989second,elliott1989nonconforming,zhang2010nonconforming},
and spectral methods~\cite{chen1998applications,he2007large}.
Most investigations using the CHC model are two-dimensional, although there are
several which investigate three-dimensional systems~\cite{zheng2015parallel}.
In this work a method to model the Cahn-Hilliard-Cook system on an arbitrary two-dimensional surface
in three-dimensional space is presented. Prior works in modeling the phase-segregation on
curved surfaces have been performed at the nanometer level, including 
those based on Molecular Dynamics (MD)~\cite{de2004molecular,shinoda2010zwitterionic,marrink2003molecular},
where atomistic level forces can be incorporated. A limit of these MD simulations 
is the length and time scales which can be investigated. 
To allow for longer time-scales and larger domain sizes to be investigated,
coarse grained methods such as Dissipative Particle Dynamics (DPD) have been used~\cite{bagatolli2009phase}.
An alternate technique is to examine phase segregation on surfaces using a continuum-based method,
which is the approach taken here. Samples of this type of work in the absence of noise
have been presented in the
past~\cite{marenduzzo2013phase,funkhouser2014dynamics,greer2006fourth,wodo2011computationally}.

The work here is based on a splitting method previously used to model the
Cahn-Hilliard equation~\cite{lowengrub2007surface,Li2012}. 
The coarsening rates for the CH and CHC systems are compared using both constant and variable
mobility, in addition to varying noise levels. While in an actual system
thermal fluctuations may also influence the shape of the underlying interface,
this is not considered here.

In the following section of the paper the governing equations of the system is described. 
The numerical tools and techniques
that are used and the overall algorithm is then explained.
Phase segregation on
a sphere is examined and systematically investigated. A statistical
analysis on growth rate of the domains that appear on the sphere is done.
Further, the effect of the underlying geometry is also shown as the statistical
analysis of the growth rate on a dumbbell is presented.

\section{\label{sec:GovEq} Mathematical Formulation}

In this work, the dynamics of phase separation and coarsening of a two-component system 
on a surface is described by the Cahn-Hilliard
equation. This equation captures how a system will change over time to reduce the 
overall free energy of a multicomponent system. Further, the segregation process
is restricted to a co-dimension one interface and thus will involve surface derivatives.
Let $\Gamma(\vec{x},t)$ define the interface at any point in time. The concentration field
$f(\vec{x},t)$ is defined on $\Gamma(\vec{x},t)$ such that $0\leq f(\vec{x},t)\leq 1$ is the concentration of one surface phase
while the concentration of the remaining surface phase is $1-f(\vec{x},t)$.
Using this, the CH equation is derived from the mass-continuity equation, 
\begin{equation} 
	\frac{\partial f}{\partial t} +\nabla_s \cdot \vec{J}_s=0, 
\end{equation} 
where $\vec{J}_s$ is the surface
flux and $\nabla_s \cdot$ is the
surface divergence while 
$\nabla_s=\vec{P}\nabla$ is the surface gradient \cite{jeong2015microphase}  and $\vec{P}=\vec{I}-\vec{n}\otimes\vec{n}$ is the Laplace-Beltrami operator 
for an outward pointing unit normal to the interface $\vec{n}$.
Using Fick's law the surface flux is related to the chemical potential $\mu$ via the surface gradient and a mobility $\nu(f)$\cite{Li2012},
\begin{equation} 
	\vec{J}_s=-\nu(f)\nabla_s \mu.
\end{equation} 
From these two equations, 
\begin{equation} 
	\frac{\partial f}{\partial t} - \nabla_s \cdot (\nu(f)\nabla_s \mu) = 0. \label{CH}
\end{equation} 

Two types of mobility are considered in this work. The first type
is constant mobility,
\begin{equation}
	\nu(f)=\nu_0,
	\label{eq:constantMobility}
\end{equation}
where $\nu_0$ is the surface mobility
constant. This type of mobility is appropriate when molecules can freely
move through bulk phases.
The second type is more appropriate for situations where the majority 
of molecular motion occurs at the interface between phases.
In this case a degenerate mobility is defined as
\begin{equation} 
	\nu(f)=4\nu_0 f(1-f).
	\label{eq:variableMobility}
\end{equation} 
It should be noted that for the variable mobility case, the overall mobility
in the bulk is very small,
which is often the case in realistic 
system~\cite{cahn1961spinodal,lacasta1993front}. It is also possible to utilize
other mobilities, as has been recently investigated~\cite{Dai2016}.

The chemical potential $\mu$ can be derived by applying the variational derivative to
the free energy of the surface phase field~\cite{gomez2008isogeometric}. Intuitively, it is
expected that the local free energy will depend on the homogeneous free energy
and the energy due to the interface separating the phases~\cite{cahn1958free}. This energy
functional can thus be written as,
\begin{equation}
	E[f] = \int_{\Gamma(\vec{x},t)} \left(g(f)+ \frac{\epsilon^2}{2}(\nabla_s f)^2 \right)\;dA.
	\label{eq:freeEnergy}
\end{equation}
The first term is the free energy of
the homogeneous solution, while the second term is the interfacial energy,
defined using the surface gradient of the concentration field, 
where $\epsilon$ is a constant associated with the domain interface energy. 
This form of energy functional is also known as the Landau-Ginzberg free energy
functional~\cite{wodo2011computationally}. Taking
the variational derivative of the free energy functional with respect
to a change in the concentration variable results in the chemical potential
field~\cite{Lowengrub2009,elliott1989second},
\begin{equation}
	\mu = g'(f)-\epsilon^2\Delta_s f,
	\label{eq:chemicalPotential}
\end{equation}
where $\Delta_s=\nabla_s\cdot\nabla_s$ is the surface Laplacian and $g'(f)$ is the derivative
of the mixing energy with respect to argument $f$.

\begin{figure}
	\centering
	\includegraphics[width=8cm]{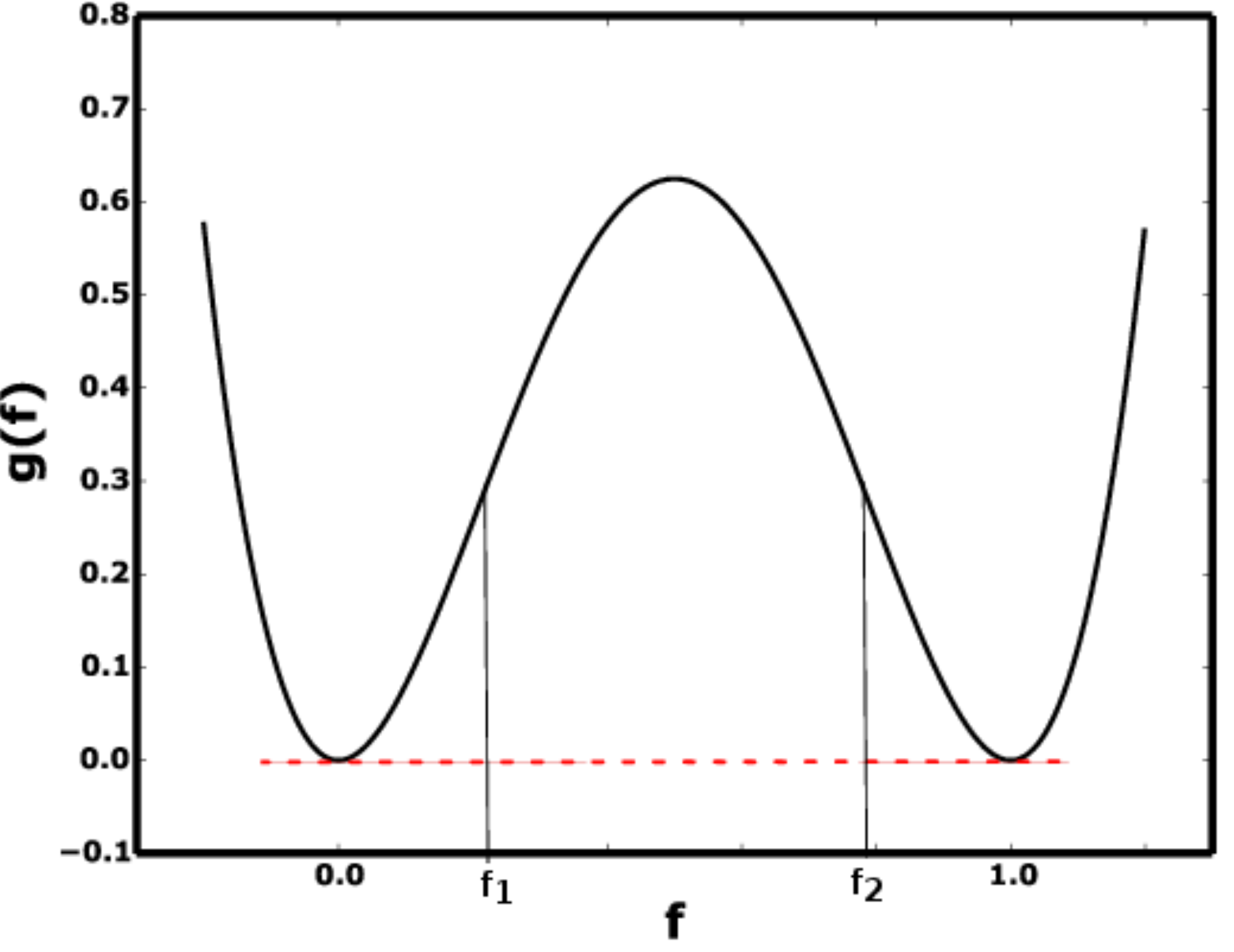}	
	\caption{Homogeneous free energy of mixing}
	\label{fig:doubleWell}
\end{figure}

The homogeneous free
energy is usually described using a double well potential.
In this work a simple mixing energy of 
\begin{equation}
	g(f)=f^2(1-f)^2,
\end{equation}
as shown 
in Fig \ref{fig:doubleWell}, is used.
The points $f_1$ and $f_2$ are called the spinodes, and are defined by
$\partial^2g/\partial^2 f=0$. 
The region between by $f_1$ and $f_2$, given by $\partial^2g/\partial f^2<0$,
is known as the
spinodal region, where a single phase decomposes into two phases.
In this simple mixing energy the equilibrium concentration of the two phases
are defined by the two wells at concentrations $f=0$ and $f=1$.

To define the Cahn-Hilliard-Cook equation, a white Gaussian 
noise~\cite{fan2010influence,rogers1988numerical} is added to the deterministic
Cahn-Hilliard equation,
\begin{align}
	\frac{\partial f(\vec{x},t)}{\partial t} &= \nabla_s \cdot(\nu\nabla_s\mu)+\xi(\vec{x},t)
	\label{eq:CHC}
\end{align} 
where $\xi$ denotes a stochastic Gaussian white noise
dictated by the Fluctuation
Dissipation Theorem with mean
$\langle\xi\rangle=0$ and variance
$\langle\xi(\vec{x},t),\xi(\vec{x}',t')\rangle= -2\nu k_{B}T
\delta(t-t')\Delta_s \delta(\vec{x}-\vec{x}')$, 
where $\delta$ is the Dirac delta function~\cite{fan2010influence,zheng2015parallel}. The Fluctuation
Dissipation Theorem implies that the random
noise that is added is uncorrelated in time but partially correlated, specifically conserved, in space.
The Laplacian in front of the Dirac delta function appears as this Langevin
force term is present in a system that is characterized by
conserved fields. These kinds of stochastic fluctuating partial differential equations are currently under 
investigation \cite{delong2013temporal}.

\subsection{Non-Dimensional System}
The dimensionless CHC equation that governs the evolution of domains is
\begin{equation} \frac{\partial f}{\partial
	\hat{t}}=\frac{1}{\Pe}\hat{\nabla}_s\cdot\left(\hat{\nu} \hat{\nabla}_s
	\hat{\mu}\right)+\hat{\xi},
	\label{eq:diffusion} 
\end{equation}
where the dimensionless units are represented by $ \hat{(\cdot)}$ and $\Pe$ is the
surface Peclet number, which relates the strength of any surface advection to diffusion. 
The dimensionless parameters are defined as follows.
\begin{align}
	\hat{t} = \frac{t}{t_0},\;\hat{\nu}=\frac{\nu}{\nu_0},\;\hat{\mu}=\frac{\mu}{\mu_0},\nonumber\\
	\hat{\nabla}_s=l_0\nabla_s,\; \nonumber\\
	\Pe=\frac{l_0^2}{t_0\mu_0\nu_0},\;\Cn^2=\frac{\epsilon^2}{\mu l_0^2}\nonumber \\
	\langle\hat{\xi}(\vec{x},t),\hat{\xi}(\vec{x}',t')\rangle=-\sigma \hat{\nu}
	\delta(t-t')\hat{\Delta}_s^2\delta(\vec{x}-\vec{x}'),
\end{align}
where $l_0$ is the characteristic length, $t_0$ is the characteristic time, 
$\mu_0$ is characteristic chemical potential, 
$\nu_0$ is characteristic mobility, $\Cn^2$, 
relates the ratio of the domain interface energy to the chemical potential were $\Cn$ is called the Cahn number, and $\sigma$ is the noise intensity defined
as $\sigma=\left(2k_BT/\mu_0\right)^2$. Using this notation
the dimensionless mobility is now $\hat{\nu}=1$ for the constant mobility case and 
$\hat{\nu}=4f(1-f)$ for the variable mobility case. The dimensionless chemical potential equation is,
\begin{equation}
	\hat{\mu} = \frac{\partial \hat{g}}{\partial f}-{\Cn}^2\hat{\Delta}_s f.
	\label{eq:chempot}
\end{equation}
 Using Eq.~\eqref{eq:diffusion}
and Eq.~\eqref{eq:chempot}, and dropping the $\hat{(\cdot)}$ notation the following fourth-order evolution equation for $f(\vec{x},t)$ is obtained:
\begin{equation}
	\dfrac{\partial f}{\partial t} + \frac{\Cn^2}{\Pe}\nabla_s\cdot\left(\nu\nabla_s\Delta_s f\right) =
	\frac{1}{\Pe}\nabla_s\cdot\left(\nu\nabla_s g'(f) \right)+\xi.
	\label{CHC}
\end{equation}

\section{\label{sec:NumericalMethod} Numerical Method}

In this section we discuss the numerical methods used to model phase dynamics on
a curved surface. The interface is described using a level set Jet method~\cite{seibold2012jet,Velmurugan2015}.
To solve the surface evolution
equation, we use the Closest Point Method,
which is described in section \ref{sec:PhaseField}. We discretize the system
using second-order, centered finite differencing techniques while a
semi-implicit time discretization is employed.

\subsection{Defining the Curved Surface Using Level-Sets}
The level-set method is a tool to define and track an interface.
Introduced by Osher and Sethian~\cite{osher1988fronts}, this method has been
used in a variety of applications including medical
imaging~\cite{hogea2006simulating}, crystal
growth~\cite{sethian1992crystal}, crack patterns~\cite{Salac2007}, and
semiconductor processing~\cite{adalsteinsson1997level}. The idea is to define
the interface implicitly through the use of an auxiliary mathematical function,
akin to density, which allows for complex motion and topological changes such as
merging and pinching. For details, readers can refer to Osher and Fedkiw~\cite{osher2001level} 
or Sethian and Smereka~\cite{sethian2003level}.

\begin{figure}
	\centering
	\includegraphics[width=5cm]{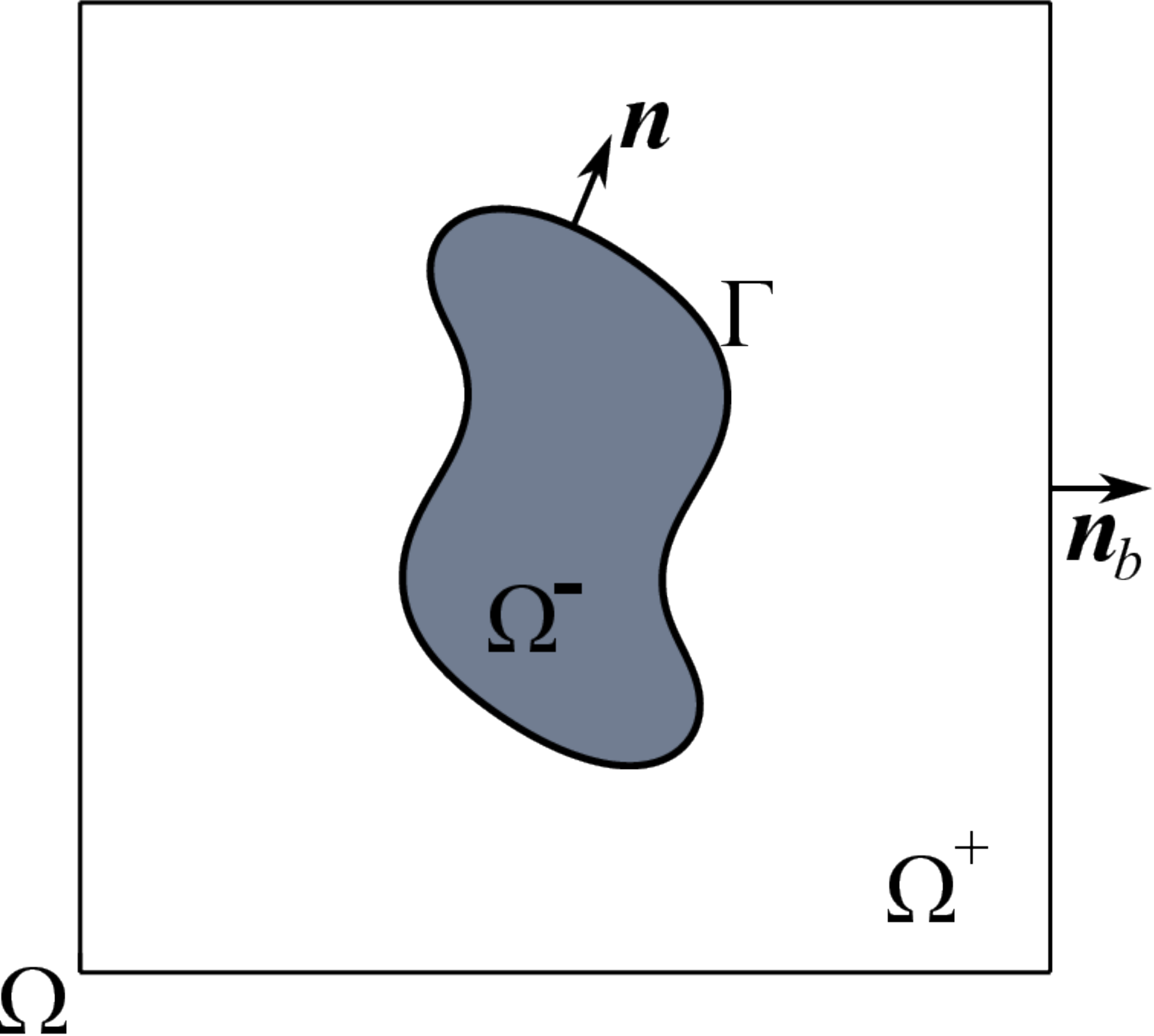}
	\caption{The computational domain.} 
	\label{fig:compDomain}
\end{figure}

Let $\Gamma(\vec{x},t)$ be the interface separating regions $\Omega^-$ and $\Omega^+$.
This interface is represented by the zero set of a higher dimensional level-set function
$\phi(\vec{x},t)$, 
\begin{equation} 
	\Gamma(\vec{x},t) = \{\vec{x}:\phi(\vec{x},t)=0\},
\end{equation}
see Fig.~\ref{fig:compDomain}. The region $\Omega^-$ is given by $\phi(\vec{x},t)<0$ while
$\Omega^+$ is defined as the region occupying $\phi(\vec{x},t)>0$.

A significant advantage of this implicit representation is that many
geometric quantities can easily be computed. For example, the normal and
total curvature (sum of the principle curvatures) of the interface can be defined as, 
\begin{equation}
	\vec{n}=\frac{\nabla \phi}{||\nabla \phi||}, \quad H= \nabla \cdot \frac{\nabla
	\phi}{||\nabla \phi||}.  
\end{equation} 


To model a surface differential equation accurate information must
exist about the location of the interface. As the interface will, in general,
not coincide with grid points interpolation schemes must be used to determine
the location of the interface. To aid in this, an extension of the base
level set method is used. The idea is to track not only a single level set
function $\phi$, but also derivatives of the level set. A grouping of this
information has been called a ``jet" of information~\cite{seibold2012jet}. Using
this jet, it is possible to define high-order interpolants without the need for
derivative approximations.  For example, using a jet which consists of the level
set function, $\phi$, in addition to gradient vector field, $\phi_x$ and
$\phi_y$, and the first cross-derivative, $\phi_{xy}$, it would be possible to
define a cubic Hermite interpolant on a two-dimensional Cartesian grid without
the need for derivative approximations. Additional information about the Jet
level-set method can be found in the work of Seibold, Rosales, and Nave~\cite{seibold2012jet}.

\subsection{Phase Field Solver\label{sec:PhaseField}} 

The Cahn-Hilliard-Cook system, Eq. \eqref{CHC}, can be written as a pair of coupled, second-order differential
equations~\cite{lowengrub2007surface,Li2012}, 
\begin{align}
	\frac{\partial f}{\partial t} - \frac{1}{\Pe}\nabla_s \cdot(\nu \nabla_s \mu) -\xi = 0 \quad \textnormal{and} \quad \mu+\Cn^2\Delta_s f =g'(f).
\end{align}
Here, a second-order backward-finite-difference (BDF2) scheme \cite{Fornberg1988} is used to discretize
in time. The system can then be written as
\begin{equation}
	\begin{bmatrix}
		\vec{I} & \Cn^2\vec{L}_s\\
		-\frac{2\Delta t}{3\Pe}\vec{L}^{\nu}_s & \vec{I}
	\end{bmatrix}
	\begin{bmatrix}
		\vec{\mu}^{n+1}\\
		\vec{f}^{n+1}
	\end{bmatrix}
	=
	\begin{bmatrix}
		g'(\hat{\vec{f}})\\
		\frac{4}{3} \vec{f}^n -\frac{1}{3} \vec{f}^{n-1}+\frac{2}{3}\Delta t\xi^n
	\end{bmatrix},
	\label{eq:CHC_block}
\end{equation}
where $\Delta t$ is a fixed time step.
In the above block system $\vec{I}$ is the identity matrix, the
constant-coefficient surface Laplacian is given by $\vec{L}_s\approx\Delta_s$, and the 
variable-coefficient surface Laplacian is given by $\vec{L}_s^{\nu}\approx\nabla_s\cdot(\nu\nabla_s)$.
The solutions $\vec{f}^n$ and $\vec{f}^{n-1}$ are at times
$t^n$ and $t^{n-1}$, respectively, and the approximation to the solution at time $t^{n+1}$ is given by $\hat{\vec{f}}=2\vec{f}^n-\vec{f}^{n-1}$.

As this is a surface partial differential equation, specialized methods
are required to evolve it properly. In this work the Closest Point Method is
used.
The Closest Point Method was first developed and analyzed by Ruuth and Merriman~\cite{ruuth2008simple} and has been modified to
increase numerical stability and accuracy~\cite{macdonald2009implicit}.  The basic idea is to extend the solution
to a surface differential equation away from the interface such that it is
constant in the normal direction. With this extension, it is possible to write a
surface differential equation as a standard differential equation in the
embedding space. It has been previously shown that the surface Laplacian
operator can be computed with second order accuracy using linear and cubic
polynomial interpolations\cite{doi:10.1137/130929497}.

Let $\vec{E}_1$ be a linear polynomial interpolation operator and $\vec{E}_3$ be a cubic polynomial interpolation operator.
For any point $\vec{x}$ not on the interface these operators return
the value of a function at the interface point closest to $\vec{x}$. For example, the operation $\vec{E}_3\vec{f}$ 
returns the value of $f$ at the point on the interface closest to $\vec{x}$
using the cubic interpolation function. Using this notation, the block matrix in
Eq. (\ref{eq:CHC_block}) is re-written as
\begin{align}
	\begin{bmatrix}
		\vec{I} & \Cn^2\left[\vec{E}_1\vec{L}+\alpha\left(\vec{E}_3-\vec{I}\right)\right]\\
		-\frac{2\Delta t}{3\Pe}\left[\vec{E}_1\vec{L}^{\nu}+\alpha\left(\vec{E}_3-\vec{I}\right)\right] & \vec{I}
	\end{bmatrix}
	&
	\begin{bmatrix}
		\vec{\mu}^{n+1}\\
		\vec{f}^{n+1}
	\end{bmatrix} \nonumber \\
	&=
	\begin{bmatrix}
		g'(\hat{\vec{f}})\\
		\frac{4}{3} \vec{f}^n -\frac{1}{3} \vec{f}^{n-1}+\frac{2}{3}\Delta t\tilde{\xi}^n
	\end{bmatrix},
	\label{eq:CP_CHC_block}
\end{align}
with $\alpha=6/h^2$ where $h$ is the uniform grid spacing and $\vec{L}\approx\Delta$ represents
the Cartesian finite difference approximation to the constant standard Laplacian and 
$\vec{L}^{\nu}\approx\nabla\cdot(\tilde{\nu}\nabla)$ represents the
Cartesian finite difference approximation to the variable-coefficient Laplacian.
Quantities denoted with $(\tilde{\cdot})$ indicate that the value has been extended off the interface.
The addition of the $\alpha$ term, also known as a side condition, ensures that the solutions
are constant in the normal direction. If this extension holds then surface operators can be replaced
with standard Cartesian operators. See Chen and Macdonald for complete
details~\cite{doi:10.1137/130929497}. 

The block system shown in Eq. (\ref{eq:CP_CHC_block}) is solved using the preconditioned 
Flexible GMRES algorithm available in PETSc~\cite{petsc-web-page,petsc-user-ref,petsc-efficient}.
The preconditioner is based on an incomplete Schur complement.
Let $\vec{L}_E=\vec{E}_1\vec{L}+\alpha\left(\vec{E}_3-\vec{I}\right)$
and $\vec{L}^\nu_E=\vec{E}_1\vec{L}^\nu+\alpha\left(\vec{E}_3-\vec{I}\right)$. The preconditioner is then
\begin{equation}
	\vec{P}=
	\begin{bmatrix}
		\vec{I} & - \Cn^2\vec{L}_E\\
		0 & \vec{I}
	\end{bmatrix}
	\begin{bmatrix}
		\vec{I}& 0\\
		0 & \vec{\hat{S}}^{-1}
	\end{bmatrix}
	\begin{bmatrix}
		\vec{I} & 0 \\
		\frac{2\Delta t}{3\Pe }\vec{L}^\nu_E & \vec{I}
	\end{bmatrix}.
\end{equation}
The Schur complement is written as $\vec{S}=\vec{I}+\frac{2\Cn^2\Delta t}{3\Pe}\vec{L}_E\vec{L}^\nu_E$. 
The application of the approximate Schur complement inverse, $\vec{\hat{S}}^{-1}$, is obtained via 5 iterations 
of an algebraic multigrid preconditioning method~\cite{ml-guide}.

\subsection{Noise Calculation}

The noise term is calculated based on the Fluctuation Dissipation Theorem as
follows,
\begin{equation}
	\xi(\vec{x},t) =\mathcal{N}(0,-\sigma \nu \delta(t -t')\Delta_s \delta(\vec{x}-\vec{x}')).
\end{equation}
Writing the mean and the variance in discrete form,
\begin{align}
	\langle \xi^n_{\vec{x}_\Gamma}\rangle &=0, \\
	\langle \xi^n_{\vec{x}_\Gamma} \xi^{n+1}_{\vec{y}_\Gamma}\rangle&= -\sigma \nu(f^n_{\vec{x}_\Gamma}) \frac{\delta_{(n)(n+1)}}{|t^n-t^{n+1}|}\Delta_s^h \frac{\delta_{(\vec{x}_\Gamma)(\vec{y}_\Gamma)}}{h^2},
\end{align}
where $\vec{x}_\Gamma$ and $\vec{y}_\Gamma$ are two different points on the interface and $n$ defines the time
step~\cite{zheng2015parallel}. As we
are considering a two-dimensional surface, the grid size $h$ is raise to the
second power~\cite{shen2007effect}. The proof of the discretization in the above
can be found in reference~\cite{lloyd1982implementation}.

To compute the random forcing term, $\tilde{\xi}$, the following procedure is used.  To
ensure consistency of the scheme this forcing term must be constant in the
normal direction. This can be accomplished by computing the random force
contribution at the closest point of any grid point and extending this quantity
outwards.  At a closest point, a random tangential vector is determined by
choosing two random numbers, $\rho_1$ and $\rho_2$, from a Gaussian distribution with zero mean and unit
variance. A random surface vector is then determined by
$\vec{\rho}=\rho_1\vec{t}+\rho_2\vec{b}$, where $\vec{t}$ and $\vec{b}$ are two
orthonormal vectors on the surface, such as the principle directions.  Once
these tangential random vectors are calculated in a region around the interface,
it is possible to define the random force through
\begin{equation}
	\tilde{\xi}=\sqrt{\frac{\sigma \nu(f)}{h^2\Delta t}}\nabla_s\cdot\tilde{\vec{\rho}},
	\label{eq:discreteRandomForce}
\end{equation}
where a constant time step, $\Delta t$, is assumed.
Note that as the surface Laplacian is approximated numerically, it may fail to preserve the fluctuation dissipation
balance in the exact sense.

\subsection{Conservation of Surface Phase Concentration}
After solving the system of the partial differential equation,
Eq. (\ref{eq:CHC_block}), there will be certain amount of loss of surface phase
concentration due to numerical diffusion. The accumulative
effect may have a drastic change on the average surface concentration over time.
There have been numerous attempts to fix this issue in the
past, see Refs. \cite{enright2002hybrid,sussman1998improved,sussman2000coupled} for examples.
In this work a correction method is implemented. This method was introduced by Xu et al~\cite{xu2006level},
with the idea of adjusting the surface phase concentration 
at the end of every time step to ensure mass conservation.
Let $f_h$, $\phi$, and $\Gamma$ be the solution of the discrete surface
phase concentration equation Eq. \eqref{eq:CP_CHC_block}, level set and 
interface at a given point in time, and let $f_0$, $\phi_0$, and $\Gamma_0$ be
the initial phase concentration, initial level set function and initial
interface, respectively. Then a surface phase concentration conservation
parameter, $\beta$, is chosen such that the following condition is true,
\begin{equation}
	\int_{\Gamma}\beta f_h\;dA =\int_{\Gamma_0}f_0\;dA.
\end{equation}
Hence, $\beta$ is computed as
\begin{equation}
\beta = \frac{ \int_{\Gamma_0}f_0\;dA}{ \int_{\Gamma} f_h\;dA}=
\frac{ \int_{\Omega}f_0\delta(\phi_0)\;dV}{
    \int_{\Omega}f_h\delta(\phi)\;dV}
\end{equation}
where $\delta$ is the Dirac delta function and the integrals
are now performed over the embedding domain. The surface
phase concentration is then modified at each time step as $f=\beta f_h$.
For further details, we refer the reader to Xu et al~\cite{xu2006level}.

\section{\label{sec:Results} Results on a Sphere}

In this section qualitative and quantitative results are presented
using the method described in previous sections. In this section
the surface will be a unit sphere. First, the sample evolution of
phase dynamics is examined. Following that, a
quantitative analysis on the domain dynamics is performed.
This includes a convergence
study to justify the grid size and time step used for the analysis. The
growth rate of the domains is examined, and in particular the impact of
variable and constant mobility in the system is considered. Finally, the role of
noise in the system is investigated.

For simplicity, the shape considered 
is a unit sphere in a computational domain spanning $[-1.25,1.25]^3$. 
Unless otherwise stated, the average concentration is set to $0.3$, with an initial random perturbation
of 0.01. The Peclet number is set to $\Pe=1.0$, the 
Cahn number is $\Cn=0.015$,
and when noise is present, has intensity of $\sigma=10^{-5}$. See
Fig.~\ref{fig:sampleEvolSphere} for
a sample evolution.

Surface dynamics will be quantified by a characteristic length,
$\bar{R}(t)$, for the domains present on the surface. 
This surface characteristic length is defined as
\begin{equation}
	\bar{R}(t) =\frac{A(t)}{L(t)}\label{eq:CL},
\end{equation}
where $A(t)=\int_\Gamma f\;dA$ is the total area of the domains and $L(t)=\int_\Gamma \|\nabla_s f\|\;dA$ is the corresponding
total interface length of the domains. 

\subsection{Sample Evolution of Phase Dynamics}

In this section, the dynamics on a smooth spherical surface is
examined for four cases:
\begin{inparaenum}[a)]
	\item Cahn-Hilliard with constant mobility,
	\item Cahn-Hilliard with variable mobility,
	\item Cahn-Hilliard-Cook with constant mobility,
	\item and Cahn-Hilliard-Cook with variable mobility.
\end{inparaenum}
In all four cases, the initial condition is a random perturbation with a
magnitude of $0.01$ about the average concentration of $0.3$. There are three
expected regimes. Initially, very rapid phase segregation will occur and a large
number of domains will appear. This will be followed by slow coarsening of the
domains, which results in an increasing average domain size. The final regime
will be characterized by a very slow coarsening process. In the simulations
performed in this work, approximately 400 domains are seen during the initial
phase segregation process. These domains coarsen in time, and only 5 to 6
domains remain in the final slow coarsening stage.
See Fig.~(\ref{fig:sampleEvolDumbbell}) for a sample
evolution. 

\begin{figure}
	\centering
	\begin{tabular}{>{\centering}m{0.8cm} >{\centering}m{2.5cm}
    >{\centering}m{2.5cm} >{\centering}m{2.5cm} >{\centering}m{2.5cm}}
    & \multicolumn{2}{c}{$\sigma=0$}     & \multicolumn{2}{c}{$\sigma=10^{-5}$} \tabularnewline
		\cmidrule(l){2-3} \cmidrule(l){4-5}
		$t$ & $\nu=1$ & $\nu=4f(f-1)$ & $\nu=1$ & $\nu=4f(f-1)$\tabularnewline
		0.1 & \includegraphics[width=2.2cm]{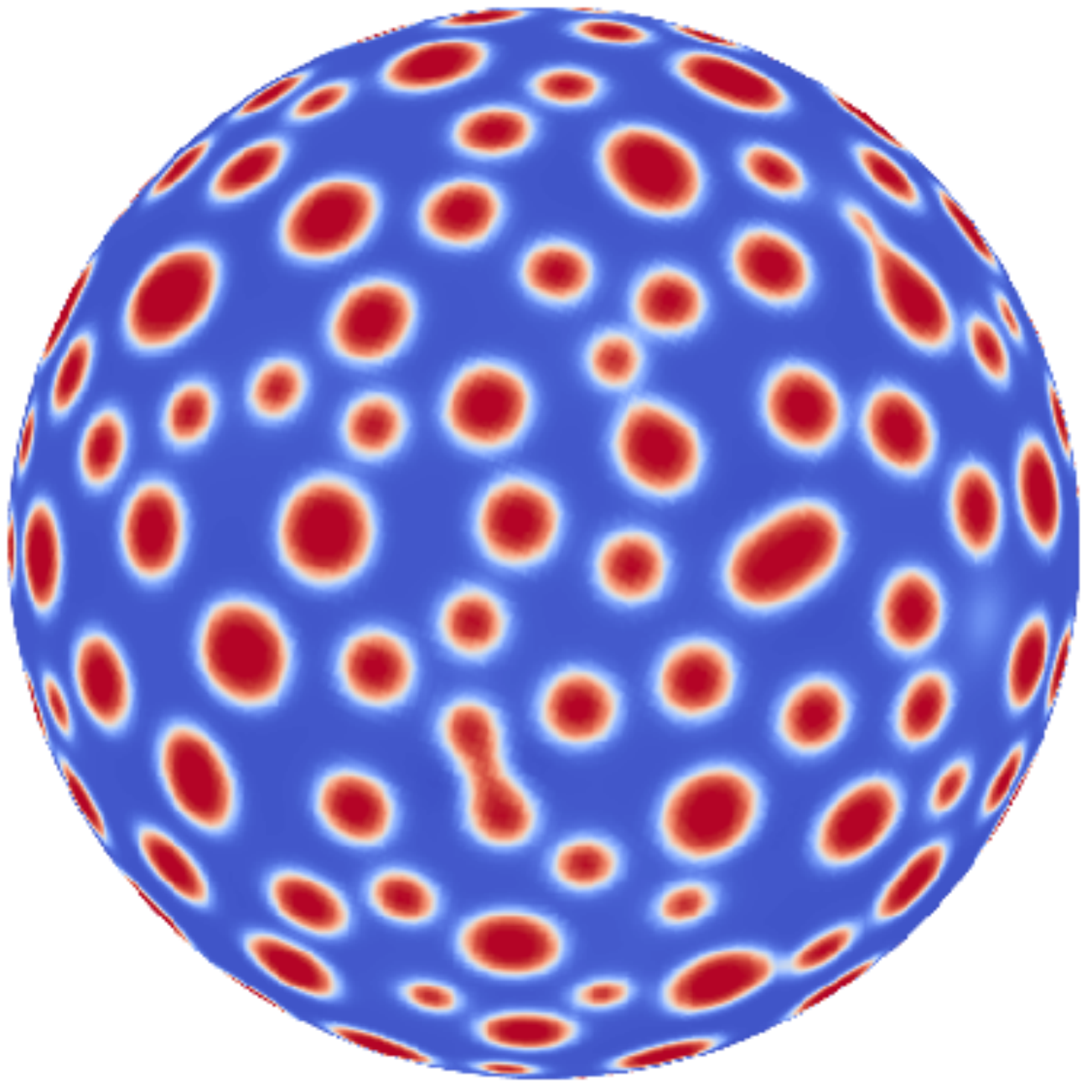} &\includegraphics[width=2.2cm]{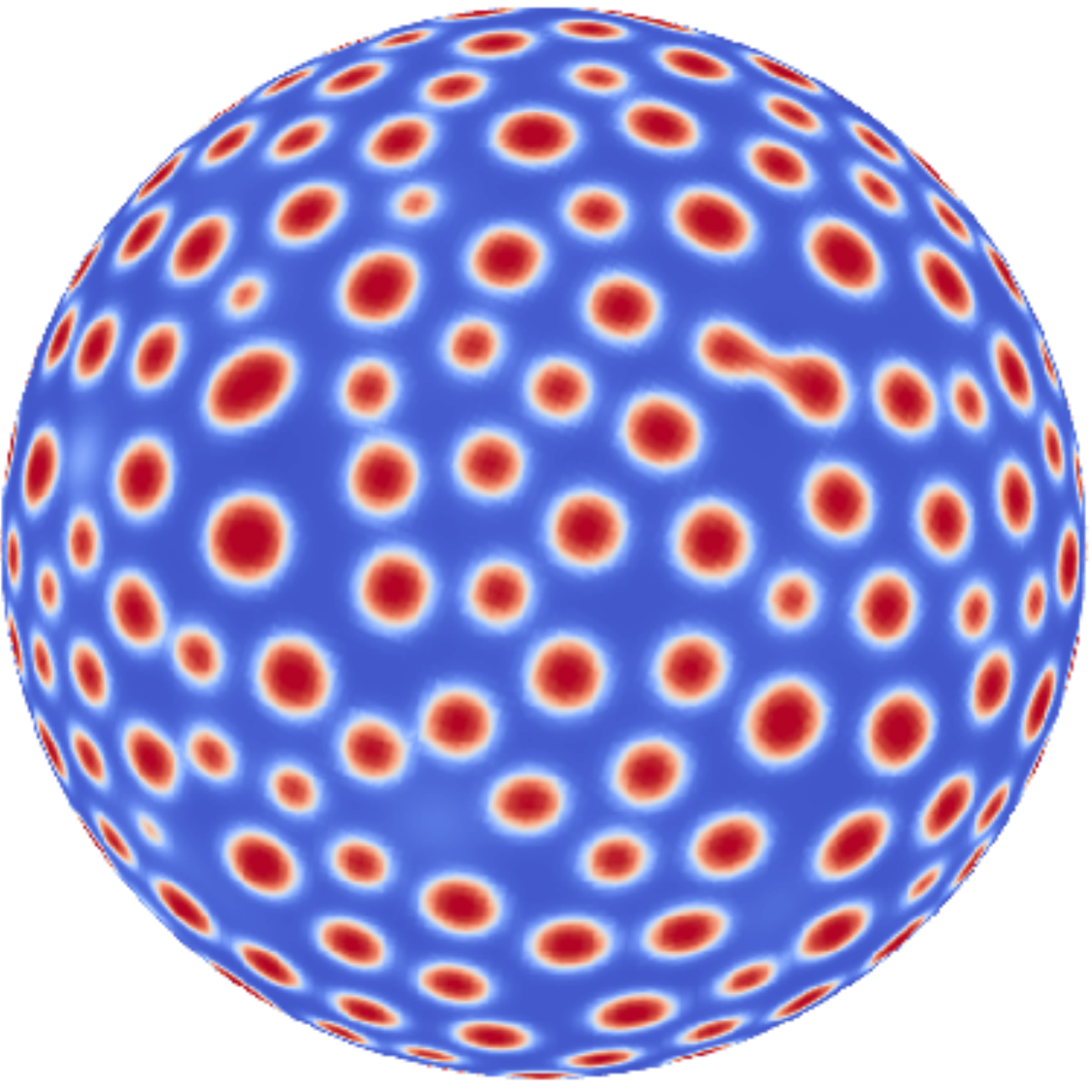}&
				\includegraphics[width=2.2cm]{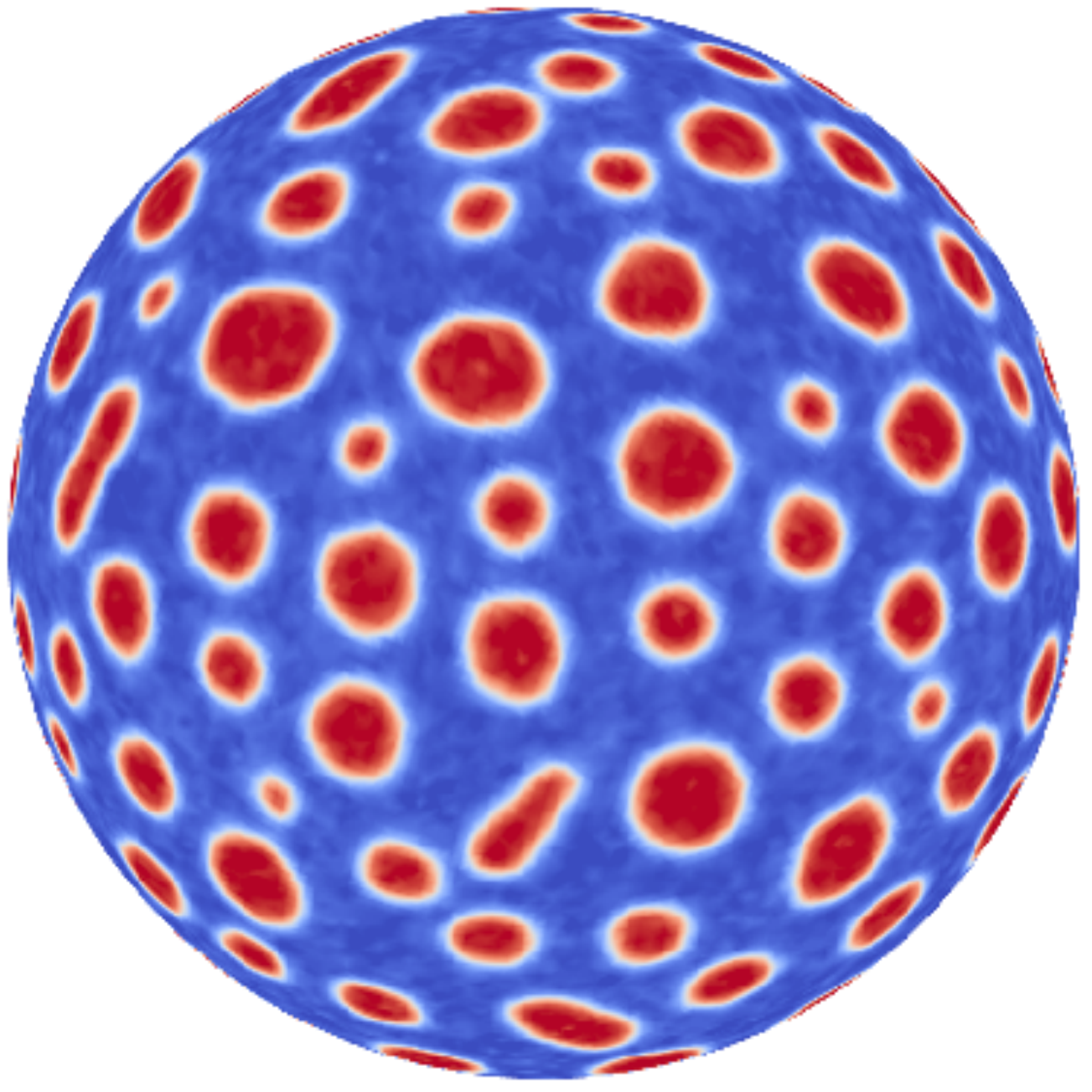}&\includegraphics[width=2.2cm]{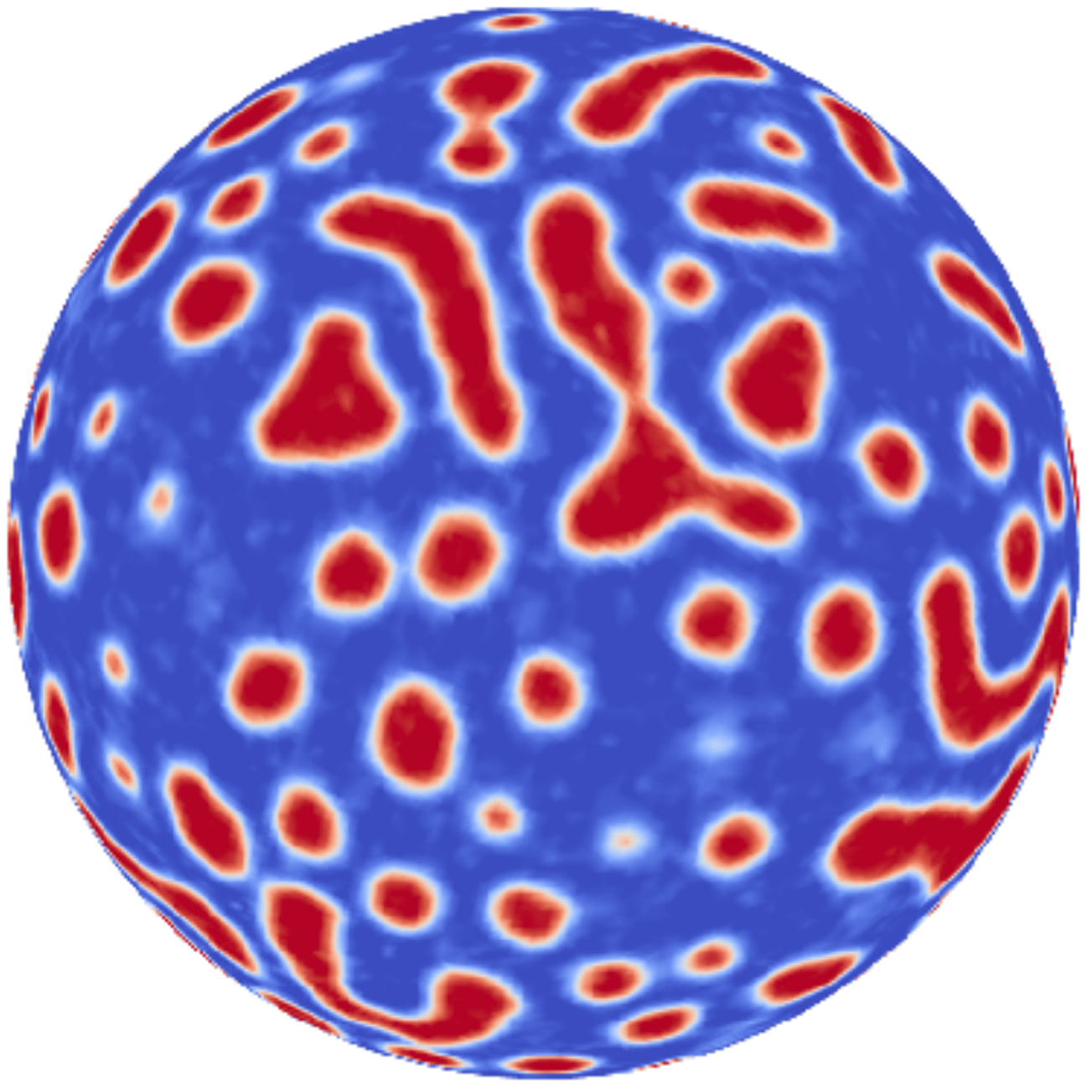}\tabularnewline
		1.0 & \includegraphics[width=2.2cm]{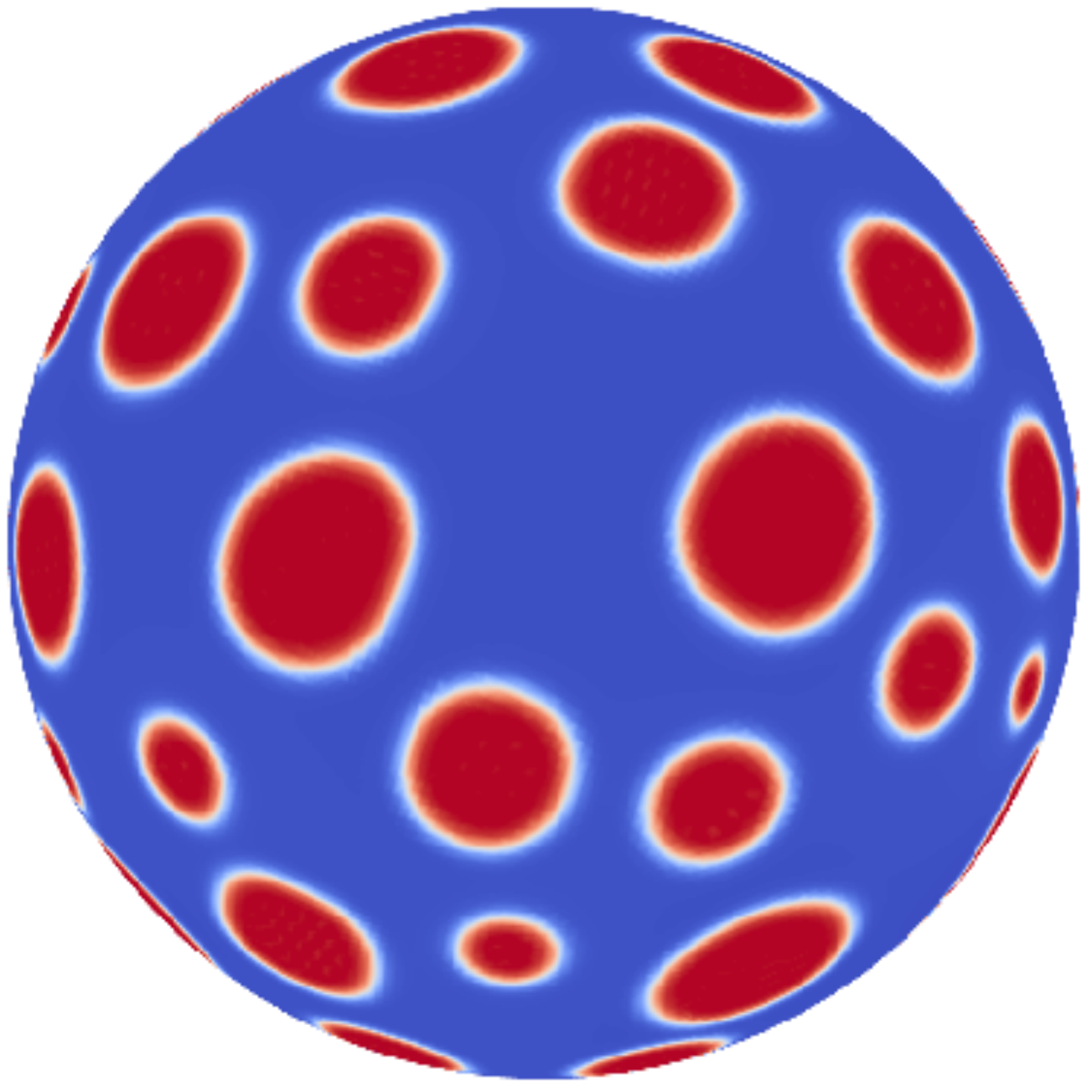} & \includegraphics[width=2.2cm]{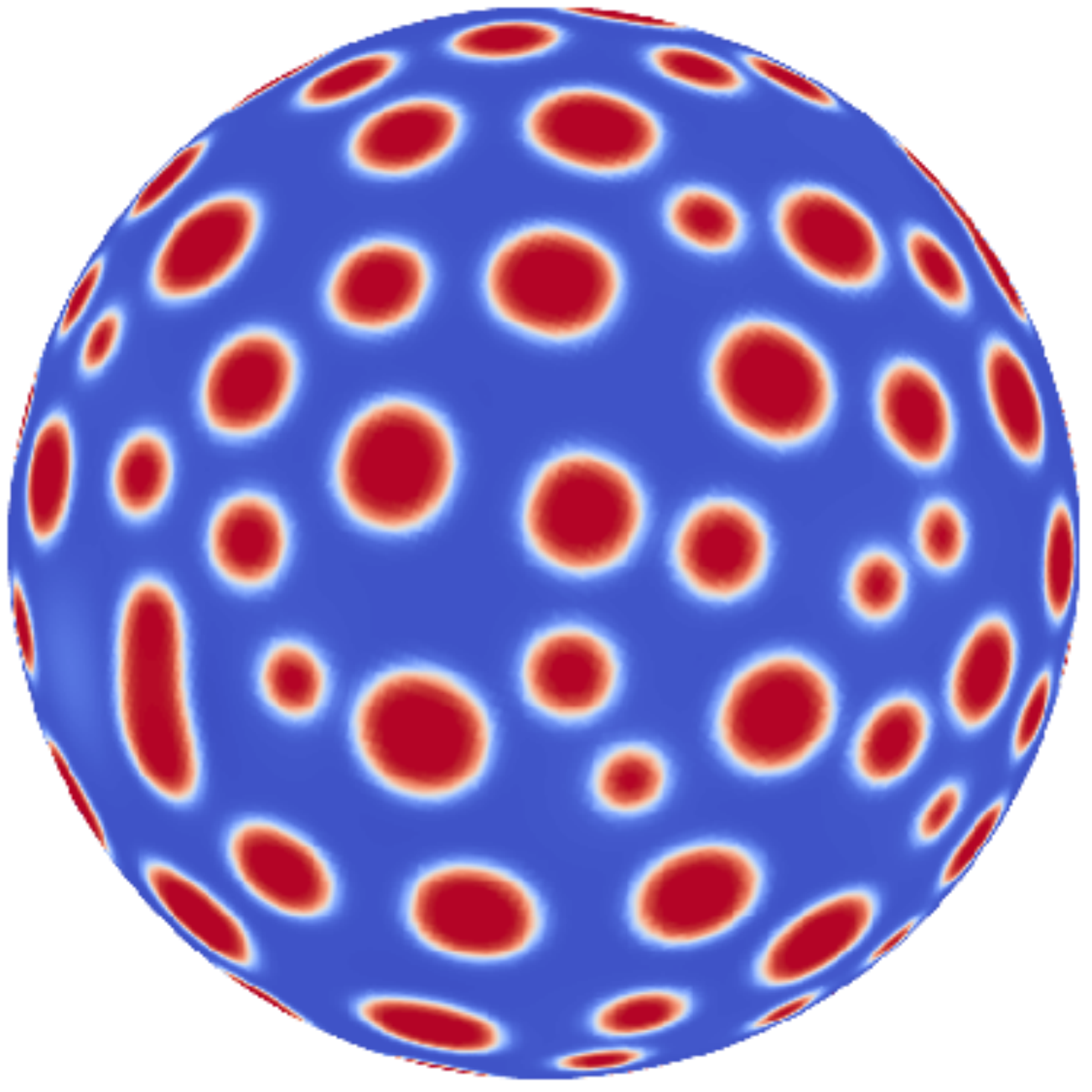}&
				\includegraphics[width=2.2cm]{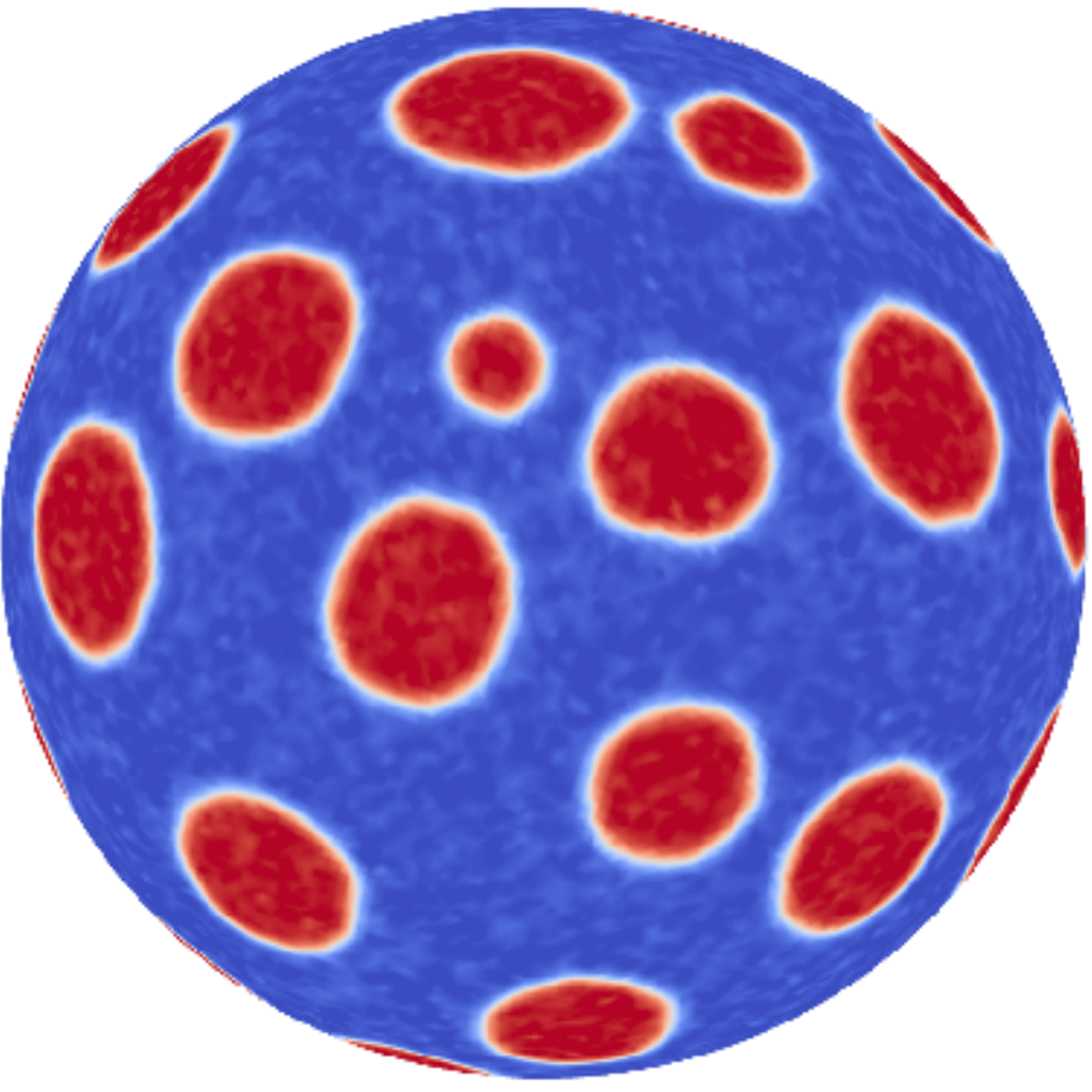}&\includegraphics[width=2.2cm]{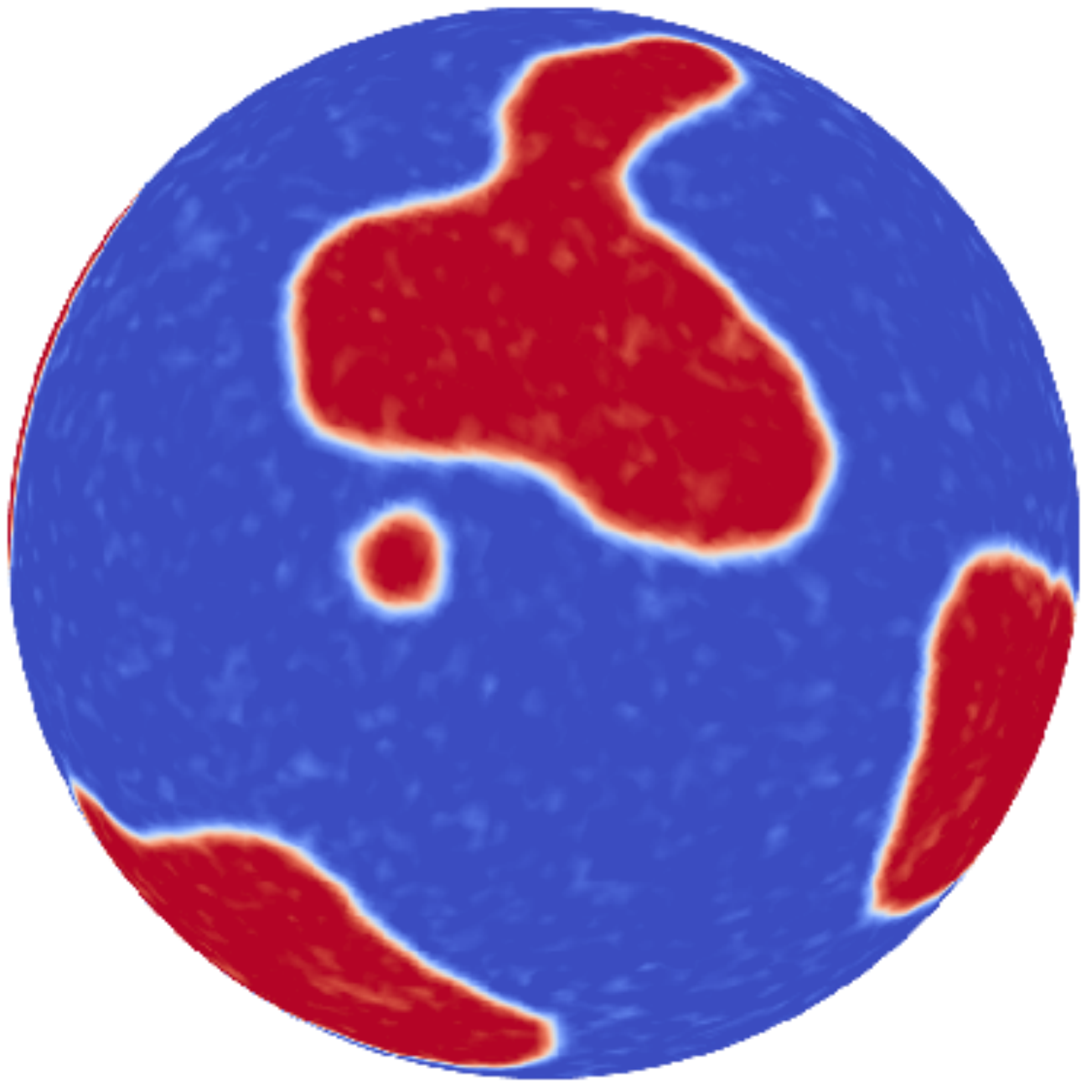}\tabularnewline
		5.0 & \includegraphics[width=2.2cm]{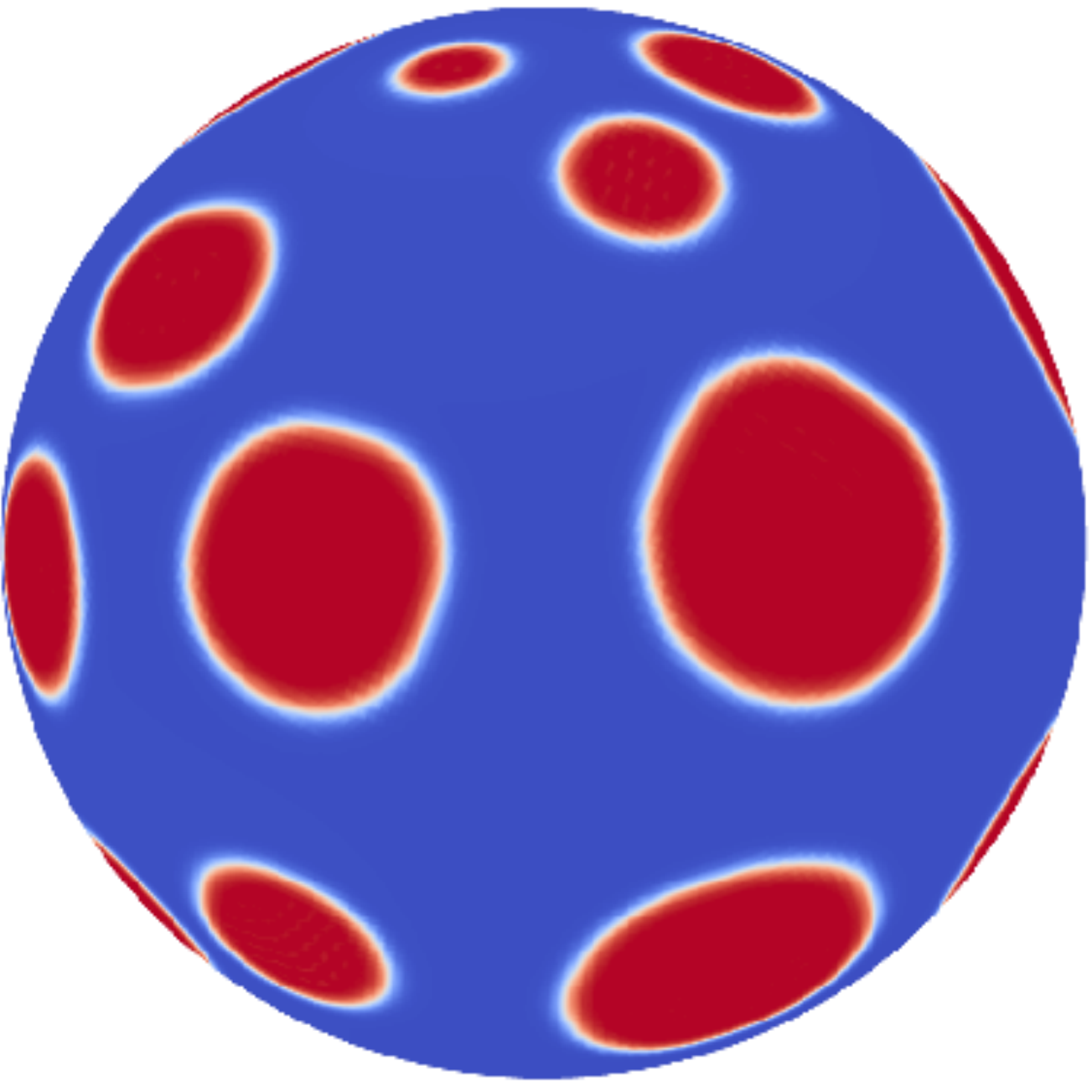}&\includegraphics[width=2.2cm]{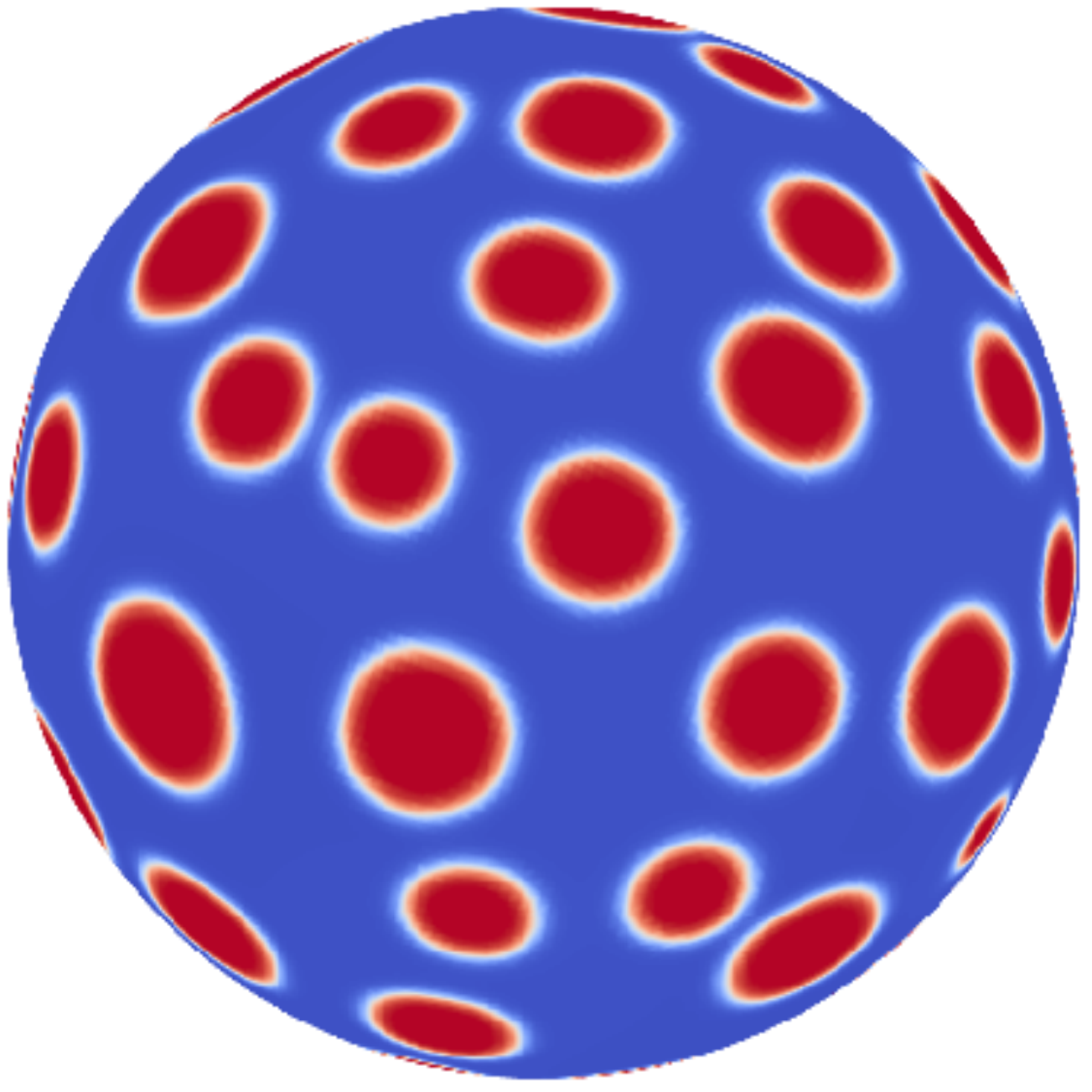}&
				\includegraphics[width=2.2cm]{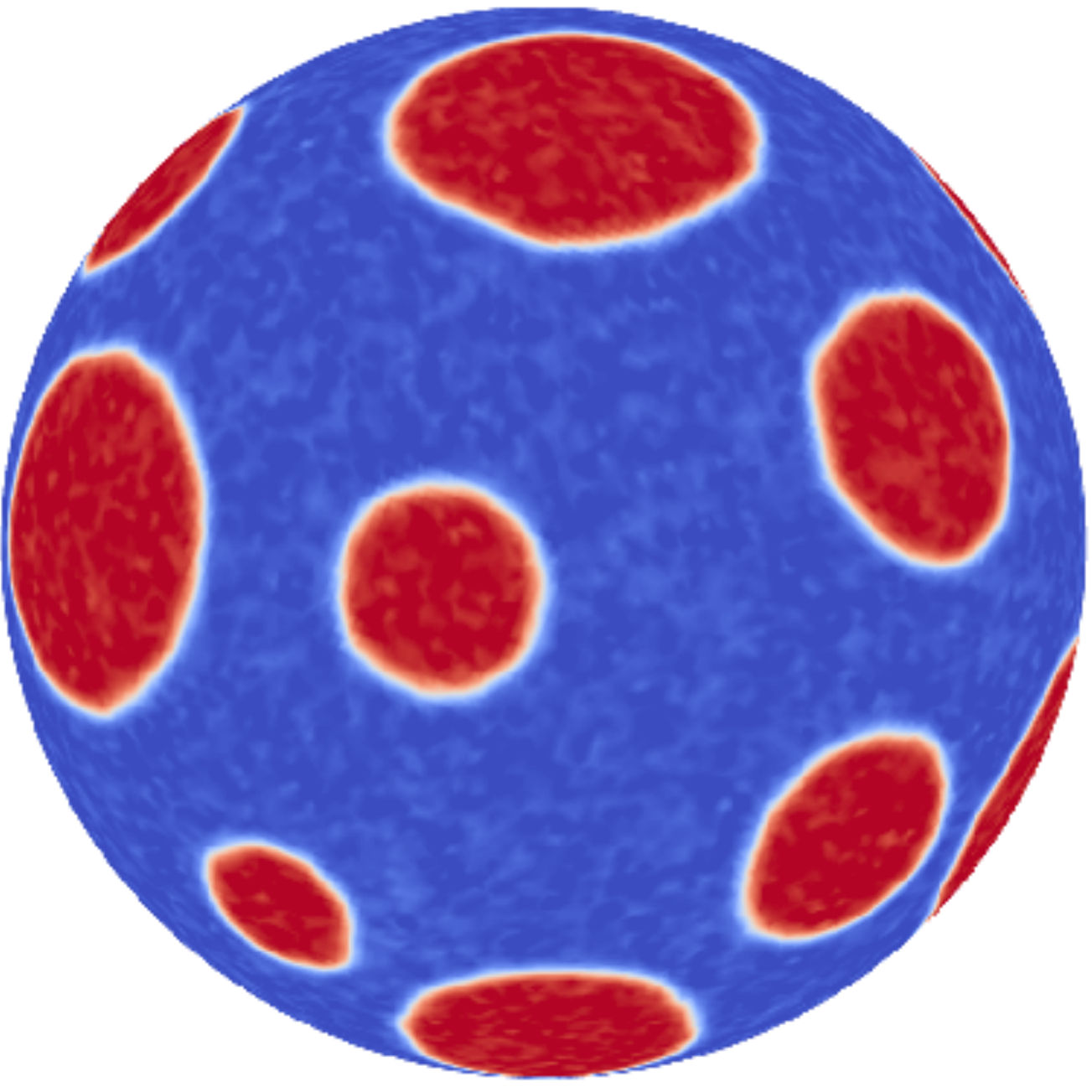}&\includegraphics[width=2.2cm]{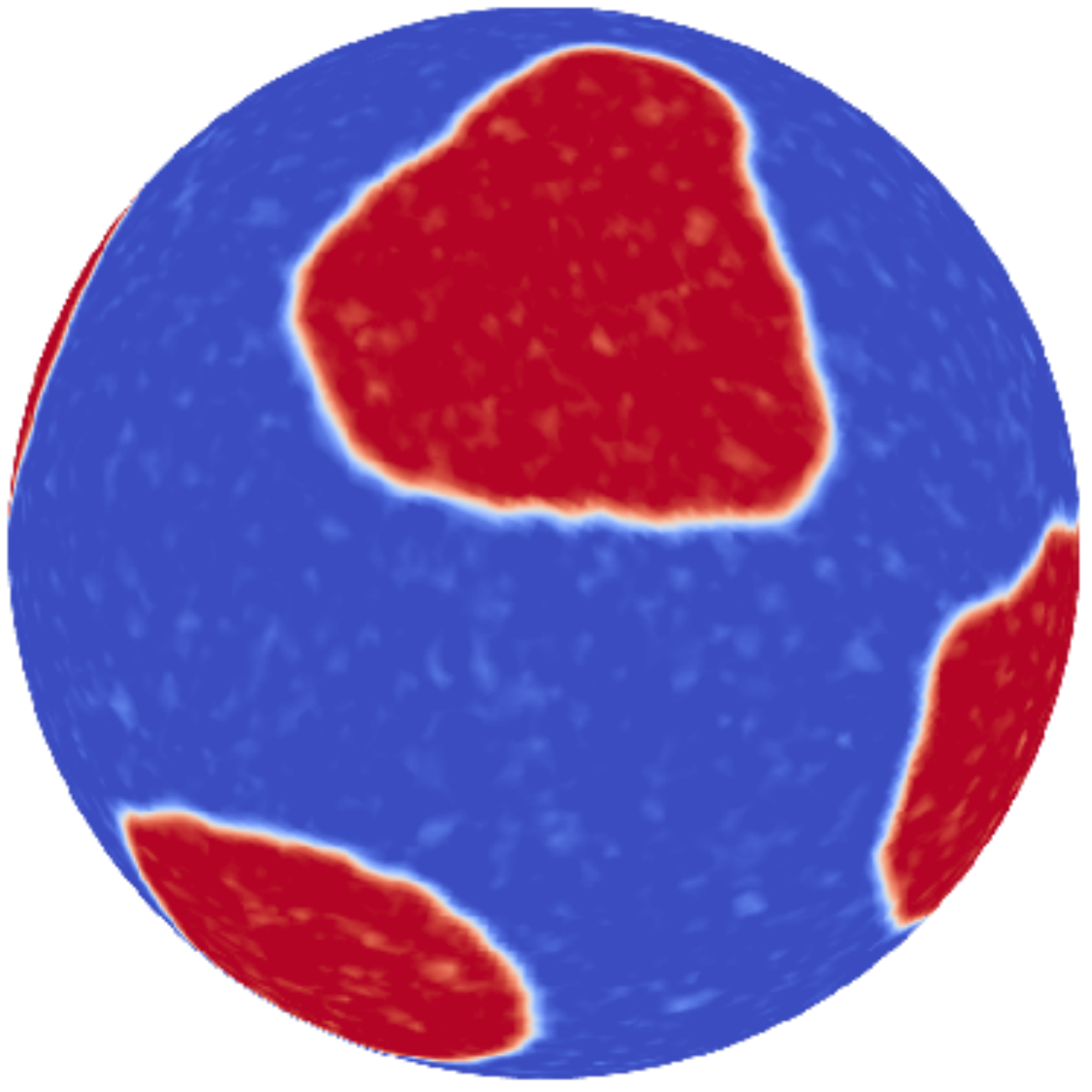}\tabularnewline
    	10.0 & \includegraphics[width=2.2cm]{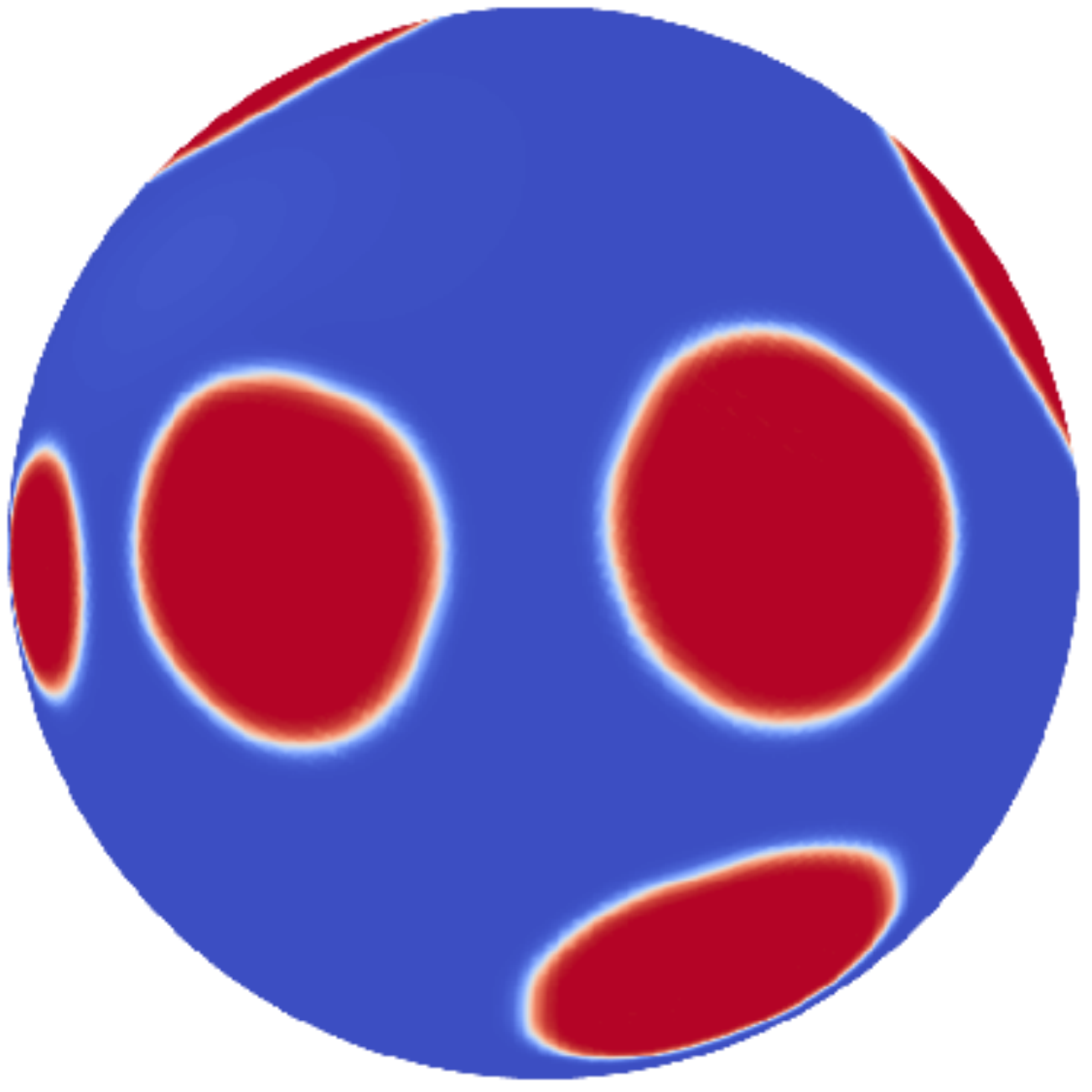} &\includegraphics[width=2.2cm]{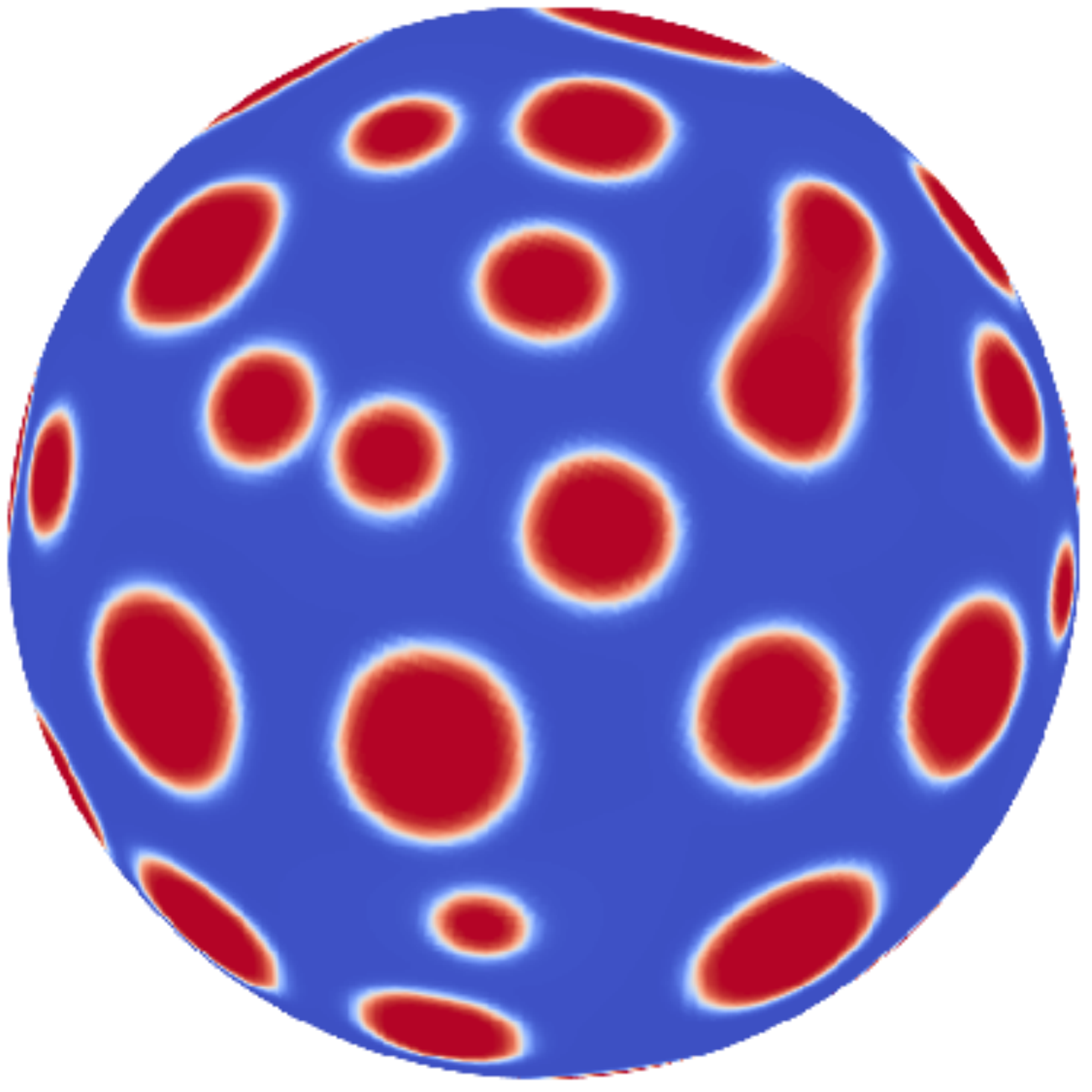}&
				\includegraphics[width=2.2cm]{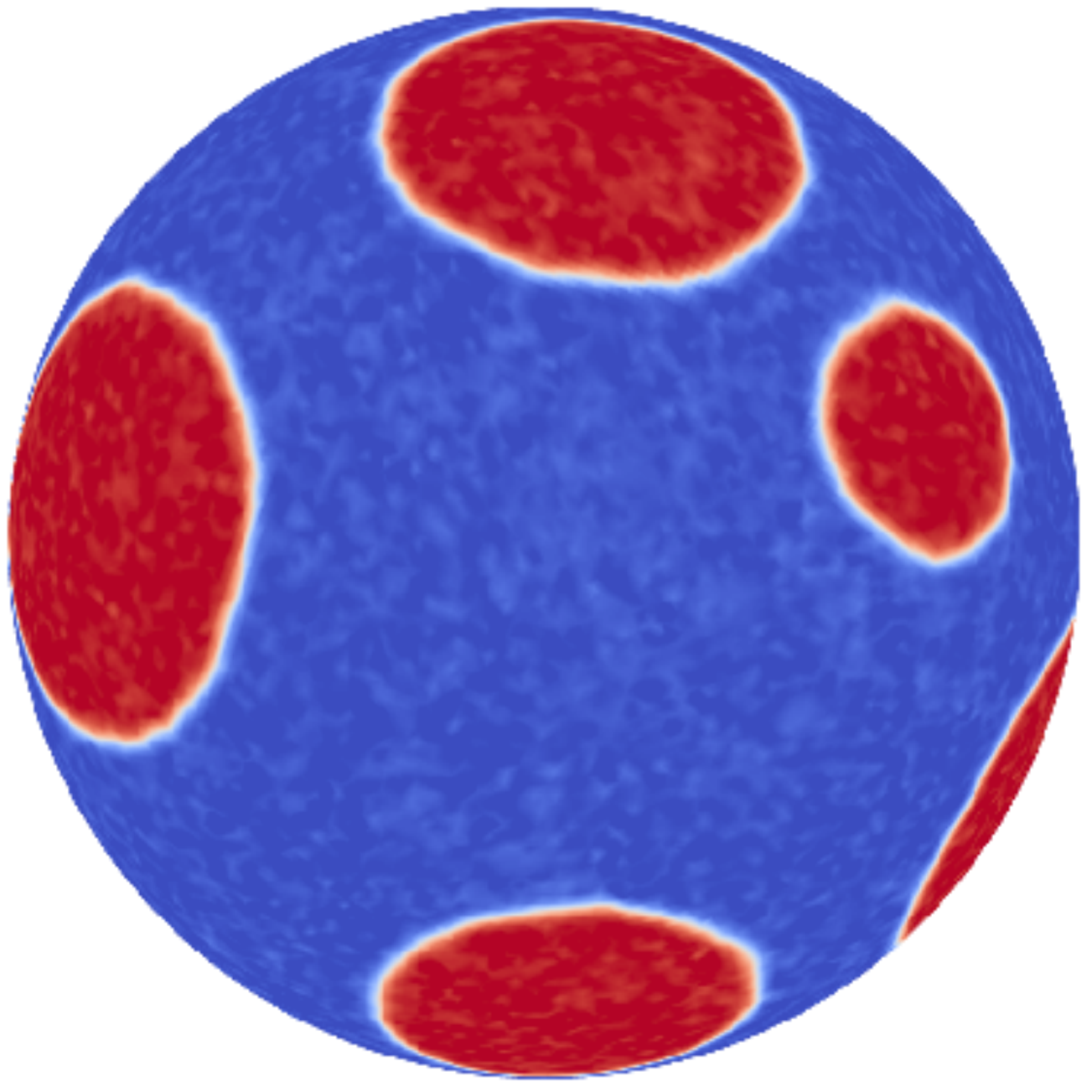}&\includegraphics[width=2.2cm]{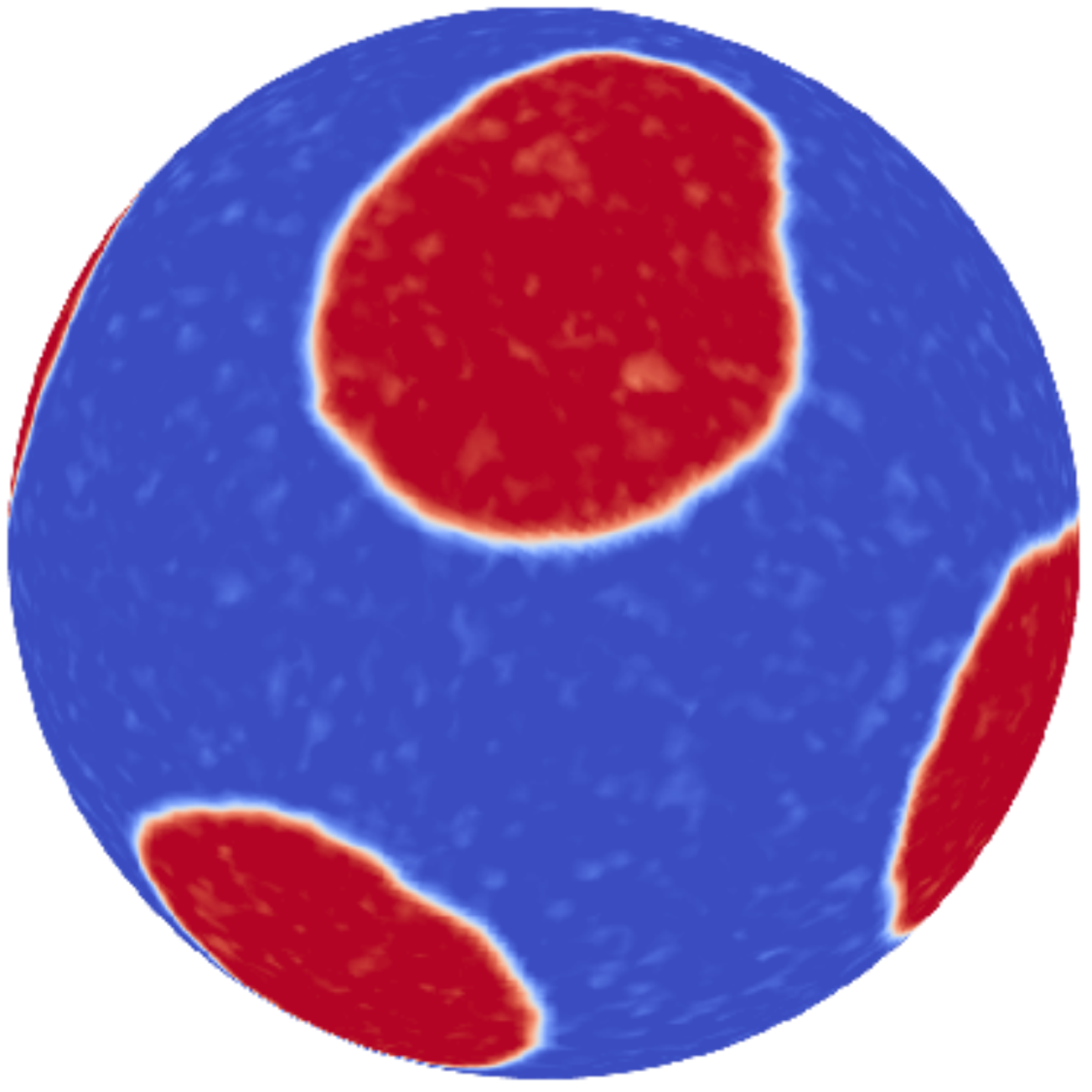}\tabularnewline
	\end{tabular}
  \caption{Evolution for Cahn-Hilliard ($\sigma=0$) and Cahn-Hilliard-Cook ($\sigma=10^{-5}$) model with
  constant and variable mobility.}
  \label{fig:sampleEvolSphere}
\end{figure}%

For the case of the Cahn-Hilliard model with constant mobility, the fast phase segregation occurs 
up to a time of $t=0.1$, see Fig.~\ref{fig:sampleEvolSphere}, thereafter the domains start to slowly coarsen in
time. The primary means of coarsening
in this case is spinodal decomposition, where a domain large in size grows at
the expense of smaller, nearby, domains. In this type of behavior the 
center of each domain remains relatively fixed. 
The use of degenerate mobility 
decreases this coarsening rate but does not change the coarsening mechanism, 
which can be seen by comparing the first two columns of
Fig.~\ref{fig:sampleEvolSphere}. 

The combination of CHC and variable mobility has the effect of increasing
the coarsening rate. This can be observed
by visually comparing the sizes of the domains at a time of $t=1$,
as the general size of the domains in the CHC plus variable mobility
are larger than the other three cases. 
This behavior will be further explored in subsequent sections.

\subsection{Convergence Study}
\begin{figure}
		\centering
		\subfigure[Constant Mobility]{
			\label{fig:Conv_CHC_M1}
			\includegraphics[width=6.0cm]{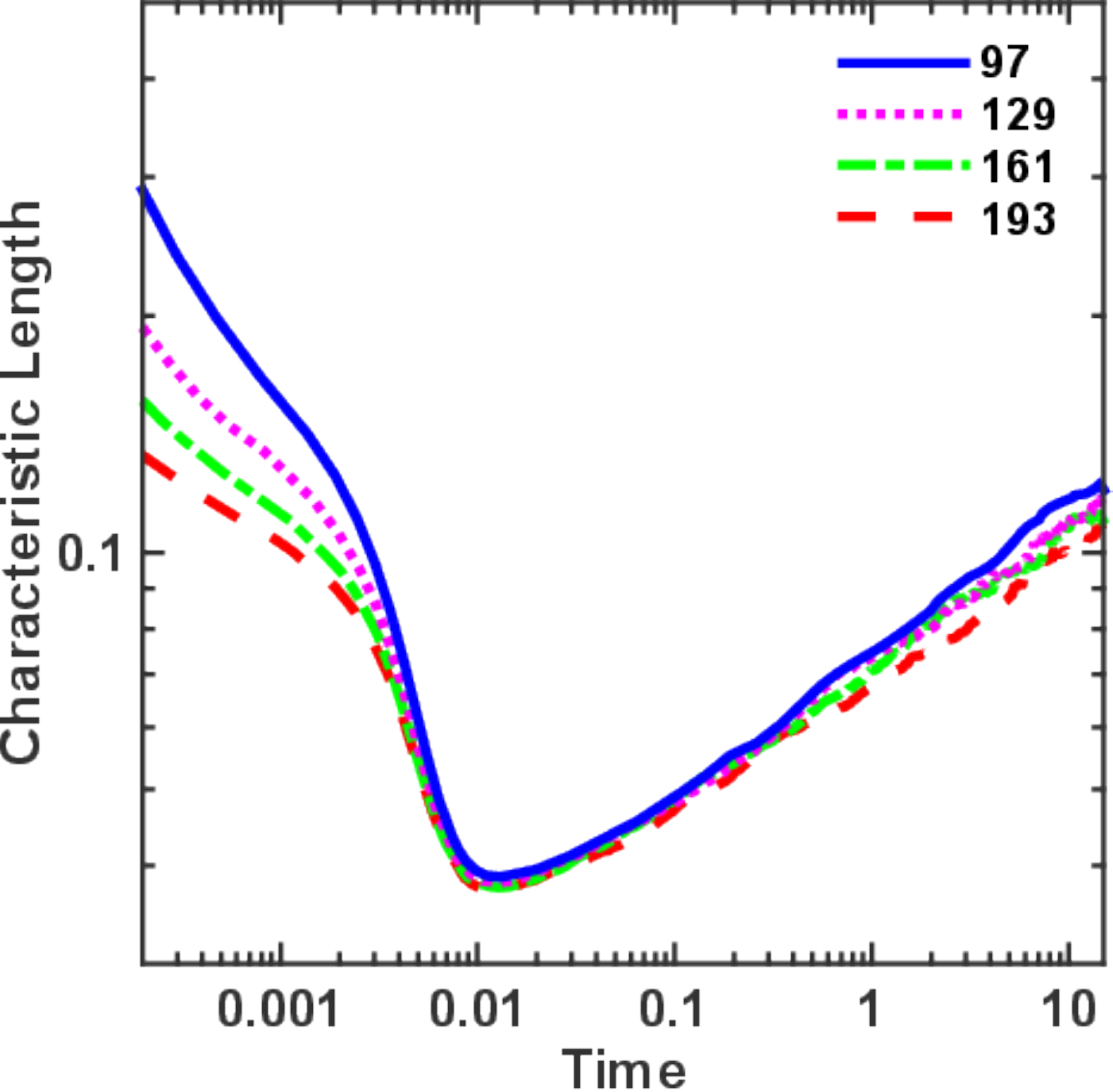} 
		}  
		\hfill
		\subfigure[Variable Mobility]{
			\label{fig:Conv_CHC_MV}
			\includegraphics[width=6.0cm]{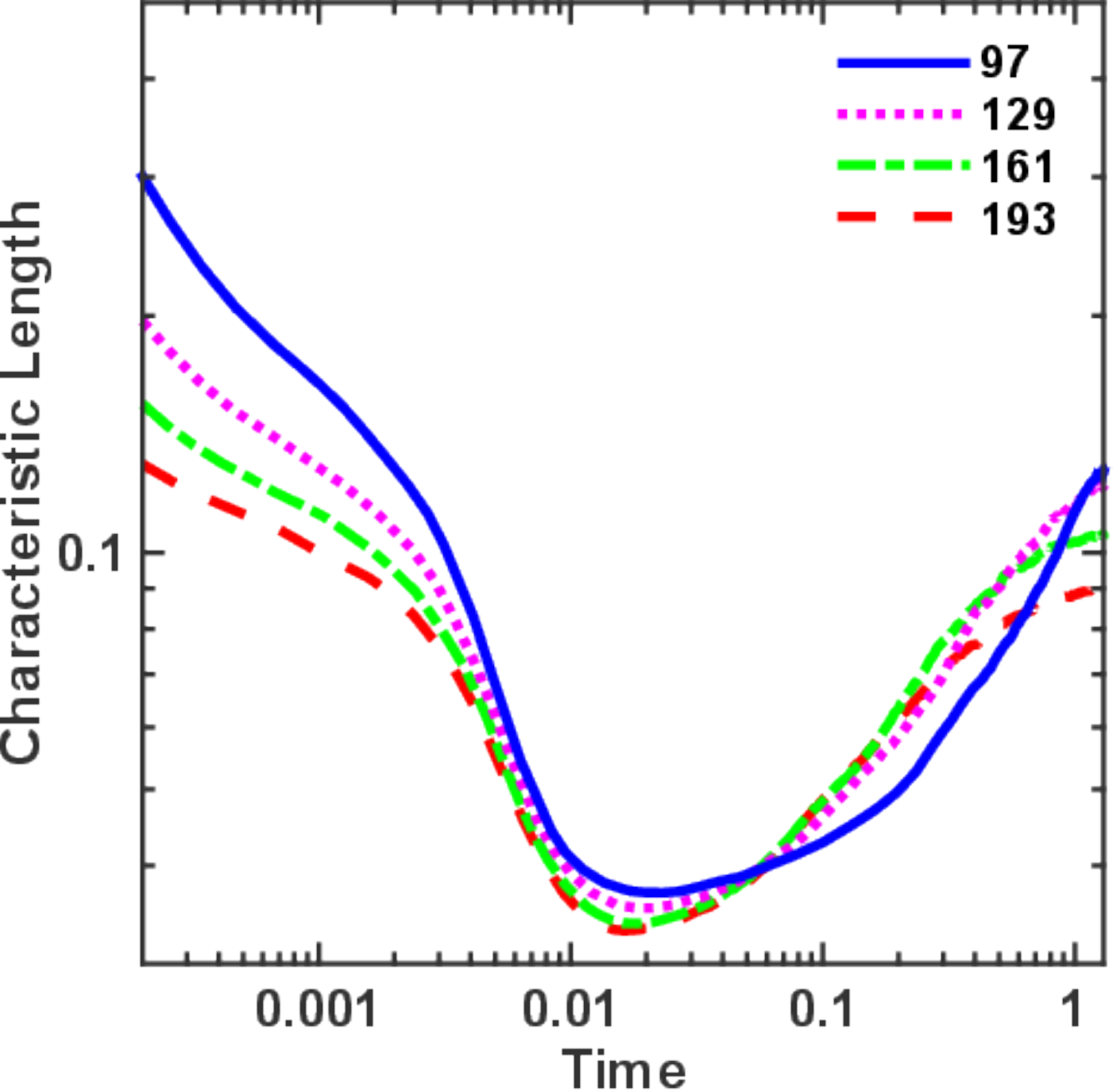} 
		}
	  \caption{The characteristic length over time using the Cahn-Hilliard-Cook model on a sphere with constant and variable mobilities for 
			various grid sizes.}
	  \label{fig:Conv_CHC}
	\end{figure}

The numerical convergence of the Cahn-Hilliard system in the absence of noise has been previously investigated by
the authors~\cite{Gera2016}. In this section a qualitative convergence 
study is performed with regard to the change of the characteristic length over time.
Sample plots of the characteristic length over time for both the constant and 
variable mobility care are shown in Fig.~\ref{fig:Conv_CHC}. 
Here four different grid sizes are considered:
$N=97$, $N=129$, $N=161$, and $N=193$. For each case the time
step is set to $\Delta t=5.12\times 10^{-3}h$, where $h=2.5/(N-1)$ is the grid spacing.

Initially there is a rapid 
decrease in the characteristic length as the system undergoes
rapid phase segregation from a well-mixed interface to 
one with many, small domains. After this point the domains coarsen
at a given rate, before reaching the near-equilibrium configuration.
This middle region, after initial coarsening and before the near-equilibrium
dynamics, is the region of interest. 

For the constant mobility case, Fig.~\ref{fig:Conv_CHC_M1}, the growth in this middle region, from 
approximately $t=0.1$ to $t=10$, the rate is similar across all 
grid spacings. With variable mobility, Fig.~\ref{fig:Conv_CHC_MV},
the region of interest is only from $t=0.1$ to $t=1$, as the final very slow coarsening stage is achieved 
sooner. A larger difference between the $N=97$ grid compared to the 
the others is seen. There is little qualitative difference in the growth rates 
using grid sizes larger than $N=129$, and thus that is the size chosen
for the further analysis.

\subsection{Characteristic Length and Energy Evolution}

In this section sample evolution curves for the characteristic length, Eq.~(\ref{eq:CL}), and 
the total energy of the system, Eq.~(\ref{eq:freeEnergy}), are examined
over time for the CH and CHC systems, assuming both constant and variable mobility.
To explore the CHC systems, 3 simulation results for each mobility case is shown.
These will then be compared to a single CH simulation.
As was mentioned earlier, the noise intensity level for the results in this section
is $\sigma=10^{-5}$.

\begin{figure}
	\centering
	\subfigure[Constant Mobility]{
		\label{fig:EP_CM}
		\includegraphics[width=6.0cm]{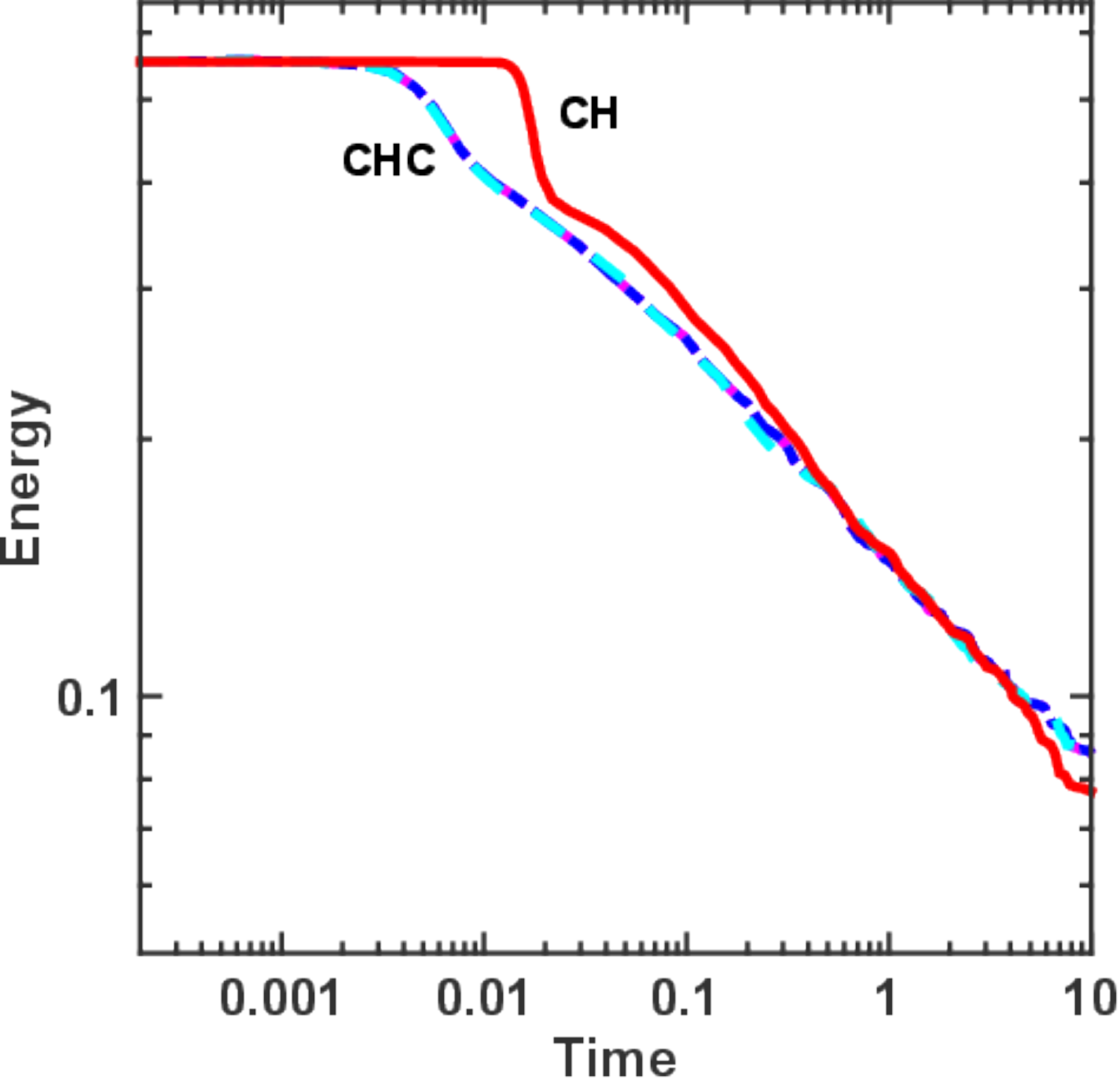} 
	}
	\hfill
	\subfigure[Variable Mobility]{
		\label{fig:EP_MV}
		\includegraphics[width=6.0cm]{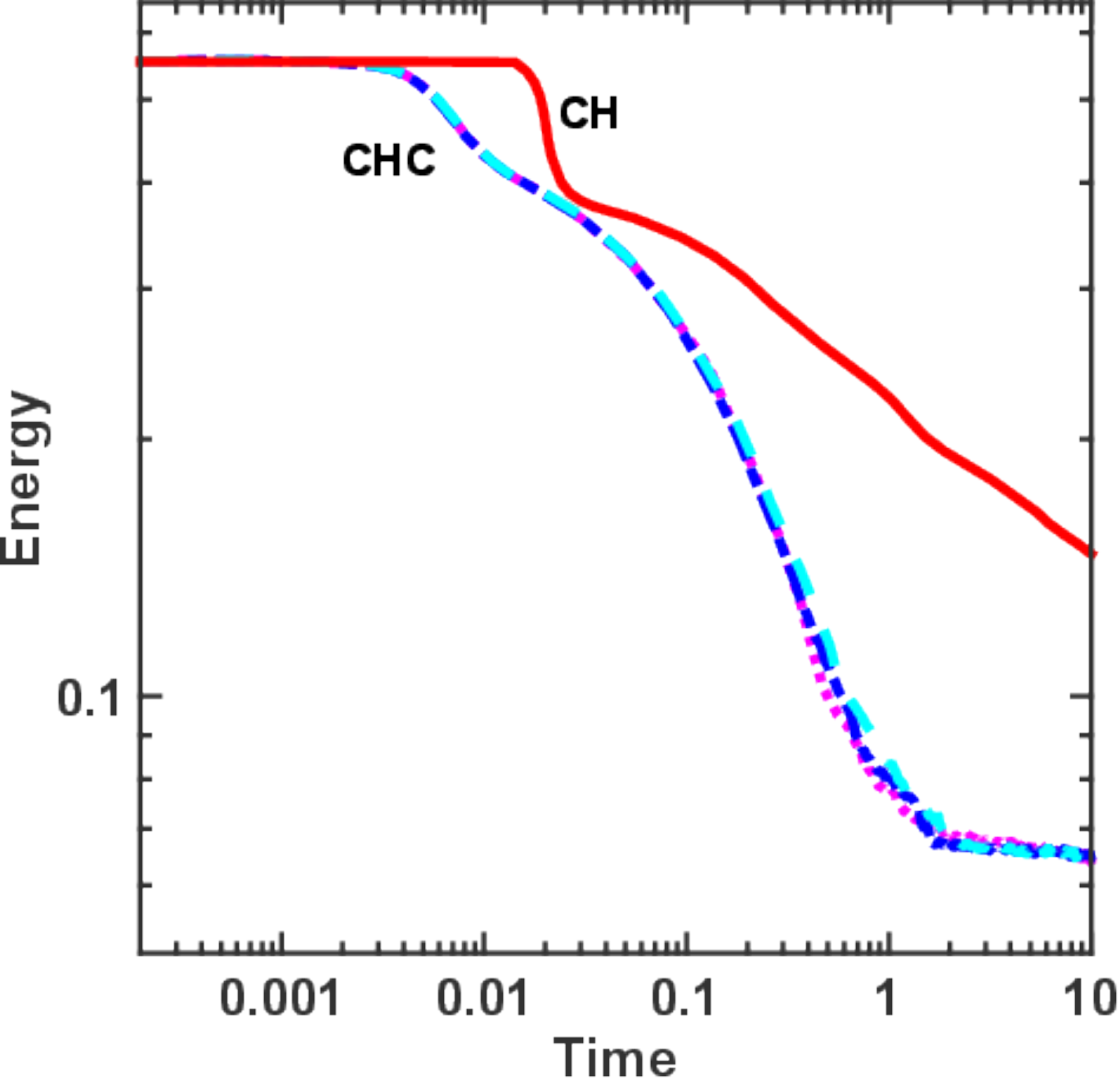} 
	}
	\caption{The change of total energy in the Cahn-Hilliard model along with
	the three sample runs of a Cahn-Hilliard-Cook model for constant and variable
	mobilities.}
	\label{fig:EnergyPlot}
\end{figure}

\begin{figure}
	\centering
	\subfigure[Constant Mobility]{
		\label{fig:CL_M1}
		\includegraphics[width=6.0cm]{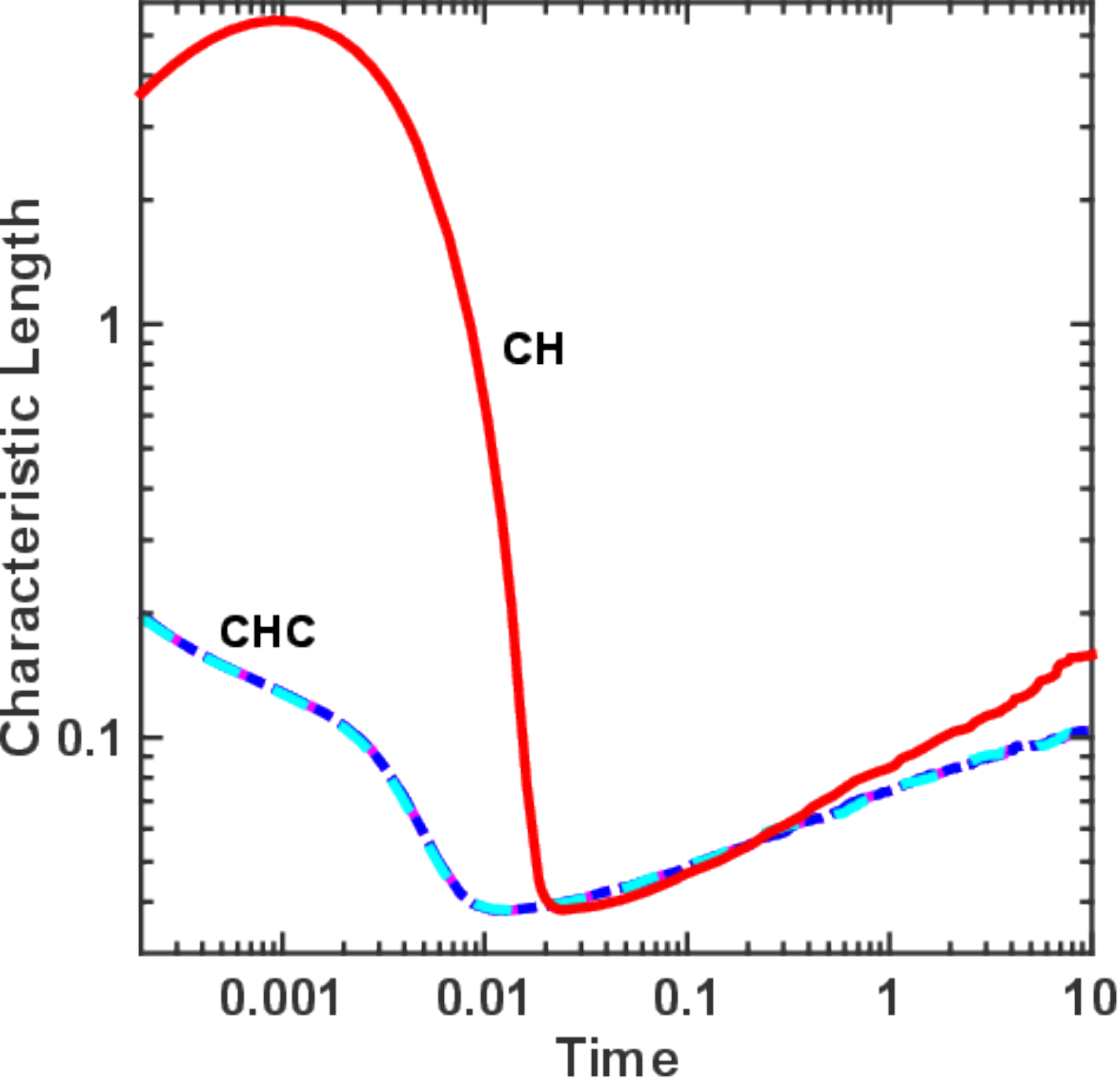} 
	}
	\hfill
	\subfigure[Variable Mobility]{
		\label{fig:CL_MV}
		\includegraphics[width=6.0cm]{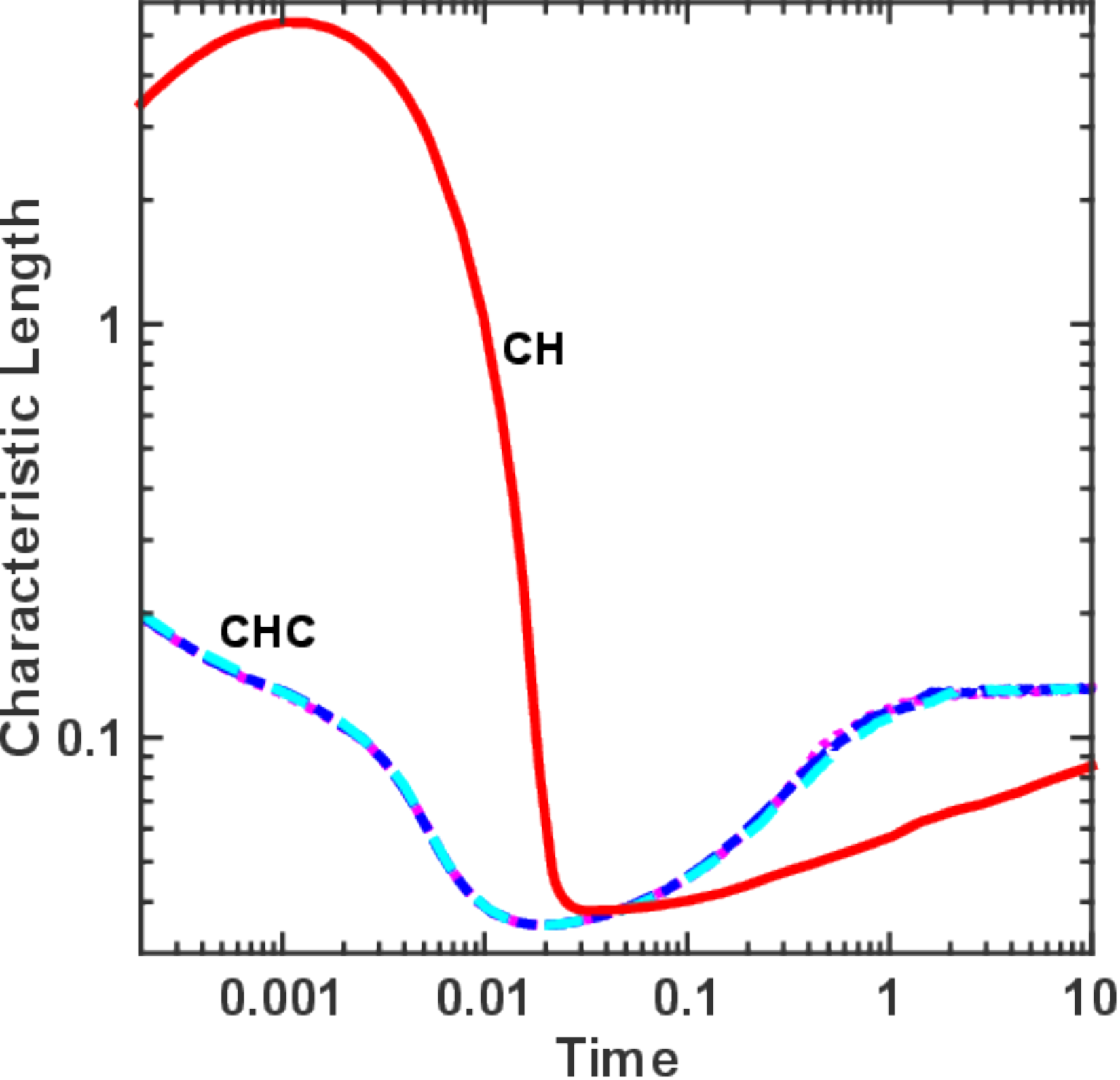} 
	}
	\caption{The characteristic length over time for the Cahn-Hilliard
	model along with three sample runs of the Cahn-Hilliard-Cook model for
	constant and variable mobilities.}
	\label{fig:CL}
\end{figure}

The energy, Fig.~\ref{fig:EnergyPlot}, and characteristic length, Fig.~\ref{fig:CL},
are shown over time for the single CH simulation and three representative CHC simulations.
During the initial stage when the system is in a
homogeneous state, the total energy is large and remains constant for both the CH and CHC systems.
After some time the initial state segregates into many, small domains, see
Fig.~\ref{fig:sampleEvolSphere} for an example.
During this rapid phase segregation regime both the CH and CHC simulations see an overall decrease in the 
characteristic length, eventually reaching a minimum length, while a rapid decrease
in the overall energy occurs. In both mobility cases, the CHC model begins the segregation
process earlier, as is evident from the earlier decrease of the energy.
It is also interesting to note 
that the CH system has a brief increase in the characteristic length for both constant 
and variable mobility. This can be attributed to the CH system remaining near the well-mixed 
initial condition longer than the CHC system. Instead of quickly segregating
to well-defined domains, the CH simulations have many, small amplitude fluctuations.
This results in a relatively small value of $L(t)$,
which quickly increases as the domains form.

After this rapid phase segregation, the system undergoes a steady and much slower coarsening process.
For the constant mobility case, Figs.~\ref{fig:EP_CM} and \ref{fig:CL_M1}, it is clear that the 
growth rate for the CHC system is below that of the CH system. At a time of $t=10$
the energy of the system is higher and the characteristic length is smaller for the CHC system
as compared to the CH system.
Thus, while noise promotes the early 
start of the coarsening process when constant mobility is assumed, 
it inhibits the process during the slower, second coarsening regime.

The assumption of variable mobility dramatically changes the influence of the noise,
see, Figs.~\ref{fig:EP_MV} and \ref{fig:CL_MV}. While the CHC system begins 
to segregate earlier than the CH system, similar to the constant mobility case,
the rate of change of CHC with variable mobility is much higher
than CH with variable mobility. It is suspected that the different scalings
of the noise magnitude with respect to the mobility is the cause of this behavior.
This will be further explored in Sec.~\ref{sec:StatOnSphere}.


\subsection{Discussion and Analysis}
\label{sec:StatOnSphere}
\begin{figure}
		\centering
		\subfigure[Constant Mobility]{
			\label{fig:CL_MMM_M1}
			\includegraphics[width=6.0cm]{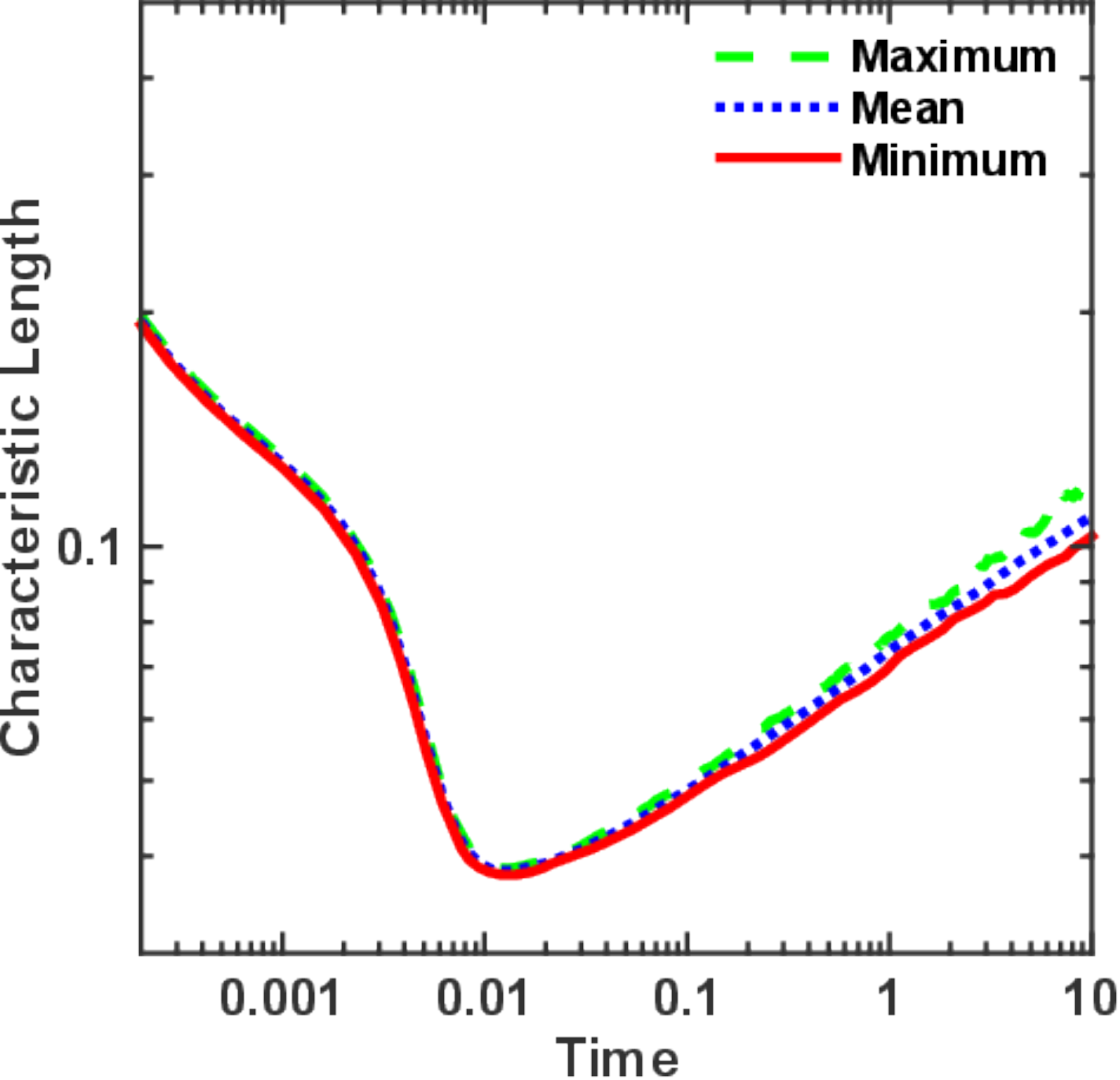} 
		}
		\hfill
		\subfigure[Variable Mobility]{
			\label{fig:CL_MMM_MV}
			\includegraphics[width=6.0cm]{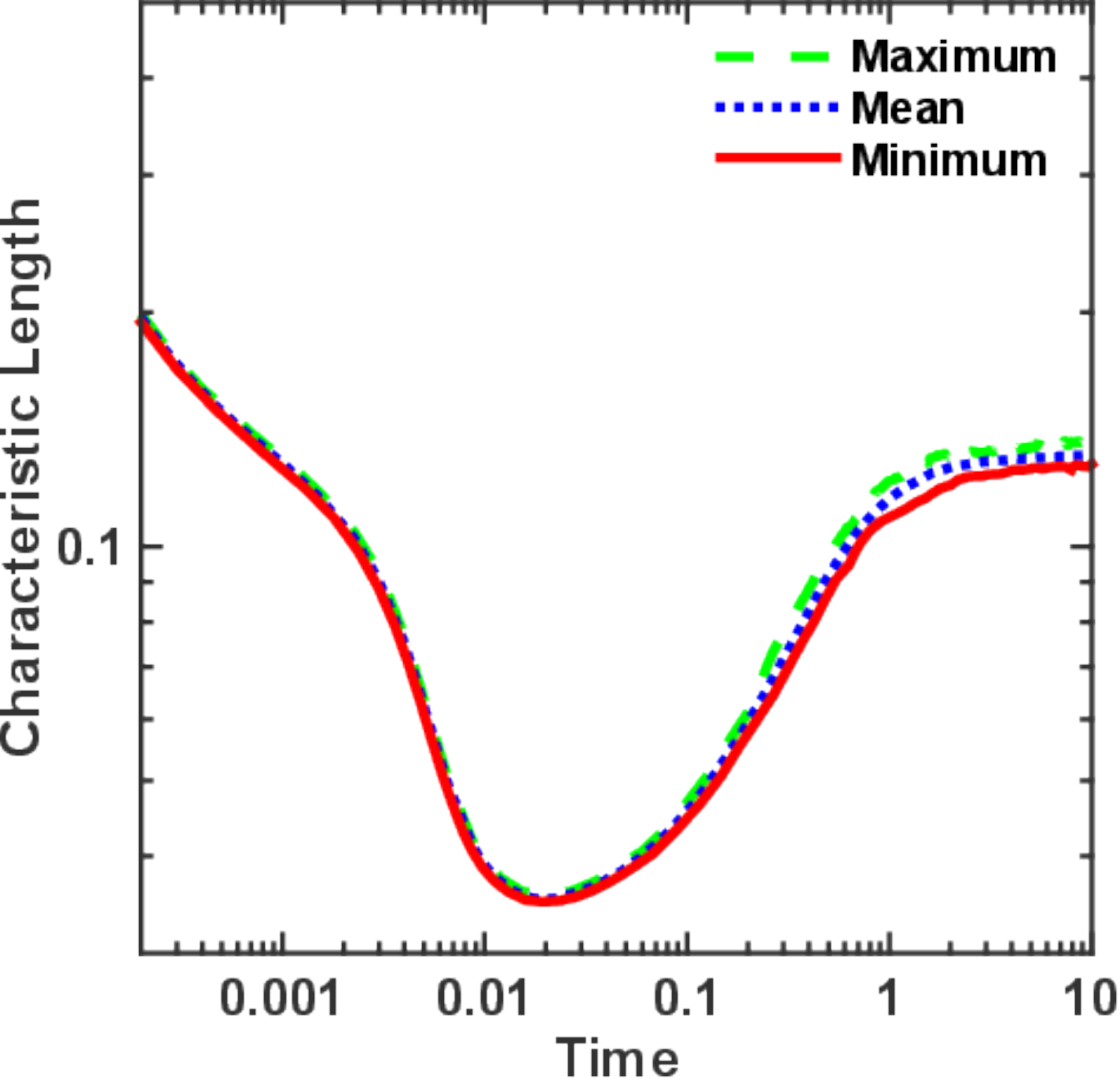} 
		}
	  \caption{The minimum, maximum, and mean characteristic lengths for the Cahn-Hilliard-Cook
      system for the 64 realizations using constant and variable mobilities each.}
	  \label{fig:CL_MMM}
	\end{figure}

To further explore the CH and CHC systems, a total of 64 realizations (simulations) per noise level and mobility type 
have been performed. For the CHC system with $\sigma=10^{-5}$ the minimum, maximum, and average
characteristic lengths for each time step have been determined from the 64 realizations,
as shown in Fig.~\ref{fig:CL_MMM}. 
Due to the cumulative effects of the noise during the course of the simulation, 
the spread of the characteristic length increases
as time progresses. 

After the initial segregation phase, it is expected that the characteristic 
length grows at a particular growth rate, $\bar{R}(t)\propto t^{\alpha}$, where
$\alpha$ is the growth rate. For the constant CH and CHC with constant mobility, in addition
to the variable mobility CHC model, this region extends from approximately
a time of $t=0.1$ to $t=10$. Due to the faster dynamics of the variable mobility CHC model,
this region exists approximately from $t=0.1$ to $t=0.8$. After these times 
the system enters the long-term, slow growth phase.
To determine the growth rate a linear fit is made to the appropriate 
region. The slope of this fit is taken to be the growth rate
parameter $\alpha$. The complete results for the mean, standard deviation, and coefficient
of variation for all considered systems is presented in Table~\ref{table:mean_SD_CV}.
	\begin{table}
		\caption{Statistics on the growth rate for Cahn-Hilliard and Cahn-Hilliard-Cook
			model with constant and variable mobilities.}
		\label{table:mean_SD_CV}
		\begin{center}
			\begin{tabular}{|>{\centering}m{1.2cm} | >{\centering}m{2.0cm} | >{\centering}m{1.2cm} | >{\centering}m{2.0cm} | >{\centering}m{2.0cm} | >{\centering}m{3.0cm} |  }
				\hline
				Model & Mobility & Noise & Mean & Standard Deviation  & Coefficient of Variation\tabularnewline				      
				\hline
				\hline
				\multirow{2}{*}{CH} & 1 & -- & $0.2814$ & $0.0215$ & $0.0764$\tabularnewline
				\cline{2-6} 
					 & $f(1-f)$ & -- & $0.1759$ & $0.0062$ & $0.0352$\tabularnewline
				\hline
				\multirow{6}{*}{CHC} & \multirow{3}{*}{1} & $10^{-9}$ & $0.2825$ & $0.0197$& $0.0697$ \tabularnewline
				\cline{3-6} 
				 &  & $10^{-7}$ &  $0.2672$ & $0.0241$& $0.0902$ \tabularnewline
				\cline{3-6} 
				 &  & $10^{-5}$ &  $0.1760$ &$0.0134$& $0.0761$ \tabularnewline
				\cline{2-6} 
				 & \multirow{3}{*}{$f(1-f)$} & $10^{-9}$ &$0.1765$&$0.0092$&
                 $ 0.0521$\tabularnewline
				\cline{3-6} 
				 &  & $10^{-7}$ & $0.3719$ & $0.0148$& $0.0398$ \tabularnewline
				\cline{3-6} 
				 &  & $10^{-5}$ & $0.4278$ & $0.0154$& $0.0360$ \tabularnewline
				\hline
			\end{tabular}
		\end{center}		
	\end{table}
	
For the CH system, the growth rate for constant mobility was determined to be
$\bar{\alpha}=0.2814$, which differs from the the theoretical growth rate of $\alpha=1/3$
for flat surfaces~\cite{KAHLWEIT19751,lifshitz1961kinetics}. This deviation from the theory
indicates the underlying geometry does have an impact on the rate at which
phase segregation occurs. In this case, the curvature of the
sphere has played a role in retarding the rate of coarsening.  When a
degenerate mobility is used, this growth rate decreases to a value of
$\bar{\alpha}=0.1759$.  As mentioned earlier, this decrease should be expected
as the evolution process is now limited to only occur near the interface. 

The CHC system is explored by not only varying the mobility type, but also
the intensity of the noise, $\sigma$. First consider the constant mobility case.
Using a noise intensity of $\sigma=10^{-5}$, 
the average growth rate decreased to a
value of $\bar{\alpha}=0.1760$.
As the noise intensity decreases, the mean approaches the that of the CH system,
with values of $\bar{\alpha}=0.2672$ for $\sigma=10^{-7}$ 
and $\bar{\alpha}=0.2825$ for $\sigma=10^{-9}$. This trend of
recovering the Cahn-Hilliard system as the noise intensity is lowered has been
also observed in the past~\cite{rogers1988numerical}.

When variable mobility is employed the mean growth rate of
for a Cahn-Hilliard-Cook model increases to $\bar{\alpha}=0.4278$
when $\sigma=10^{-5}$. 
As the noise intensity level decreases, the growth rate also decreases,
with $\bar{\alpha}=0.3719$ for $\sigma=10^{-7}$ and $\bar{\alpha}=0.1765$
for $\sigma=10^{-9}$. As with the constant mobility case, this growth rate approaches the 
Cahn-Hilliard value.

The fact that the growth rate increases for CHC and variable mobility is quite surprising.
One possible explanation can be obtained 
by comparing two contributions to $\partial f/\partial t$:
the diffusive contribution $\nu\Delta^2_s f$ and 
the conserved random force $\xi$. The diffusive contribution will tend to smooth out 
any oscillations which occur while the random force will push the system away from 
an equilibrium configuration.
 
\begin{figure}
	\centering	
	\includegraphics[width=8.0cm]{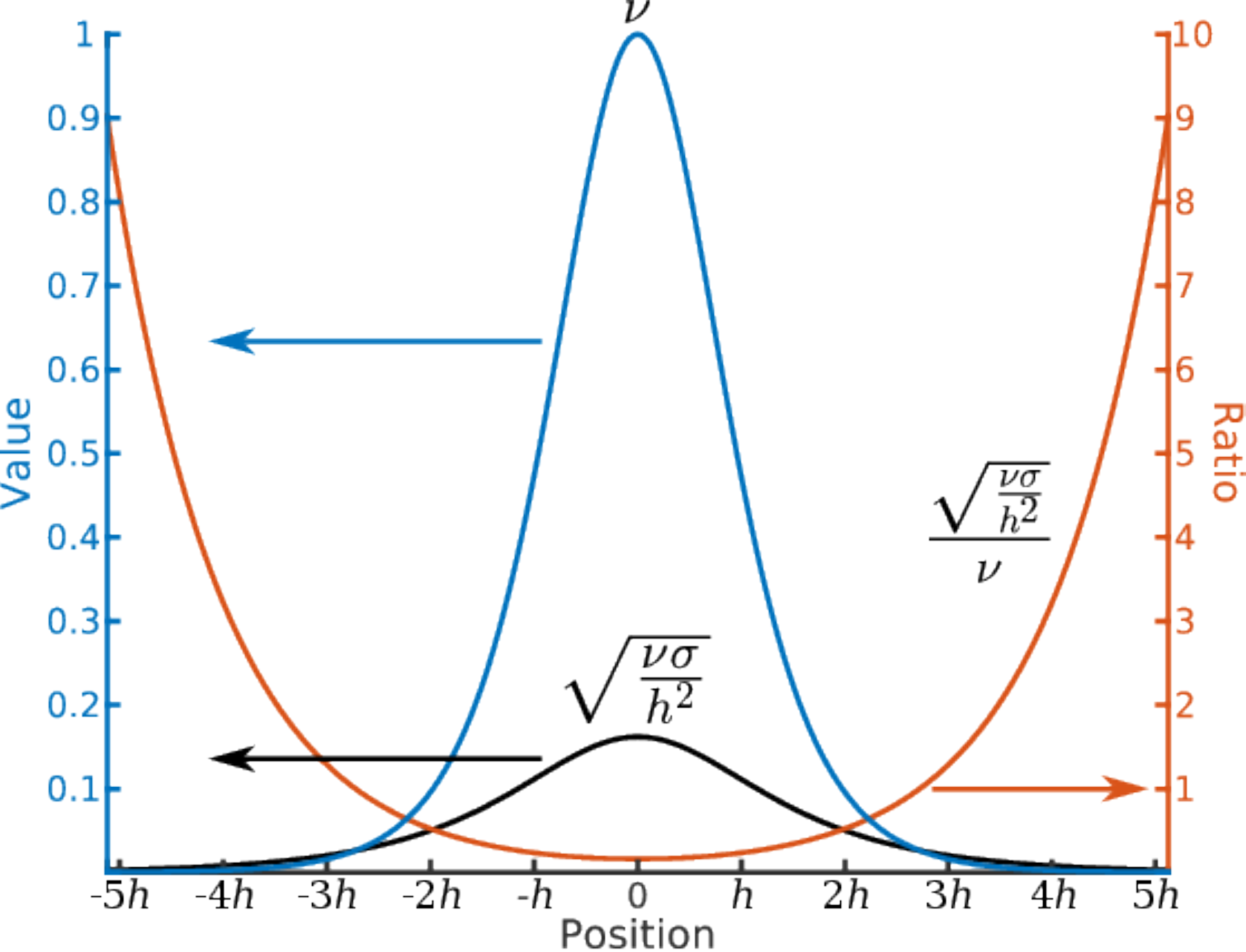} 	
	\caption{The mobility ($\nu=4f(1-f)$) and noise magnitude ($\sqrt{\nu\sigma/h^2}$) 
		where $\sigma=10^{-5}$ and $h=2.5/128$ for an equilibrium 1D profile, Eq.~(\ref{eq:equilibriumProfile}). 
			The ratio of the noise magnitude to mobility is also provided.}
	\label{fig:Scalings}
\end{figure}

Consider a simple, 1D equilibrium phase field profile, which is given by~\cite{Lee2014}
\begin{equation}
	f_{eq}\left(x\right)=\dfrac{1}{2}\left[\tanh\left(\dfrac{x}{\Cn\sqrt{2}}\right)+1\right].
	\label{eq:equilibriumProfile}
\end{equation}
The diffusive contribution scales as the mobility, $\nu$, while at the discrete level the noise
will scale as $\sqrt{\nu\sigma/h^2}$, Eq.~(\ref{eq:discreteRandomForce}).
As an example use $\Cn=0.015$ with $\sigma=10^{-5}$ and $h=2.5/128$. The value of the 
mobility and of the noise scaling is presented in Fig.~\ref{fig:Scalings}. 
The ratio between the noise scaling and the mobility is also shown in this figure.

Both the mobility and noise magnitude decrease quickly away from $x=0$, with the mobility decreasing at
a faster rate than the noise magnitude. This becomes apparent when the ratio is considered. In regions away from
$x=0$ the influence of noise becomes more larger than the diffusive term. It is suspected
that the larger influence of noise in regions away from the interface drive the system to coarsen 
faster than the other cases. This behavior also explains the oscillations (faint white patches) observed 
in the well-segregated regions in Fig.~\ref{fig:sampleEvolSphere}, as the influence
of the noise is relatively large, compared to the diffusive terms.

\begin{figure}
		\centering
		\subfigure[CH]{
			\label{fig:hist_CH}
			\includegraphics[width=0.45\textwidth]{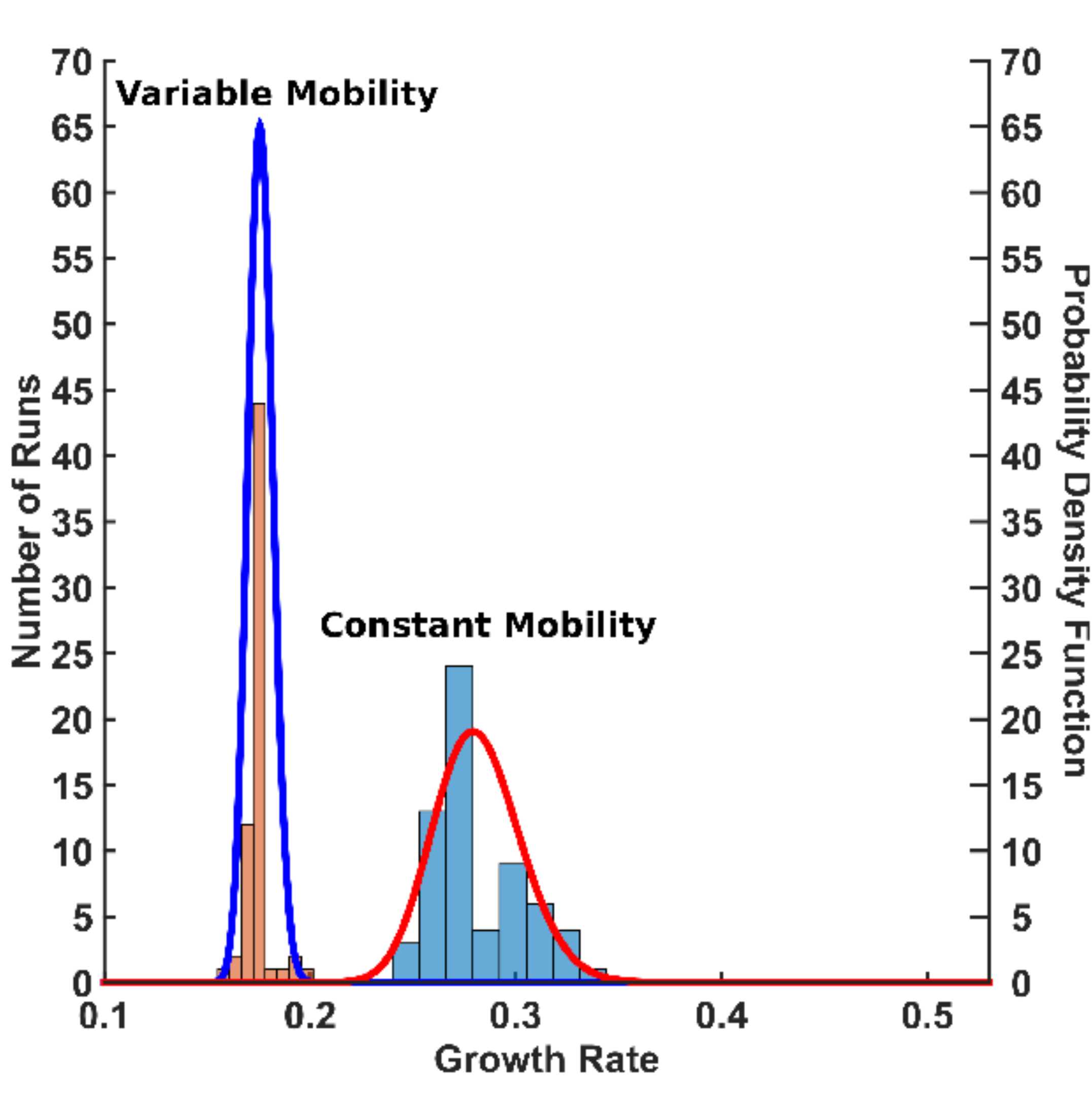} 
		} 
		\hfill
		\subfigure[CHC: $\sigma=10^{-9}$]{
			\label{fig:hist_N-9}
			\includegraphics[width=0.45\textwidth]{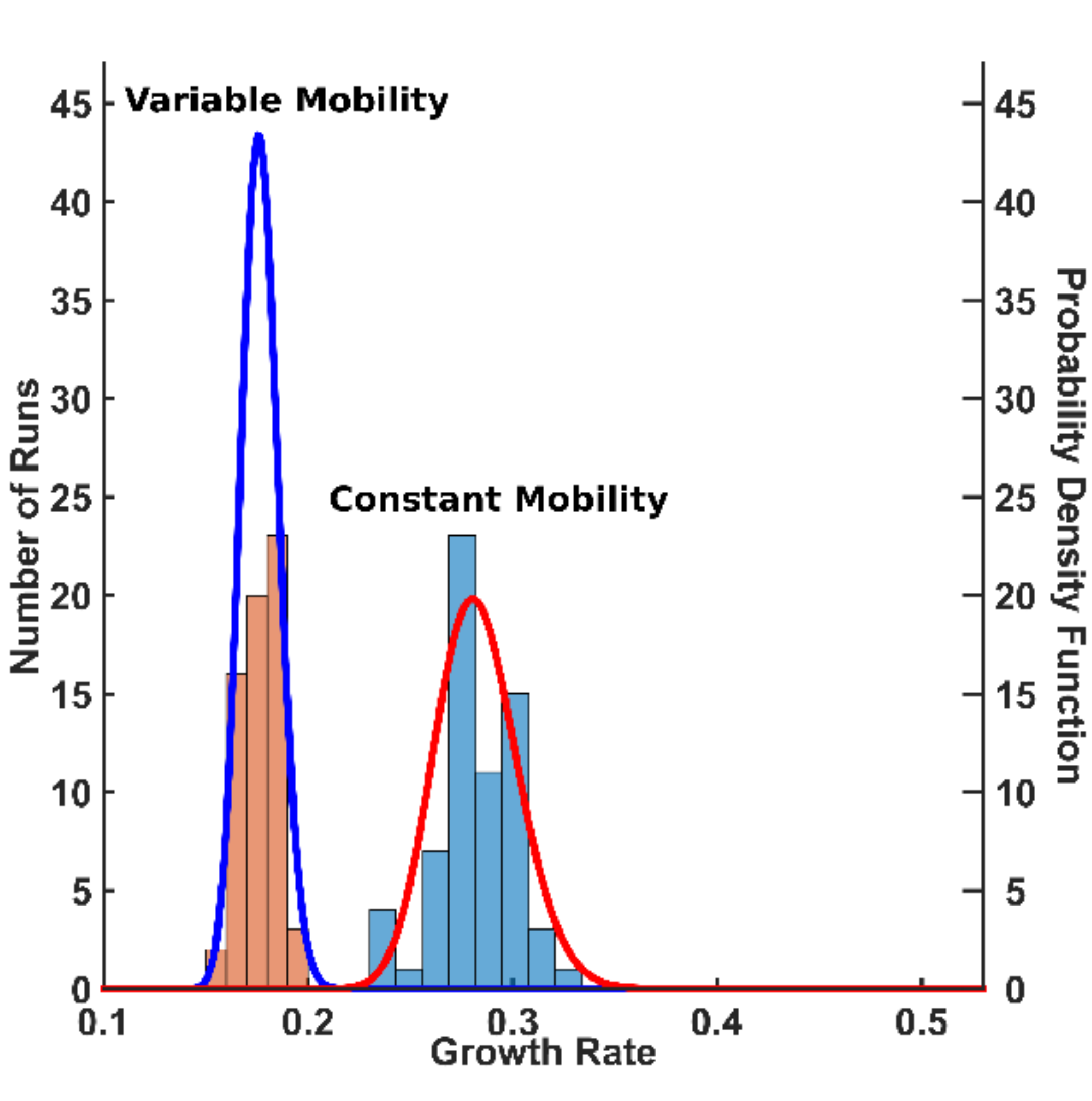} 
		}  
		\\
		\subfigure[CHC: $\sigma=10^{-7}$]{
			\label{fig:hist_N-7}
			\includegraphics[width=0.45\textwidth]{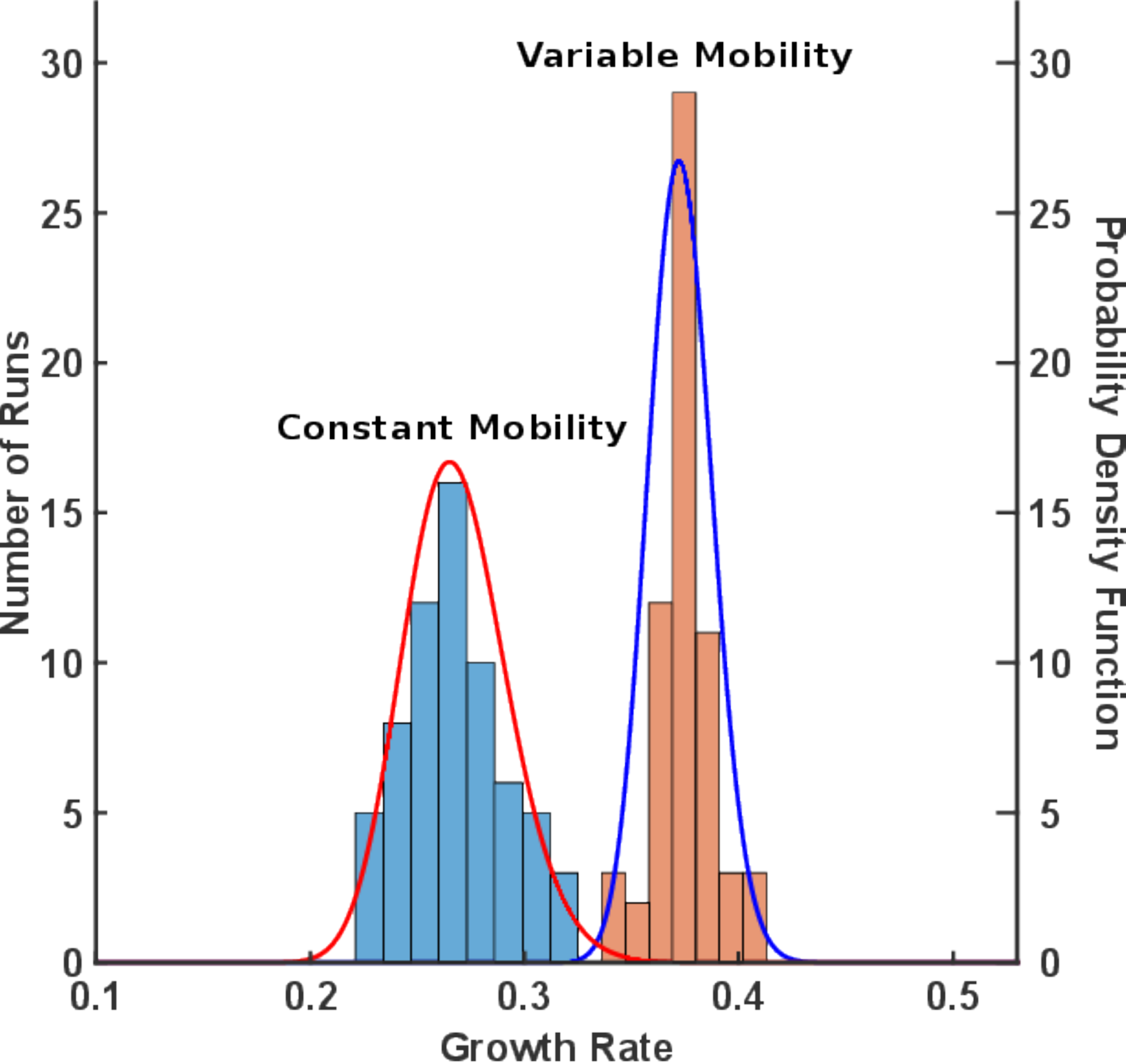} 
		}
		\hfill
		\subfigure[CHC: $\sigma=10^{-5}$]{
			\label{fig:hist_N-5}
			\includegraphics[width=0.45\textwidth]{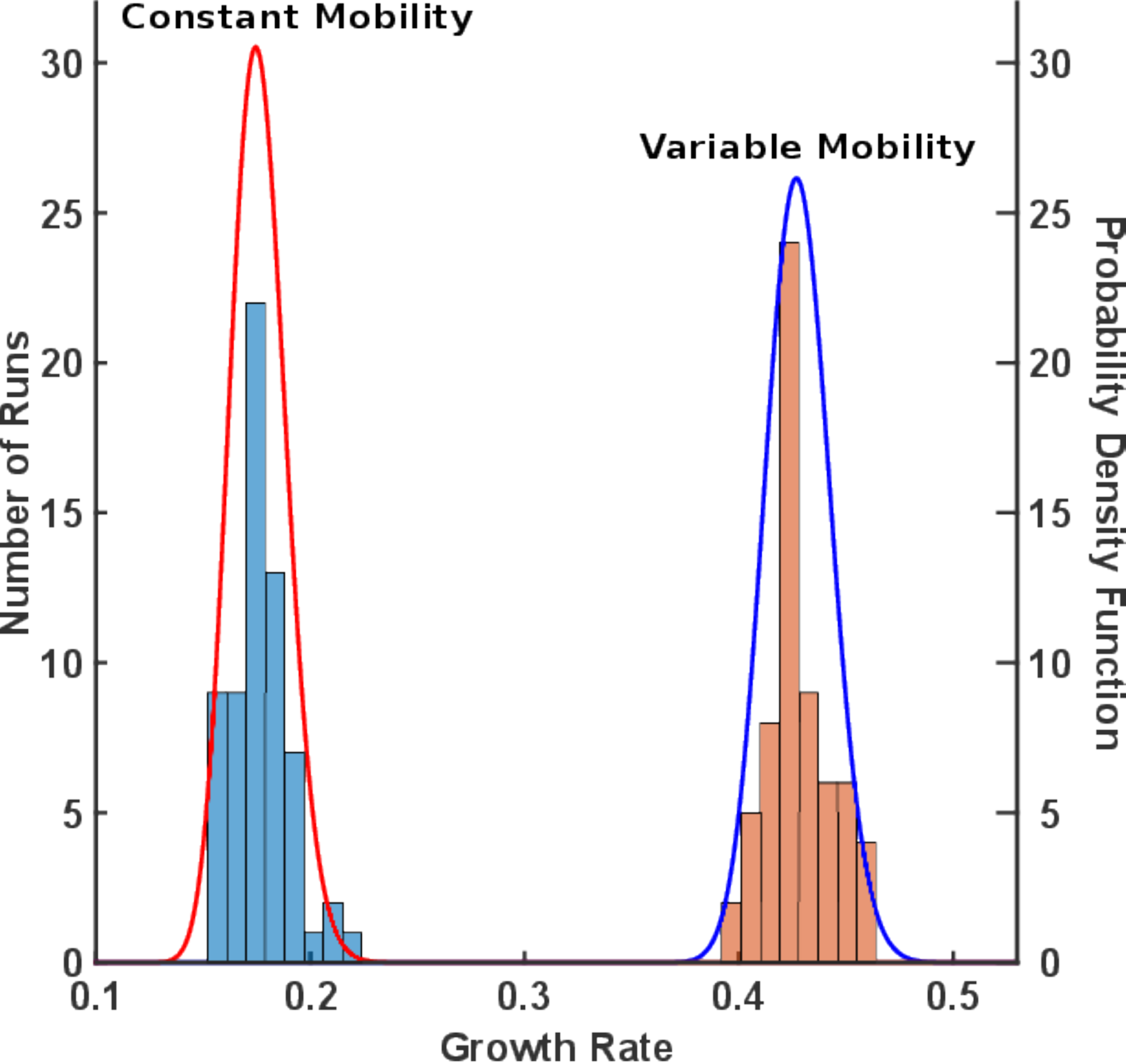} 
		}  
	  \caption{Histograms for the growth rate of the characteristic length for 
		CH and CHC systems using 64 realizations for constant and variable
		mobilities. A log-normal distribution function is fit on the
		results.}
	  \label{fig:sphereHist}
	\end{figure}

The complete CH and CHC results for various noise intensity levels can be seen
in the histograms shown in Fig.~\ref{fig:sphereHist}.  Using this data a probability
density function is fit and also shown on the histograms. As the growth rate can
never be negative, only non-negative distributions were explored. The log-normal
distribution gave a good qualitative fit and as the null hypothesis also passed
the Kolmogorov-Smirnov test at 5\% level of
significance~\cite{chakravati1967handbook} for all situations, it is chosen to
be an appropriate fit.

	\begin{table}
		\caption{Parameter estimates for the log-normal distribution
		function that fits the histogram of the growth rate for the 
		CH and CHC
		systems with constant and variable mobilities.}
		\label{table:paraestimates}
		\begin{center}			
			\begin{tabular}{|>{\centering}m{1.2cm} | >{\centering}m{2.0cm} | >{\centering}m{1.2cm} | >{\centering}m{2.0cm} | >{\centering}m{2.0cm} | }
				\hline 
				Model & Mobility & Noise & $\mu$ & $\sigma_{pdf}$\tabularnewline
				\hline 
				\hline 
				\multirow{2}{*}{CH} & 1 & -- & $-1.2707$ & $0.0748$ \tabularnewline
				\cline{2-5} 
				 & $f(1-f)$ & -- & $-1.7384$ & $0.0349$\tabularnewline
				\hline 
				\multirow{6}{*}{CHC} & \multirow{3}{*}{1} & $10^{-9}$ & $-1.2664$ & $0.0716$\tabularnewline
				\cline{3-5} 
				 &  & $10^{-7}$ & $-1.3237$ & $0.0902$\tabularnewline
				\cline{3-5} 
				 &  & $10^{-5}$ & $-1.7339$ & $0.0747$\tabularnewline
				\cline{2-5} 
				 & \multirow{3}{*}{$f(1-f)$} & $10^{-9}$ & $-1.7358$ & $0.0523$\tabularnewline
				\cline{3-5} 
				 &  & $10^{-7}$ & $-0.9900$ & $0.0402$\tabularnewline
				\cline{3-5} 
				 &  & $10^{-5}$ & $-0.8498$ & $0.0357$\tabularnewline
				\hline 
			\end{tabular}
		\end{center}
	\end{table}

The probability density function of the log-normal distribution is given by
\begin{align}
f(x|\mu,\sigma_{pdf})=\frac{1}{x\sigma_{pdf}\sqrt{2\pi}}\exp \left(\frac{-(\ln
x-\mu)^2}{2\sigma_{pdf}^2}\right)\qquad\text{with}\qquad x>0,
\end{align}
where $x$ is the data, $\mu$ is the log mean, and $\sigma_{pdf}$ is log
standard deviation with the support $-\infty<\mu<\infty$ and $\sigma_{pdf} \geq
0$. The log mean and the log standard deviation can be estimated from the
probability density function. See Table~\ref{table:paraestimates} for the fitted parameters. 
It is noted as the noise intensity goes down the log mean value approaches 
the result for the Cahn-Hilliard system. 

\section{Dumbbell Interface\label{sec:StatOnDumbbell}} 
In this section the phase segregation on a dumbbell is examined. 
The shape are two spheres connected by a cylinder. Each sphere has a radius of 0.75
and are centered at $(-1.125,0,0)$ and $(1.125,0,0)$ while the cylinder connecting 
the two sphere has a radius of $0.375$.
The average concentration is $0.3$ while the initial 
random perturbation has a magnitude of $0.01$.
To focus on the influence of the underlying geometry, 
only the constant-mobility case is considered.

\begin{figure}
	\centering
	\begin{tabular}{>{\centering}m{0.8cm} >{\centering}m{4.5cm} >{\centering}m{4.5cm}}
	    $t$ & $\sigma=0$ & $\sigma=10^{-5}$ \tabularnewline				
		0.1 & \includegraphics[width=4.5cm]{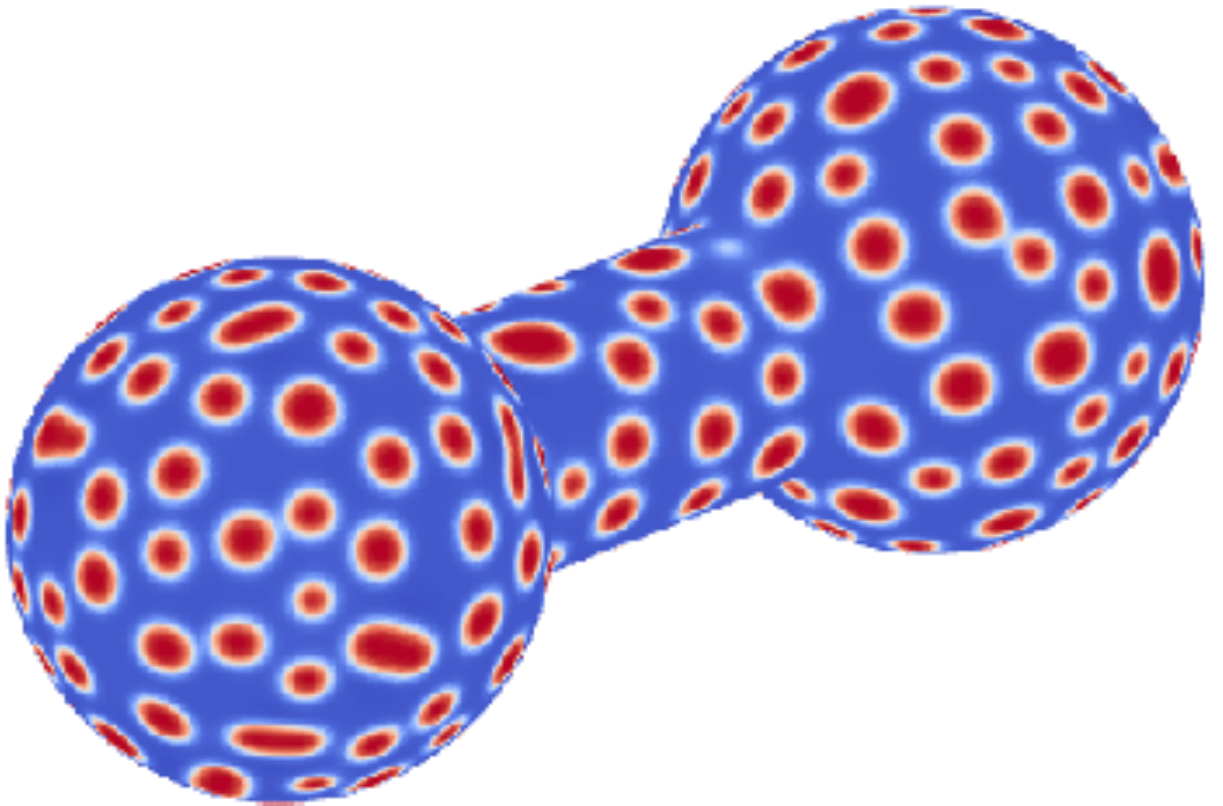} &\includegraphics[width=4.5cm]{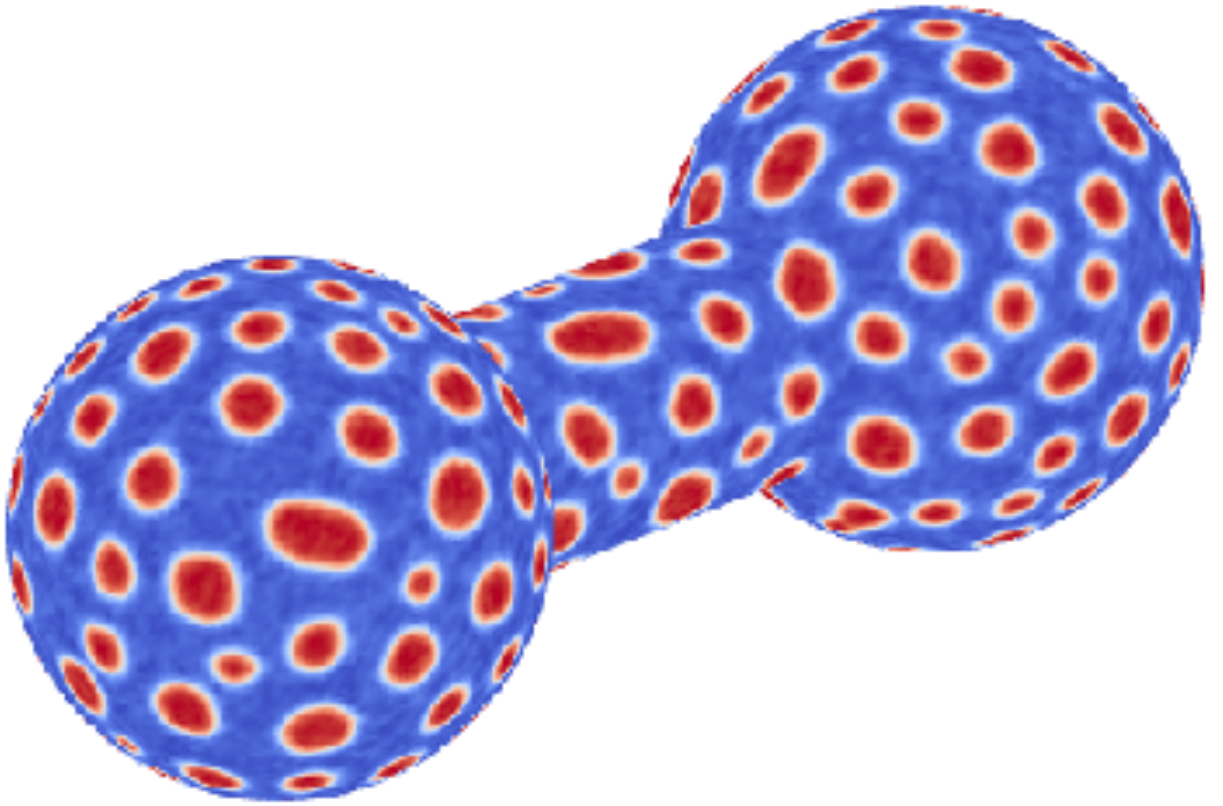}\tabularnewline
		1.0 & \includegraphics[width=4.5cm]{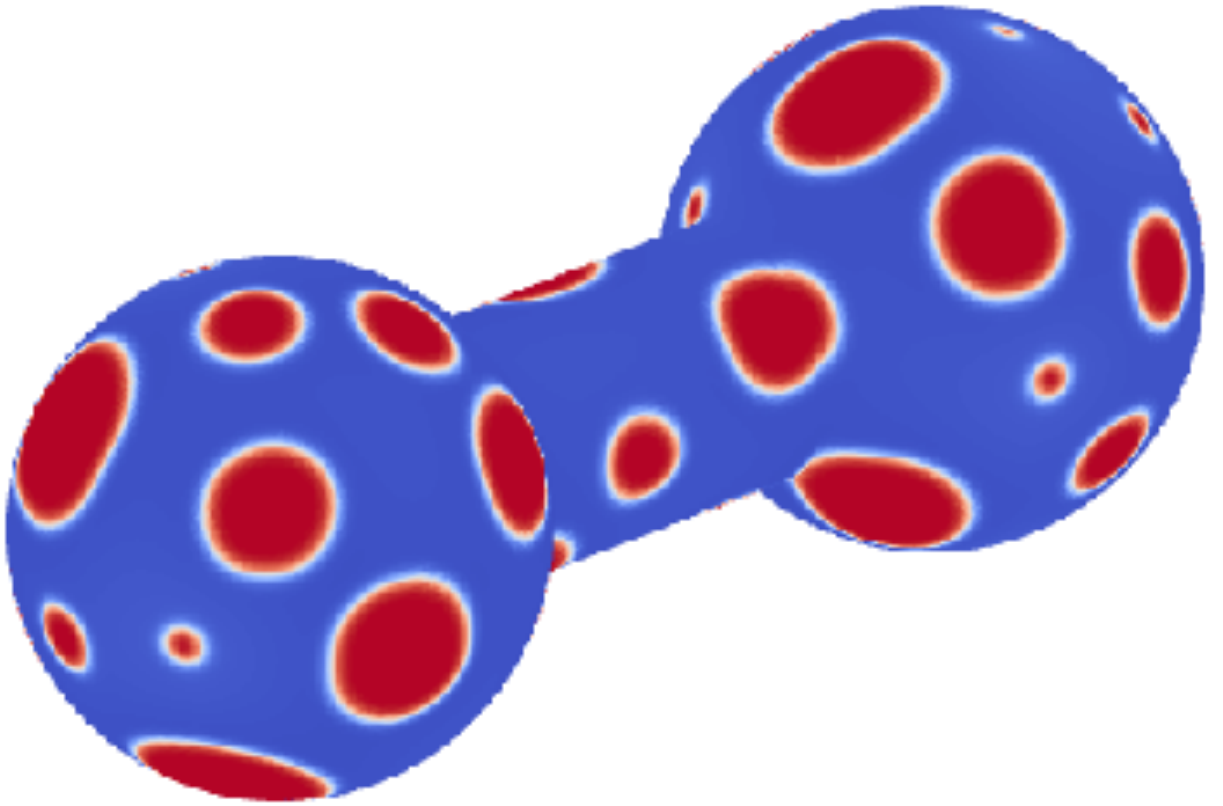} & \includegraphics[width=4.5cm]{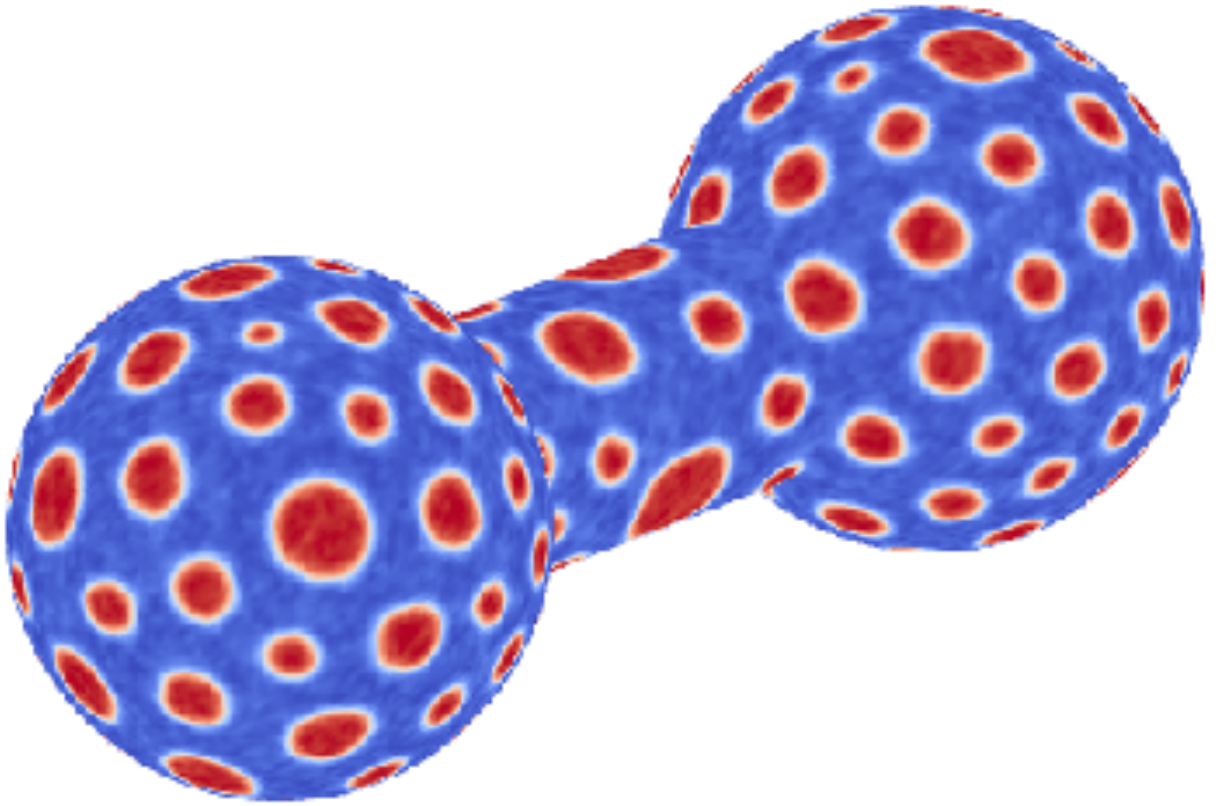}\tabularnewline
		5.0 & \includegraphics[width=4.5cm]{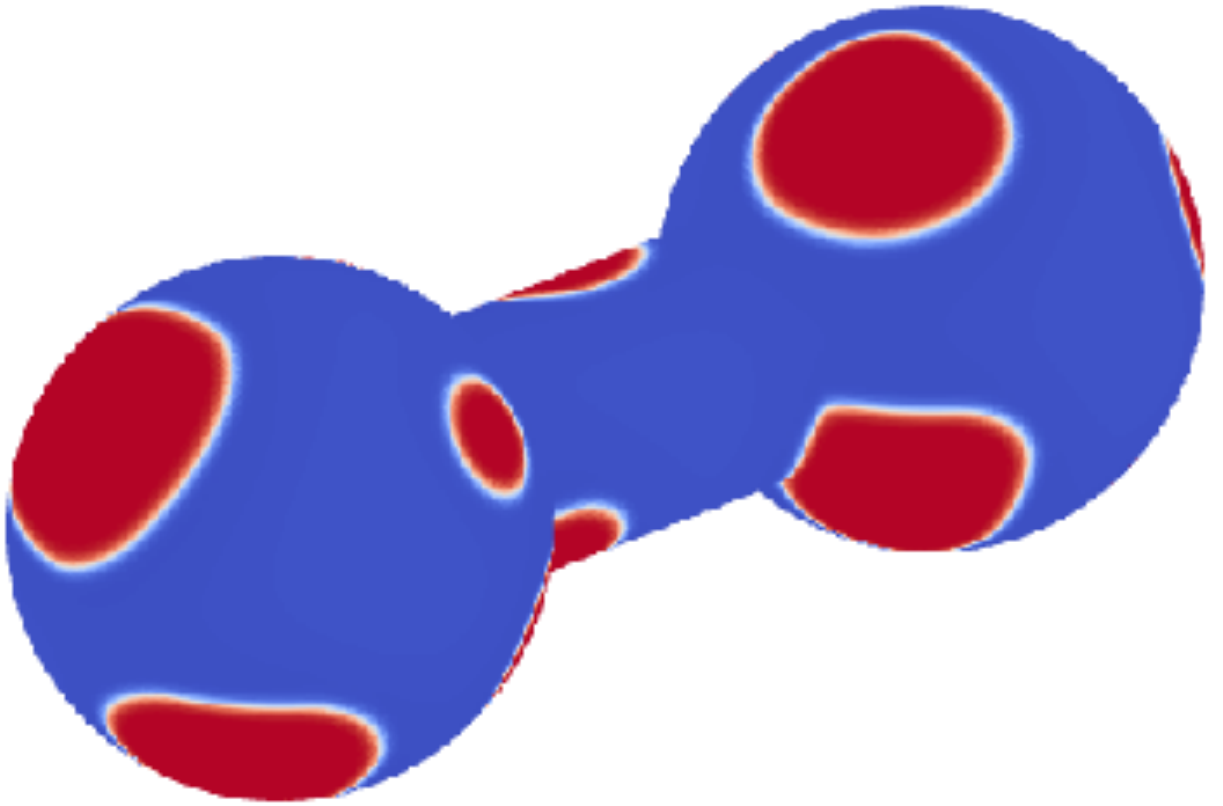}&\includegraphics[width=4.5cm]{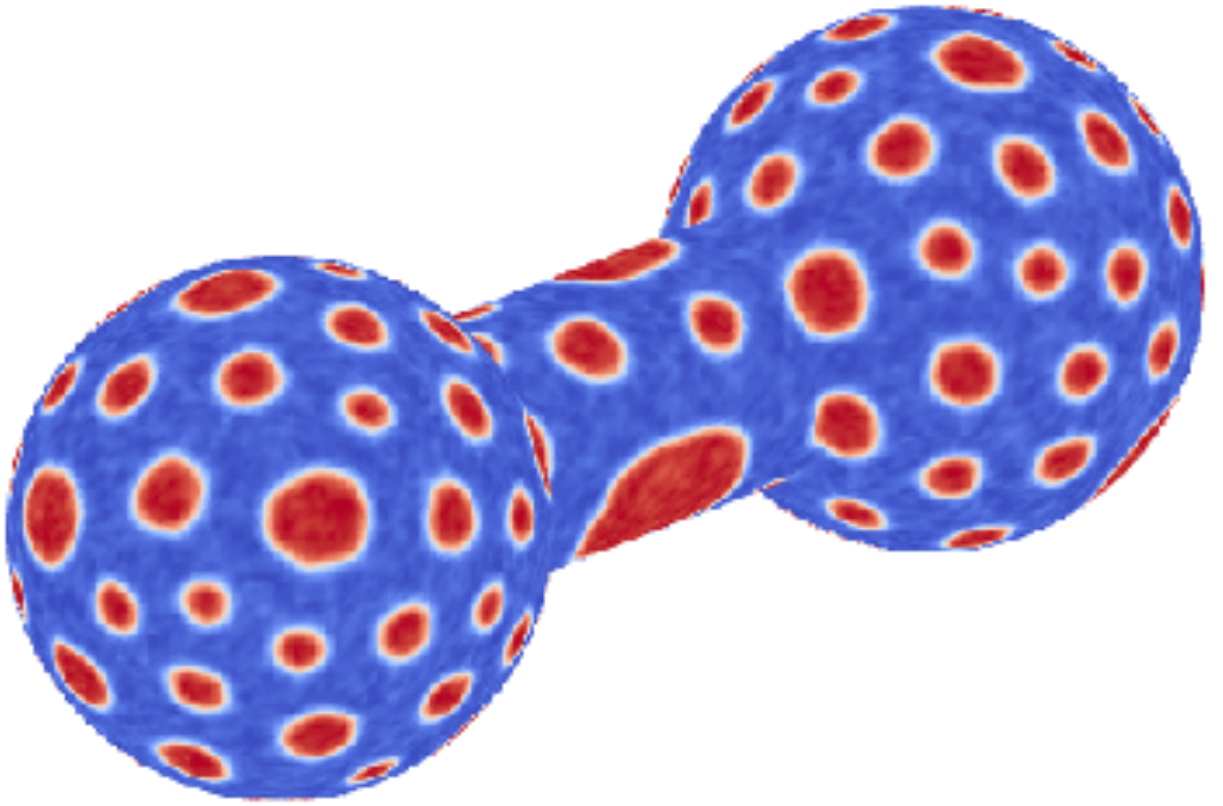}\tabularnewline
    	10.0 & \includegraphics[width=4.5cm]{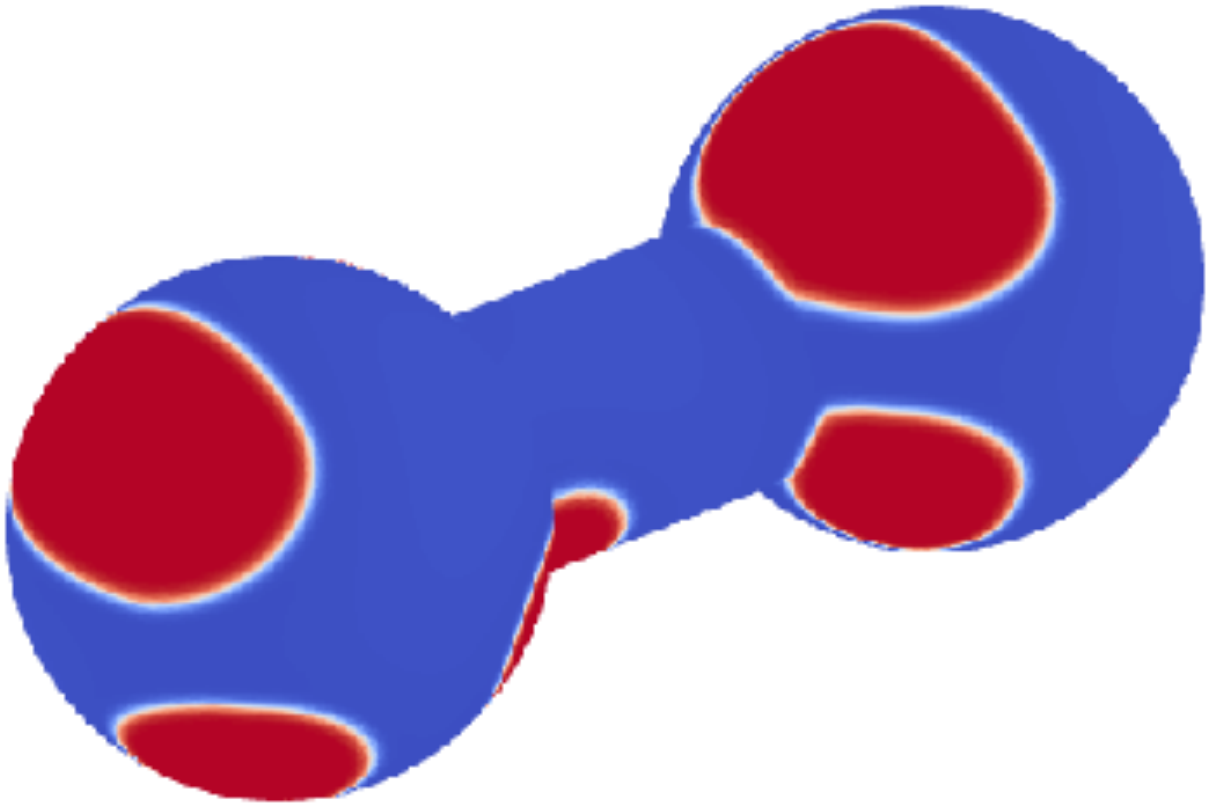} &\includegraphics[width=4.5cm]{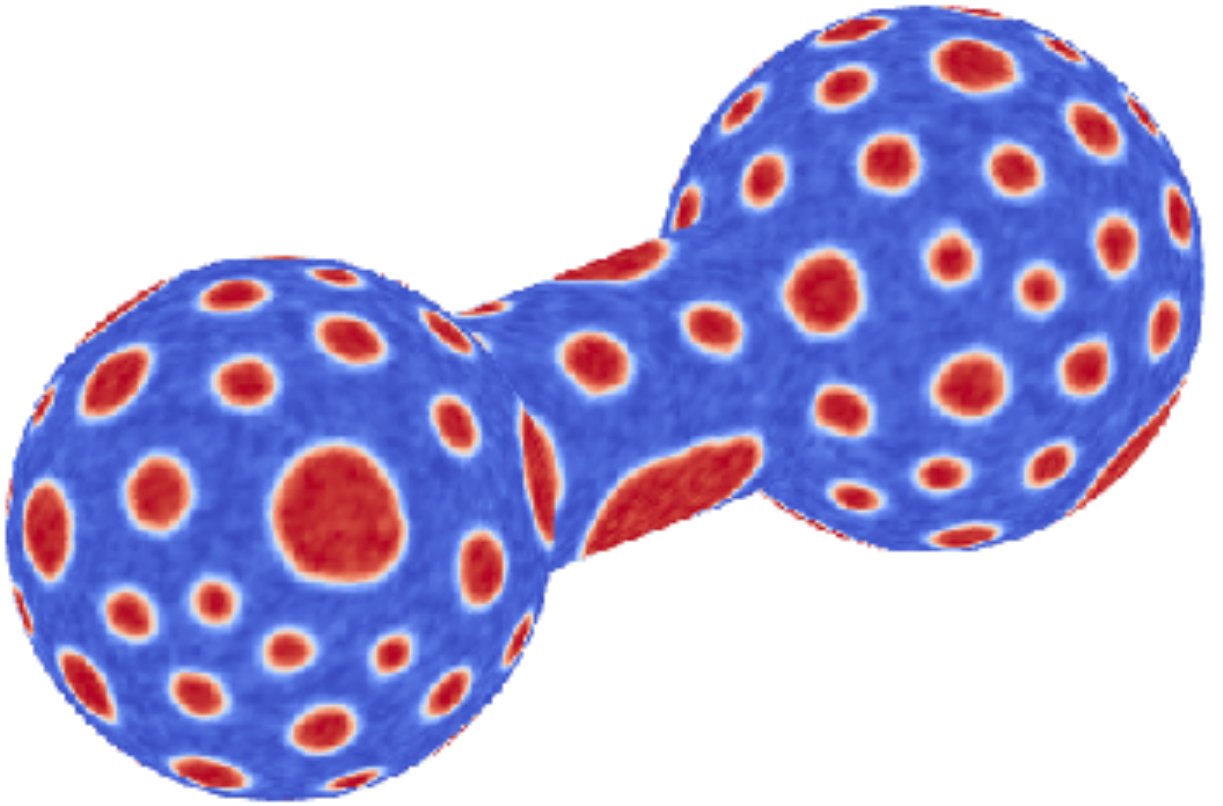}\tabularnewline
	\end{tabular}
  \caption{Evolution for Cahn-Hilliard ($\sigma=0$) and Cahn-Hilliard-Cook ($\sigma=10^{-5}$) model with
  constant mobility on a dumbbell.}
  \label{fig:sampleEvolDumbbell}
\end{figure}

First consider the Cahn-Hilliard system, Fig.~\ref{fig:sampleEvolDumbbell}. 
As before, the initially well-mixed system quickly segregates
into many, small domains. Over time, the domains begin to grow and coarsen, until a small
number of large domains exist. Next consider the Cahn-Hilliard-Cook system
with $\sigma=10^{-5}$, again Fig.~\ref{fig:sampleEvolDumbbell}. 
As with the CH system, the well-mixed system quickly
segregates into small domains. Unlike the CH case, the growth rate of the domains
using the CHC model on the dumbbell is greatly reduced, as seen by the many, small domains
at a time of $t=10$. This indicates that the underlying geometry has a large influence
on the coarsening process in the presence of noise.

\begin{figure}
	\centering
	\subfigure[Cahn Hilliard Model]{
		\label{fig:CL_MMM_CH}
		\includegraphics[width=6.0cm]{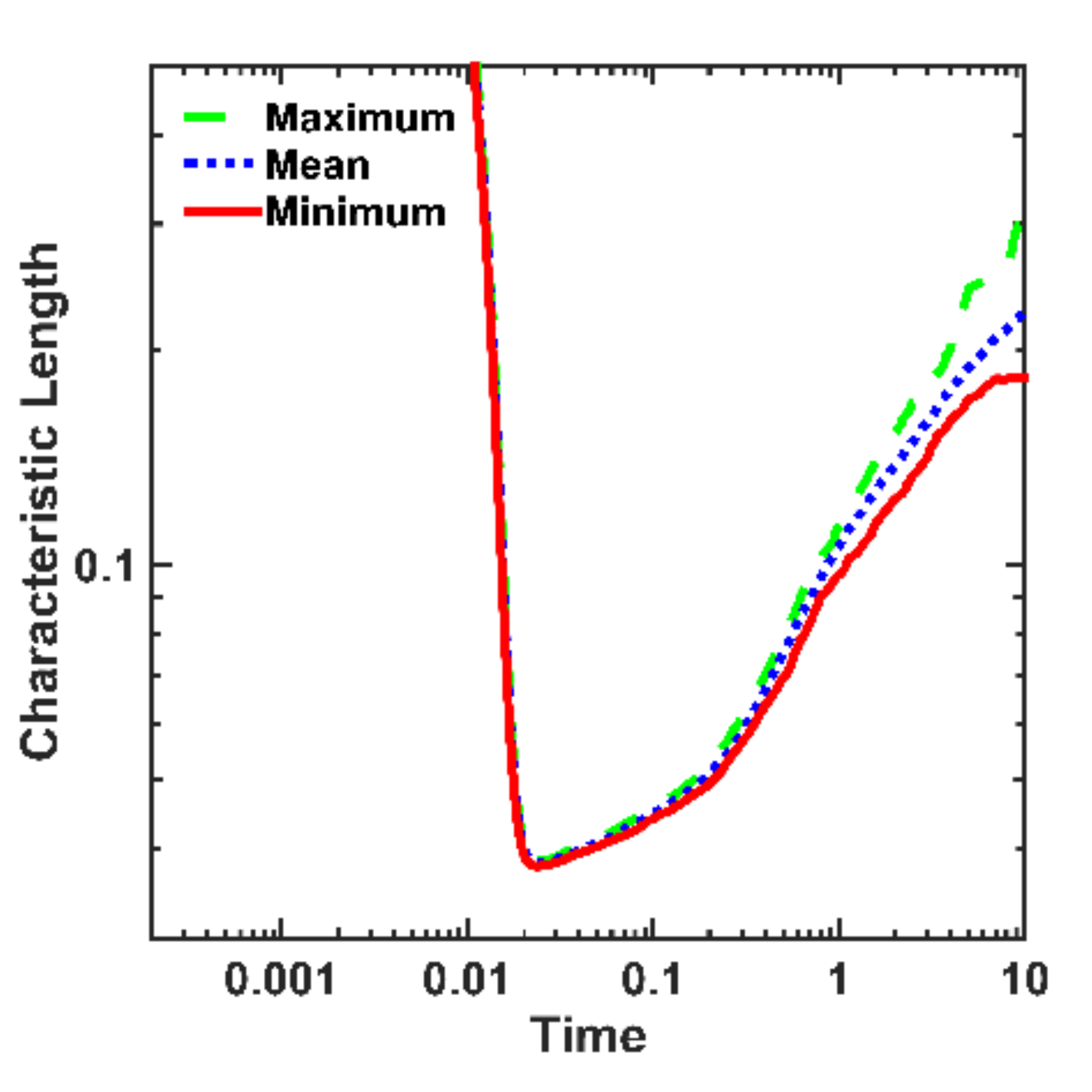} 
	}
	\hfill
	\subfigure[Cahn Hilliard Cook Model]{
		\label{fig:CL_MMM_CHC}
		\includegraphics[width=6.0cm]{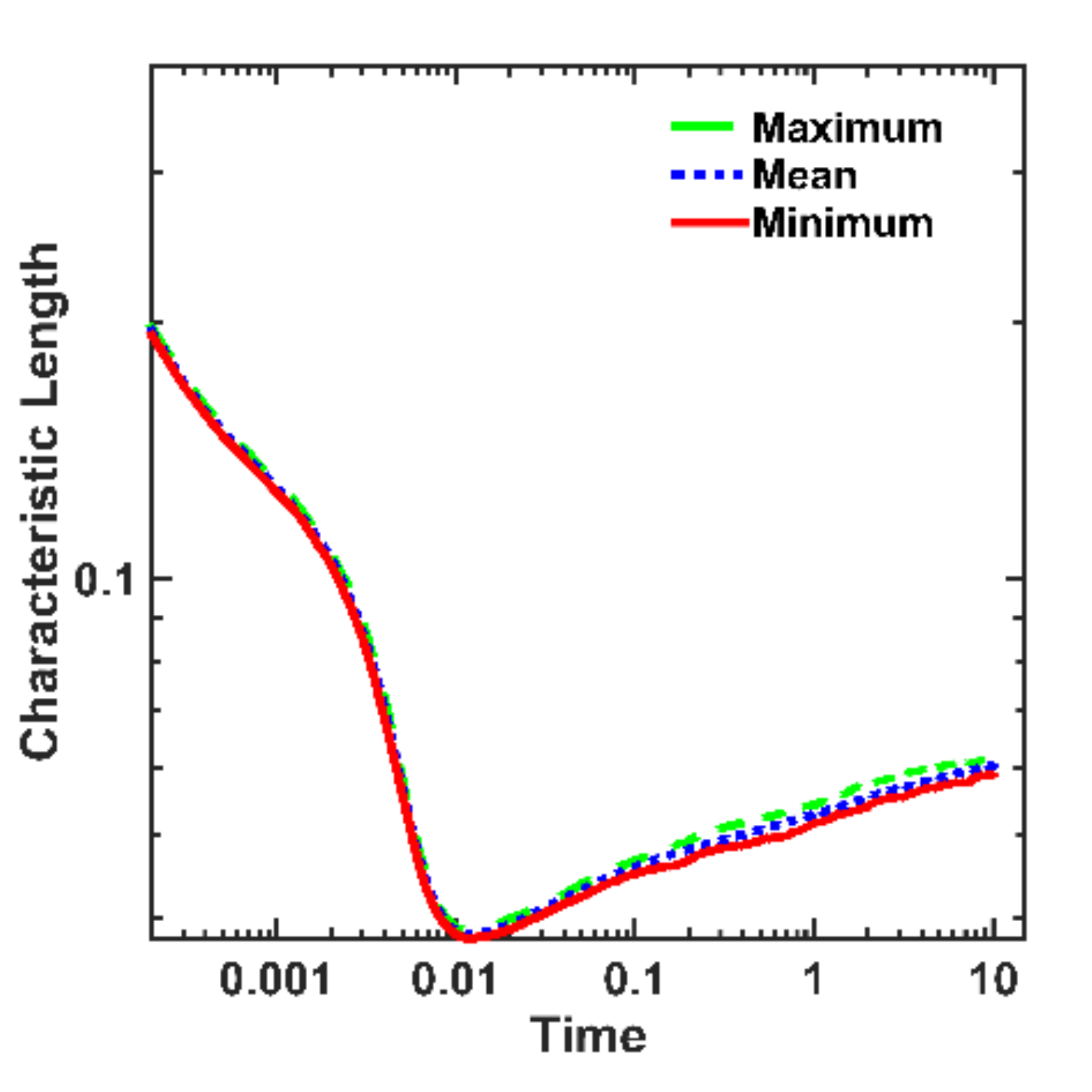} 
	}
  \caption{The minimum, maximum, and mean characteristic lengths for the Cahn-Hilliard and Cahn-Hilliard-Cook system for the 64 realizations using constant mobility on a dumbbell.}
  \label{fig:CL_MMM_dumbbell}
\end{figure}

\begin{figure}
	\centering
	\subfigure[CH]{
		\label{fig:hist_CH_dumbbell}
		\includegraphics[width=0.45\textwidth]{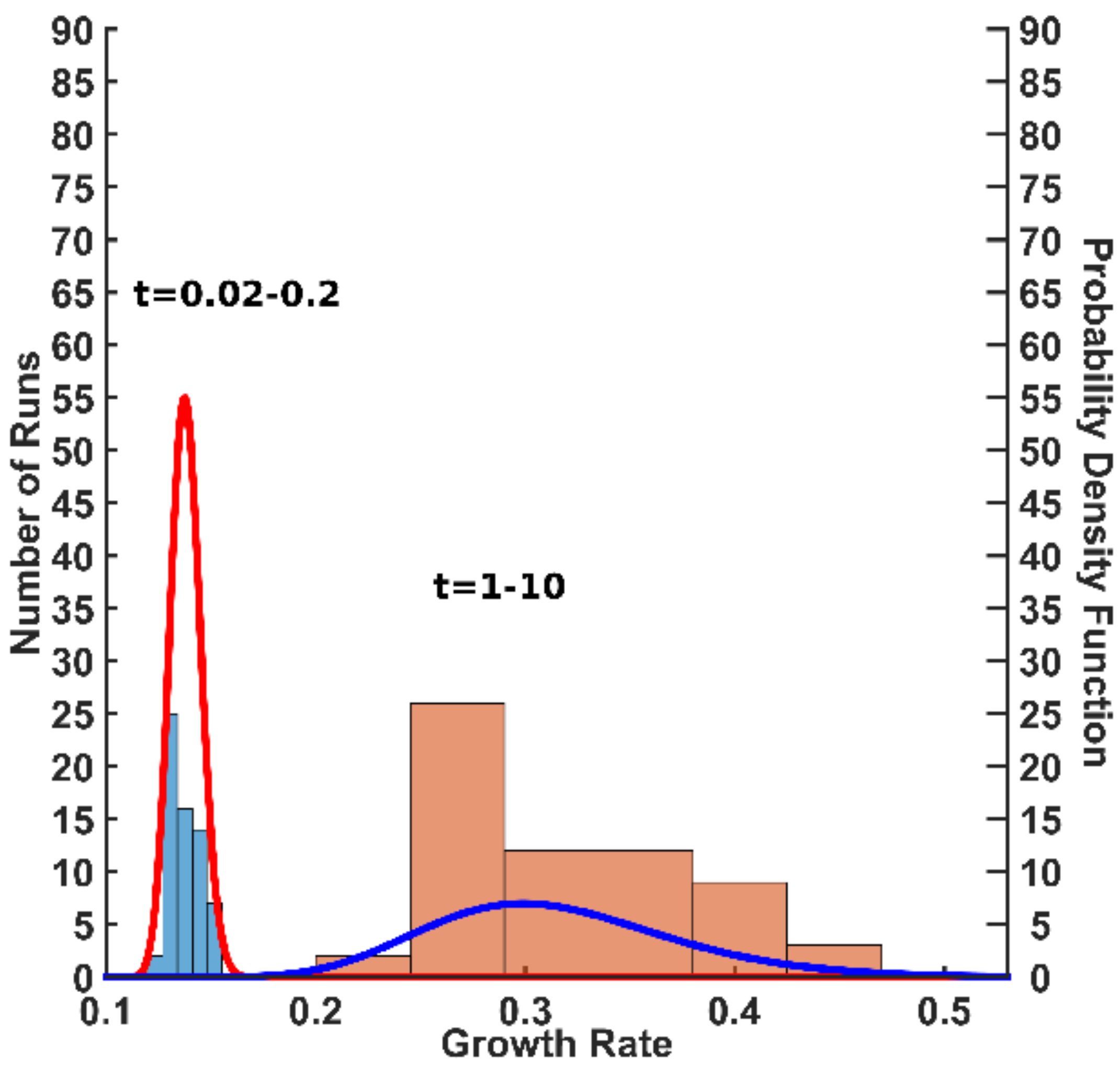} 
	} 
	\hfill
	\subfigure[CHC: $\sigma=10^{-5}$]{
		\label{fig:hist_CHC_dumbbell}
		\includegraphics[width=0.45\textwidth]{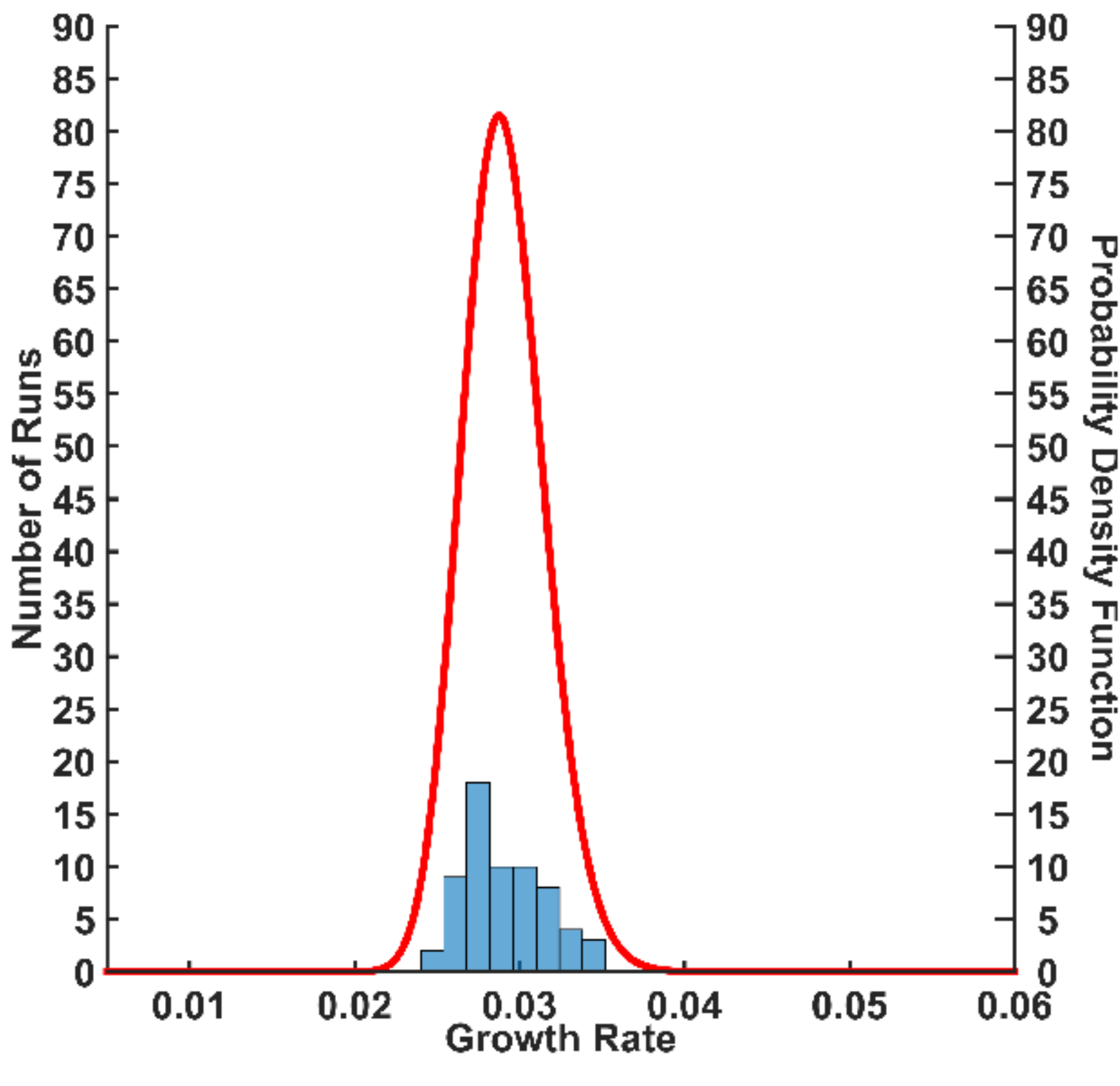} 
	}  
  \caption{Histograms for the growth rate of the characteristic length for 
	CH and CHC systems using 64 realizations for constant mobility on a
	dumbbell. A log-normal distribution function is fit on the
	results.}
  \label{fig:dumbbellHist}
\end{figure}

As before, a statistical analysis of the growth rate is done for the Cahn-Hilliard and
the Cahn-Hilliard-Cook model using $\sigma=10^{-5}$, using 64 realizations for each
model. The minimum, maximum, and average characteristic length for each time
step is plotted in Fig.~\ref{fig:CL_MMM_dumbbell}. 
As expected, the characteristic length for both the CH and CHC model
increases over time, with the spread between the minimum growth rate and the maximum
growth rate also increasing with time. 
Additionally, the mean, standard deviation, and coefficient of variation
are shown in Table~\ref{table:mean_SD_CV_dumbbell}
while the corresponding histograms are shown in
Fig.~\ref{fig:dumbbellHist}.

\begin{table}
	\caption{Statistics on the growth rate for Cahn-Hilliard and Cahn-Hilliard-Cook
		model with constant mobility of a dumbbell}
	\label{table:mean_SD_CV_dumbbell}
	\begin{center}		
		\begin{tabular}{|>{\centering}m{1.2cm} | >{\centering}m{2.0cm} | >{\centering}m{1.2cm} | >{\centering}m{2.0cm} | >{\centering}m{2.0cm} | >{\centering}m{3.0cm} |  }		
			\hline
			Model & Time & Noise & Mean & Standard Deviation  & Coefficient of Variation\tabularnewline			      						
			\hline
			\hline
			\multirow{2}{*}{CH} & 0.02 - 0.2 & -- & $0.1380$ & $0.0073$ & $0.0529$\tabularnewline
			\cline{2-6} 
				 & $1 -10$ & -- & $0.3152$ & $0.0607$ & $0.1926$\tabularnewline
			\hline
			\multirow{1}{*}{CHC} & \multirow{1}{*}{0.1-10} 
			  & $10^{-5}$ & $0.0581$ & $0.0050$& $0.0861$ \tabularnewline
			\hline
		\end{tabular}
	\end{center}		
\end{table}

\begin{figure}
	\centering
	\begin{tabular}{>{\centering}m{0.5cm} >{\centering}m{5.0cm} >{\centering}m{5.0cm}}
	Run & \multicolumn{1}{c}{$t=0.5$}     & \multicolumn{1}{c}{$t=10$} \tabularnewline		
		I & \includegraphics[width=4.5cm]{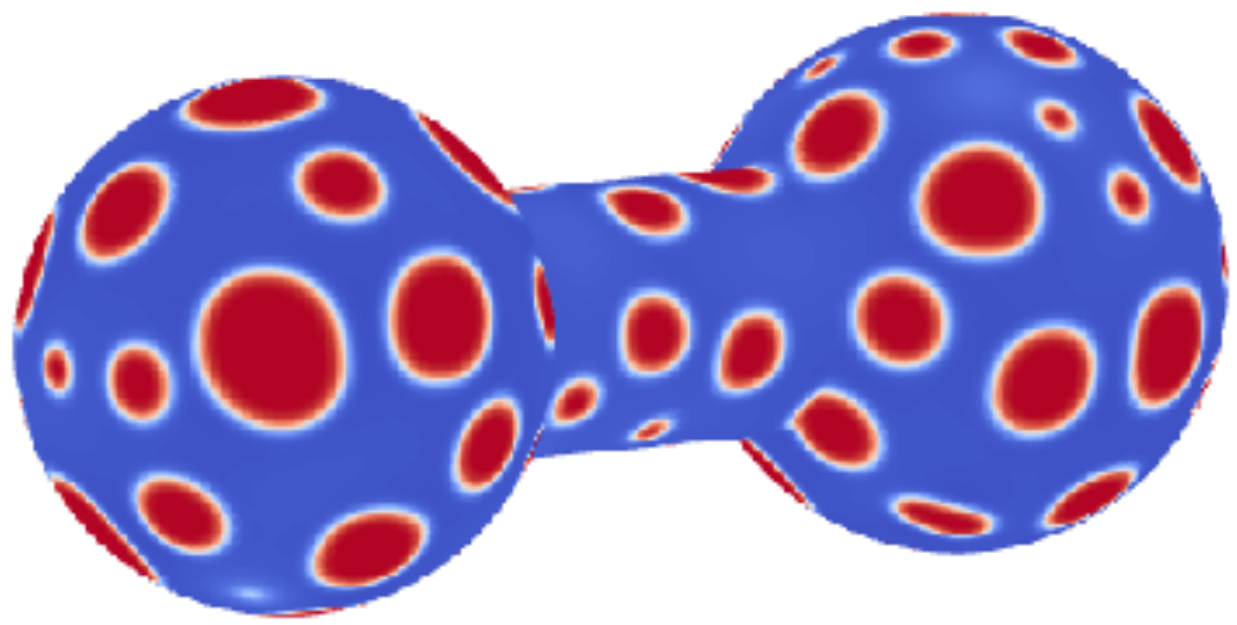} &\includegraphics[width=4.5cm]{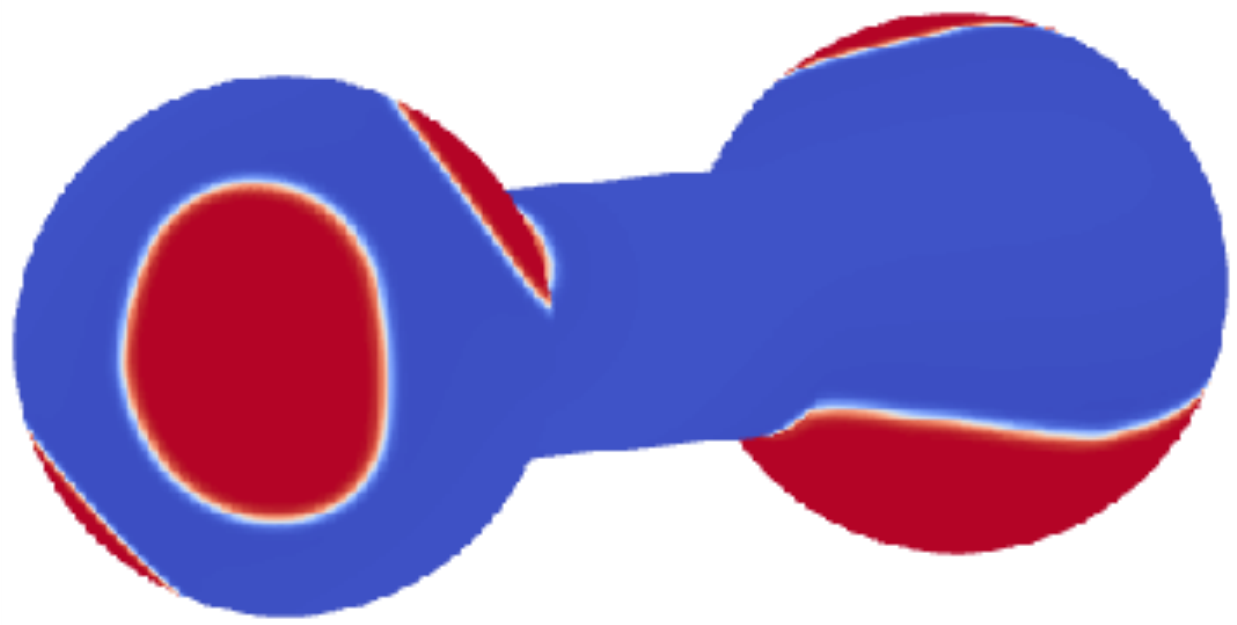}\tabularnewline
		II & \includegraphics[width=4.5cm]{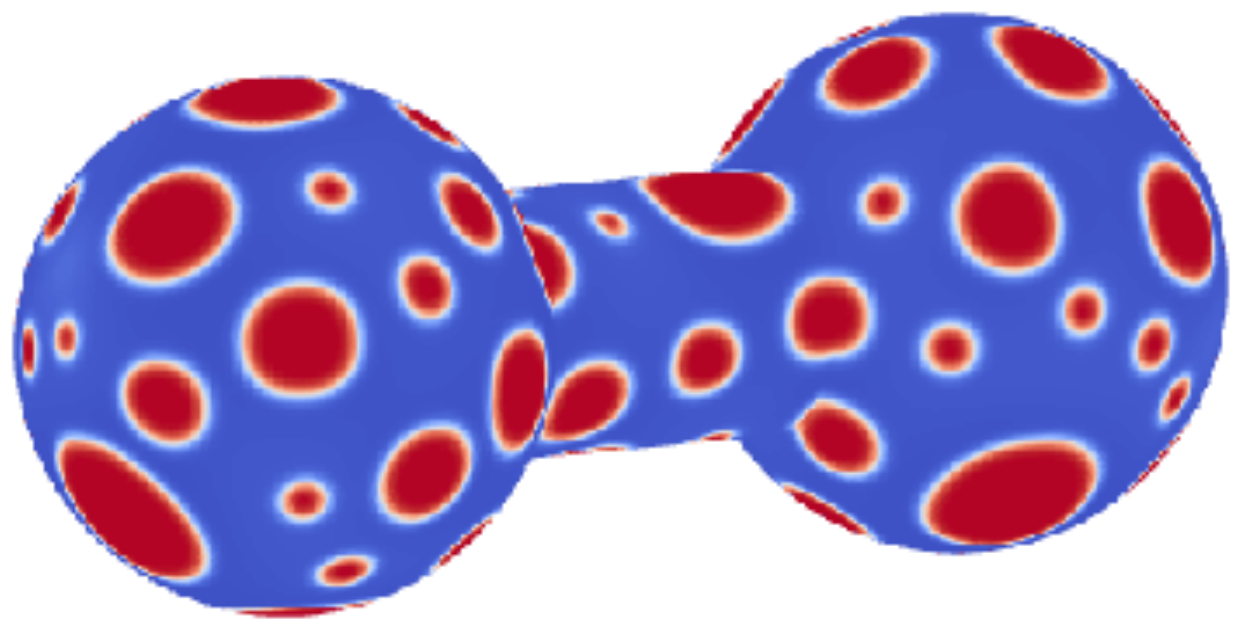} & \includegraphics[width=4.5cm]{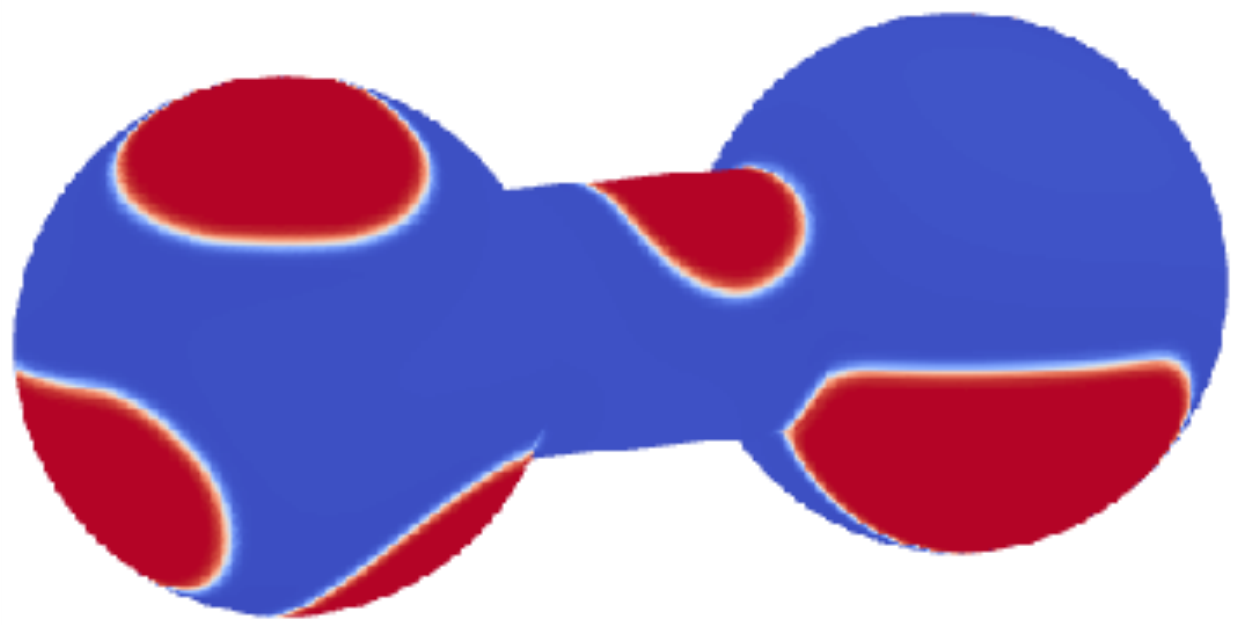}\tabularnewline
	\end{tabular}
  \caption{Evolution for Cahn-Hilliard model with constant mobility on a dumbbell, for two realizations.}
  \label{fig:DumbbellTwoRuns}
\end{figure}

When considering the CH model, 
there are two growth rates apparent in the system. 
The first extends from $t=0.02$ to $t=0.2$, while the second extends
from $t=0.2$ onward. It is therefore appropriate to consider
two growth rates, with $\bar{\alpha}=0.1380$ from $t=0.02$ to $t=0.2$
and $\bar{\alpha}=0.3152$ from $t=0.2$ onwards. When considering 
this latter regime, the standard deviation and coefficient of variation
is quite large, as shown in Table~\ref{table:mean_SD_CV_dumbbell}.
It is suspected that the underlying interface plays a large role in the large
variation in the growth rate at late time. To demonstrate this, 
consider two sample runs shown in Fig.~\ref{fig:DumbbellTwoRuns}.
At a time of $t=0.5$, both interfaces are well-covered by small domains,
with the distance between domains similar. At at time of $t=10$, 
the domains for Run I are predominantly on the two spheres, with no
domain on the connecting cylinder. In Run II, there is a domain on the
cylinder. The distance between domains plays a critical role in the coarsening
process, with a larger distance corresponding to a lower growth rate.
Unlike the spherical interface, the lack of full symmetry in the dumbbell
shape results in a growth rate which depends on the initial condition,
as that will determine where large domains will preferentially occur.

Now consider the Cahn-Hilliard-Cook system with noise intensity of $10^{-5}$.
From the sample result in Fig.~\ref{fig:sampleEvolDumbbell}
and Table~\ref{table:mean_SD_CV_dumbbell}, it is apparent
that the growth rate is very small, with a value of $\bar{\alpha}=0.0581$
and a small standard deviation of $0.005$. It is suspected that the larger
curvatures present in the dumbbell shape, coupled with the large noise 
intensity, results in this decrease in the growth rate.

%

\section{\label{sec:Conclusion} Conclusion}
In this work the Cahn-Hilliard-Cook model is solved on smooth interfaces
using a splitting method that converts the fourth-order partial differential 
equation into a two coupled second-order PDEs. The surface differential equations
are solved using the Closet Point Method, using a level-set Jet scheme to 
describe the interface.

These results indicate that the underlying surface plays a large role in the segregation
process, both in the presense and in the absence of thermal fluctuations/noise.
When assuming constant surface mobility, the presense of noise slows the 
coarsening rate of domains, with the growth rate increasing with a decrease in the 
noise magnitude. Surprisingly, the presense of noise actually increases
the growth rate when assuming a degenerate mobility. This is most likely due to the 
fact that the diffusive evolution contributions decay at a faster rate than the noise 
contributions as one moves away from the interface.

When examining a spherical interface in the absence of noise, 
the overall growth rate is slightly lower
than that predicted for flat, two-dimensional surfaces. The inclusion of noise 
further decreases the observed growth rate.
The use of a 
dumbbell shape, with a spatially varying curvature, further influences the evolution.
Assuming no noise, two growth regimes were identified on the dumbbell, with the 
final growth rate highly dependent on the initial condition.
Inclusion of noise for the dumbbell shape dramatically decreased the 
growth rate.

\begin{acknowledgments}
This work has been supported by the National Science Foundation through the Division of Chemical, Bioengineering, Environmental, and Transport Systems Grant \#1253739.
\end{acknowledgments}

\bibliography{chcos}

\end{document}